\documentclass[11pt]{article}

\usepackage[caption=false, labelformat=simple, listofformat=subsimple, labelfont=default, margin=5pt, justification=raggedright]{subfig}

	\makeatletter
		\renewcommand{\p@subfigure}{}
	\makeatother

\usepackage{amsmath,amssymb}
\usepackage{array}
\usepackage{float}
\usepackage{amsmath,amsfonts,graphicx,bm}
\usepackage{amsmath,amsfonts,graphicx,bbm,multicol}

\usepackage{graphicx}
\graphicspath{{./figures/}}
\usepackage{hyperref}
\usepackage{mcite}

\newcommand{\mP}{{\bar M}_{P}}
\newcommand{\ttbar}{\ensuremath{t\bar{t}}}
\newcommand{\pt}{\ensuremath{p_{\mathrm{T}}}}
\newcommand{\GeV}{\ensuremath{\; \mathrm{GeV}}}
\newcommand{\TeV}{\ensuremath{\; \mathrm{TeV}}}

\def\beq{\begin{equation}}
\def\eeq{\end{equation}}
\def\bea{\begin{eqnarray}}
\def\eea{\end{eqnarray}}
\def\bmat{\begin{pmatrix}}
\def\emat{\end{pmatrix}}

\newcommand{\ignore}[1]{}
\newcommand{\be}{\begin{equation}} \newcommand{\ee}{\end{equation}}
\newcommand{\ba}{\begin{eqnarray}} \newcommand{\ea}{\end{eqnarray}}
\newcommand{\nn}{\nonumber} \renewcommand{\bf}{\textbf}

\def\slashb#1{\setbox0=\hbox{$#1$}#1\hskip-\wd0\dimen0=5pt\advance
        \dimen0 by-\ht0\advance\dimen0 by\dp0\lower0.5\dimen0\hbox
          to\wd0{\hss\sl/\/\hss}}
\def\sl#1{#1 \!\!\! \!\!   \!\! \slash}

\begin{document}

\def\beq{\begin{equation}}
\def\eeq{\end{equation}}
\def\bea{\begin{eqnarray}}
\def\eea{\end{eqnarray}}
\def\bmat{\begin{pmatrix}}
\def\emat{\end{pmatrix}}

% user definitions

\def\bbbar{\ensuremath{b\bar{b}}}
\def\ttbar{\ensuremath{t\bar{t}}}

%%%%%%%%%%
%%%%%%%%%%    Title page
%%%%%%%%%%

\thispagestyle{empty}
\vspace{20pt}
\font\cmss=cmss10 \font\cmsss=cmss10 at 7pt

\begin{flushright}
\today \\
%
%UMD-PP-10... \\
%
\end{flushright}

\hfill
\vspace{20pt}

\begin{center}
{\Large \textbf
{Warped Extra Dimensional Benchmarks for Snowmass 2013}
}
\end{center}

\vspace{15pt}

\begin{center}
{\large 
Kaustubh Agashe$\, ^{a}$, Oleg Antipin$\, ^{b}$, Mihailo Backovi\'{c}$\, ^{c}$, 
Aaron Effron$\, ^{d}$, Alex Emerman$\, ^{e}$, Johannes Erdmann$\, ^{d}$, 
Tobias Golling$\, ^{d}$, Shrihari Gopalakrishna$\,^{f}$, 
Tuomas Hapola$\, ^{g}$, Shih-Chieh Hsu$\, ^{h}$,Jos\'{e} Juknevich$\, ^{c}$,  
Seung J. Lee$\, ^{i,j}$, Tanumoy Mandal$\, ^{f}$, August Miller$\, ^{k}$,
Edward Moyse$\, ^{k}$, 
Tuhin Subhra Mukherjee$\,^{f}$,
Chris Pollard$\, ^{l}$, 
Soumya Sadhukhan$\,^{f}$,
Daniel Whiteson$\, ^{m}$, Stephane Willocq$\, ^{k}$
}
\\
\vspace{15pt}
$^{a}$\textit{Maryland Center for Fundamental Physics,
     Department of Physics,
     University of Maryland,
     College Park, MD 20742, U.S.A.}

$^{b}$\textit{ CP$^{\hspace{.5mm}3}{-Origins}$ $\&$ the Danish Institute for Advanced Study {\rm{Danish IAS}},  University of Southern Denmark, Campusvej 55, DK-5230 Odense M, Denmark.}

$^{c}$\textit{Department of Particle Physics and Astrophysics, Weizmann Institute of Science,\\ Rehovot 76100, Israel}

$^{d}$\textit{Department of Physics,
    Yale University,
    New Haven, CT 06520, U.S.A.}

$^{e}$\textit{Department of Physics, Reed College, Portland, OR 97202, U.S.A.}

$^{f}$\textit{Institute of Mathematical Sciences (IMSc),
C.I.T Campus, Taramani, Chennai 600113, India}

$^{g}$\textit{Institute for Particle Physics Phenomenology,
Durham University,\\
South Rd, Durham DH1 3LE, UK}

$^{h}$\textit{Department of Physics, University of Washington Seattle, Seattle, WA 98195, U.S.A.}

$^{i}$\textit{Department of Physics, Korea Advanced Institute of Science and Technology, \\335 Gwahak-ro, Yuseong-gu, Daejeon 305-701, Korea}

$^{j}$\textit{School of Physics, Korea Institute for Advanced Study, Seoul 130-722, Korea}

$^{k}$\textit{Department of Physics, University of Massachusetts,
             Amherst, MA 01003, U.S.A.}

$^{l}$\textit{Department of Physics,
    Duke University,
    Durham, NC 27708, U.S.A.}

$^{m}$\textit{Department of Physics and Astronomy, University of California, Irvine, CA 92697, U.S.A.}

\end{center}

\vspace{20pt}
\begin{center}
\textbf{Abstract}
\end{center}
\vspace{5pt} {\small \noindent 
%The framework of a warped extra
%dimension with the Standard Model (SM) fields propagating in it is a
%very well-motivated extension of the SM since it can address both
%the Planck-weak and flavor hierarchy problems of the SM.
%%
%We consider signals at the large hadron collider (LHC)
%resulting from the direct production of the new particles (Kaluza-Klein excitations of the SM particles) in
%this framework. We propose two benchmarks for this purpose.
%%
%We also briefly discuss shifts in the couplings of 
%top quark and Higgs boson to each other and to $W/Z$ gauge bosons
%%
%%SM particles themselves
%%
%(as an indirect effect of the new particles) which can also be probed at the LHC.
%%
%Details of implementation of flavor, including
%flavor-violating signals (which are relevant for both LHC and lower-energy experiments)
%and alternate warped models which do not address flavor hierarchy,
%are
%left to a companion note.
The framework of a warped extra dimension with the Standard Model (SM) fields propagating in it is a
very well-motivated extension of the SM since it can address both the Planck-weak and flavor hierarchy problems of the SM.
We consider signals at the 14 and 33 TeV large hadron collider (LHC) resulting from the direct production of the new particles
in this framework, i.e.,Kaluza-Klein (KK) excitations of the SM particles. We focus on spin-1 (gauge boson) and spin-2 (graviton) KK particles
and their decays to top/bottom quarks (flavor-conserving) and W/Z and Higgs bosons, in particular. We propose two benchmarks for
this purpose, with the right-handed (RH) or LH top quark, respectively, being localized very close to the TeV end of the extra dimension.
We present some new results at the 14 TeV (with 300 fb$^-1$ and 3000 fb$^-1$) and 33 TeV LHC.  We find that the prospects
for discovery of these particles are quite promising, especially at the high-luminosity upgrade.

}

\vfill\eject
\noindent

%%%%%%%%%%
%%%%%%%%%%    Main Text
%%%%%%%%%%
 
%\tableofcontents
 
\section{Introduction}

The framework of a warped extra dimension a la Randall-Sundrum
(RS1) model \cite{Randall:1999ee}, but 
with all the SM fields propagating
in it 
\cite{Davoudiasl:1999tf,Grossman:1999ra,Gherghetta:2000qt}
is a very-well motivated extension of the Standard Model (SM):
for a review and further
references\footnote{i.e., the list of references given in this note is {\em not} meant
to be complete or fully updated: we apologize in advance for this!}, see \cite{Davoudiasl:2009cd}.
Such a framework
can address both the Planck-weak and the
flavor hierarchy problems of the SM, the latter without resulting in 
(at least a severe) flavor problem.
The versions of this framework
with a grand unified theory (GUT)
%
%gauge symmetry 
%
in the bulk 
can naturally lead to precision unification of the three
SM gauge couplings \cite{Agashe:2005vg} and 
%
%incorporate 
%
a candidate for the 
dark matter of the universe 
(the latter from
requiring longevity of the proton)
\cite{Agashe:2004ci}. 
The
new particles in this framework are Kaluza-Klein (KK) 
excitations of all SM fields with masses at $\sim {\rm TeV}$ scale.
In this write-up, we outline and suggest some benchmarks for the signals at the large hadron collider
(LHC) for these new particles.

\section{Review of Warped Extra Dimension}
\label{review}

The framework 
consists of a slice of anti-de Sitter
space in five dimensions (AdS$_5$), where (due to the
warped geometry) the effective $4D$ mass scale is dependent
on the position in the extra dimension.
The $4D$
graviton, i.e., the zero-mode
of the $5D$ graviton, is automatically localized
at one end of the extra dimension (called the Planck/UV brane).
If the Higgs sector is localized at the other end\footnote{In fact
with SM Higgs originating as 5th
component of a $5D$ gauge field
($A_5$) it is automatically so~\cite{Contino:2003ve}.}, then the warped geometry
naturally generates the Planck-weak hierarchy.
Specifically, we have TeV $\sim \mP
e^{ - k \pi r_c }$, where $\mP$ is
the reduced $4D$ Planck scale,
$k$ is the AdS$_5$ curvature scale and $r_c$ is the proper
size of the extra dimension. The crucial 
point is that the required
modest size of the radius (in units of the curvature radius), i.e.,
$k r_c \sim 1 / \pi \log \left( \mP / \hbox{TeV}
\right) \sim 10$ can be 
%
%obtained 
%
stabilized 
%
%(i.e., the radion given a mass) 
%
with only a corresponding
modest tuning in the fundamental or $5D$ parameters of
the theory \cite{Goldberger:1999uk}.
Remarkably, the correspondence between
AdS$_5$  
and $4D$ conformal field theories (CFT) \cite{Maldacena:1997re}
suggests that 
the scenario with warped extra dimension is 
dual to the idea of a composite Higgs in $4D$ 
\cite{Contino:2003ve, Arkani-Hamed:2000ds}.

\subsection{SM in warped bulk}

It was realized that with 
SM fermions propagating in the extra dimension, we can also
account for the hierarchy between quark 
and lepton masses and mixing angles (flavor hierarchy)
as follows~\cite{Grossman:1999ra,Gherghetta:2000qt}: 
the basic idea is that the $4D$ Yukawa coupling are given by the product of the 
$5D$ Yukawa coupling and the overlap of the profiles in the extra dimension of the SM 
fermions (which are the zero-modes of the $5D$ fermions) with that of the Higgs.
In turn, these fermion profiles are determined by 
$5D$ mass parameters, in units of $k$ (denoted by $c$). 
The crucial point is that vastly different profiles for zero-mode fermions
can be realized with small variations in the 
$5D$ mass parameters of fermions.
So, 
\begin{itemize}

\item

the light SM fermions [i.e., 1st and 2nd generations and tau lepton and its neutrino, right-handed (RH) bottom quark] can be chosen to be localized 
near the Planck brane (via $c \stackrel{>}{\sim} 1/2$), resulting in a 
small overlap with the TeV-brane localized SM Higgs, while
the top quark [including possibly left-handed (LH) bottom quark, being the weak isospin partner of the LH top quark]
is localized near the TeV brane ($c \stackrel{<}{\sim} 0$) with a large 
overlap with the Higgs.

\end{itemize}
Thus we can obtain hierarchical SM Yukawa couplings without any 
large hierarchies in the parameters of the $5D$ theory, i.e. the
$5D$ Yukawas and the $5D$ masses.~\footnote{Note that at the expanse of giving up solving flavor hierarchy, it is possible to make the light generation up-type RH fermions localized towards the TeV brane with employing flavor symmetry, and, thereby, changing the production cross section and branching ratios of KK particles, which is also motivated by the possibility of improving naturalness~\cite{Delaunay:2010dw}.}

With SM fermions emerging as zero-modes of $5D$ fermions,
so must the SM gauge fields. Hence, this scenario can be dubbed ``SM in the (warped) bulk''.
Due to the different profiles
of the SM fermions in the extra dimension, 
flavor changing neutral
currents (FCNC) are generated
by their non-universal couplings to gauge KK 
states.
However, 
these contributions to
the FCNC's are suppressed due to an analog of the Glashow-Iliopoulos-Maiani
(GIM) mechanism of the SM, i.e. RS-GIM, 
%
%approximate flavor symmetries
%
which is ``built-in''~\cite{Gherghetta:2000qt,Huber:2003tu}.
The point is that {\em all} KK modes
(whether gauge, graviton or fermion) are localized near the
TeV or IR brane (just like the Higgs) so that non-universalities
in their couplings to SM fermions are of
same size as couplings to the Higgs.
In spite of this RS-GIM
suppression, the lower limit on the KK mass scale can be
$\sim 10$ TeV, 
although these constraints can
be ameliorated by addition of flavor symmetries
\cite{flavor}.
Finally, various custodial symmetries \cite{Agashe:2003zs, Agashe:2006at}
can be incorporated such that the constraints from the various 
(flavor-preserving) electroweak precision tests (EWPT)
can be satisfied for a few TeV KK scale \cite{Carena:2006bn}.
The 
bottom line is that a 
few TeV mass scale for the KK gauge bosons can be consistent with both
electroweak and flavor precision tests.

\subsection{Couplings of KK's}

Clearly, the light fermions have small
couplings to all KK's (including graviton)
based simply on the overlaps
of the corresponding profiles, 
while the top quark and Higgs have a large coupling to the KK's.
To repeat, light SM fermions are localized near the Planck brane and photon, gluon and transverse $W/Z$ have flat profiles, whereas all KK's, Higgs (including longitudinal $W/Z$) and top quark are localized near the TeV brane. 
Schematically,
neglecting effects related to electroweak symmetry breaking (EWSB), we find the following ratio of
RS1 to SM
gauge couplings:
\begin{eqnarray}
{g_{\rm RS}^{q\bar q,l\bar l\, A^{ (1) }}\over g_{\rm SM}}
&\simeq&
- \xi^{-1}\approx - {1\over5} \nonumber \\
{g_{\rm RS}^{Q^3\bar Q^3 A^{ (1) }}\over g_{\rm
    SM}},
{g_{\rm RS}^{t_R\bar t_R A^{ (1) }}\over g_{\rm
    SM}} 
& \simeq & 
1 \; \hbox{to} \; \xi \; ( \approx 5 ) \nonumber \\
{g_{\rm RS}^{ HH A^{(1)}}\over g_{\rm
    SM}}  
& \simeq & 
\xi \approx 5 \; \; \; \left( H = h, W_L, Z_L \right)
\nonumber \\
{g_{\rm RS}^{ A^{ (0) }A^{ (0) } A^{ (1) }}\over g_{\rm
    SM}}  
& \sim & 0
\label{RScouplings}
\end{eqnarray}
Here $q=u,d,s,c,b_R$, $l =$ all leptons, $Q^3= (t, b)_L$, 
and $A^{ (0) }$ ($A^{ (1) }$) correspond
to zero (first KK) states of the gauge fields. Also, 
$g_{\rm RS}^{xyz}, g_{\rm SM}$ stands for the RS1 and the three SM (i.e.,
$4D$) gauge couplings respectively.
Note that 
$H$ includes both the physical Higgs ($h$) and 
{\em un}physical Higgs, i.e., {\em longitudinal}
$W/Z$ by the equivalence theorem
(the derivative involved in this coupling is 
similar for RS1 and SM cases and hence is not shown for
simplicity). Finally, the parameter $\xi$ is
related to the Planck-weak hierarchy: $\xi \equiv \sqrt{ k \pi r_c }$.

We also present 
the couplings of the KK
graviton to the SM particles. 
These couplings involve derivatives
(for the case of {\em all} SM particles),
but (apart from a factor from the overlap
of the profiles) it turns out that 
this energy-momentum dependence is
compensated (or made dimensionless) by the $\mP e^{ - k \pi r_c }\sim$ 
TeV scale, instead of 
the $\mP$-suppressed coupling to the SM graviton. Again, schematically:
\begin{eqnarray}
g_{ \rm RS }^{ q\bar q,l\bar l\, G^{ (1) } } & \sim & 
\frac{E}{ \mP e^{ - k \pi r_c } } \times 4D \; \hbox{Yukawa}
\nonumber \\
g_{ \rm RS }^{ A^{ (0) }A^{ (0) } G^{ (1) } } &
\sim & \frac{1}{ k \pi r_c }  \frac{E^2}{ \mP e^{ - k \pi r_c } }
\nonumber \\
g_{ \rm RS }^{ Q^3\bar Q^3 A^{ (1) } }, g_{ \rm RS }^{ t_R\bar t_R G^{ (1) } }
& \sim & \left( \frac{1}{ k \pi r_c } 
\; \hbox{to} \; 1 \right) 
\frac{E}{ \mP e^{ - k \pi r_c } }\nonumber \\
g_{ \rm RS }^{ H H G^{ (1) } } & \sim  & 
\frac{E^2}{ \mP e^{ - k \pi r_c } }
\end{eqnarray}
Here, $G^{ (1) }$ is the KK graviton
and the 
tensor
structure of the couplings is not shown
for simplicity.

\subsection{Couplings of radion}

In addition to the KK excitations of the SM, there is a particle, denoted by the
``radion'', which is  roughly the degree of freedom 
corresponding to the fluctuations of the size of extra dimension, and typically has a mass
at the
weak scale. It has Higgs-like properties: see \cite{Eshel:2011wz} for details.

\subsection{Masses}

As indicated above, masses below about $2$ TeV for gauge KK
particles 
(note that these are the same for  
gluon, $Z$, $W$ and 
denoted by $M_{ \rm KK }$)
are strongly disfavored by precision tests, whereas masses
for other KK particles are expected
(in the general framework) 
to be of similar size to gauge KK mass and hence are (in turn) also constrained to be above 
$2$ TeV. However, {\em direct} 
constraints on masses of other (than gauge) KK particles can be weaker.
The radion mass can vary from $\sim 100$ GeV to $\sim 2$ TeV.
In 
{\em minimal} models, KK graviton is actually about $1.5$ heavier than
gauge KK modes, i.e., at least $3$ TeV.

As far as KK fermions are concerned, in minimal models,
they have typically masses same as (or slightly heavier than) gauge KK and hence are constrained to be heavier than $2$ TeV (in turn, based on masses of gauge KK required to satisfy precision tests). 
However, the masses of 
the KK excitations of top/bottom (and their other gauge-group partners) in some 
non-minimal (but well-motivated) models
(where the $5D$ gauge symmetry is extended beyond that in the SM)
can be (much) smaller than gauge KK modes, 
possibly even $\sim 500$ GeV.

\section{Direct KK signals at the LHC, i.e., production}

Based on the above KK couplings 
and masses, 
we are faced with the following challenges 
in obtaining signals at the LHC
from direct production of the KK modes, namely,
\begin{itemize}
\item[(i)]
Cross-section for production of these
states is suppressed to begin with
due to a small coupling to
the protons' constituents, and due to the large mass of the new particles; 
\item[(ii)]
Decays to ``golden'' channels (leptons, photons)
are suppressed. Instead, the decays are 
dominated by top quark and Higgs
(including longitudinal $W/Z$); 

\item[(iii)]
These resonances tend to be quite
broad due to the enhanced couplings to top quark/Higgs.

\item[(iv)]
The SM particles, namely, top quarks/Higgs/$W/Z$ gauge bosons, produced in the decays of the heavy
KK particles are highly boosted, 
resulting in a high degree of collimation
of the SM particles' decay products. 
Hence,
conventional methods for identifying top quark/Higgs/$W/Z$ might no longer work 
for such a situation.

\end{itemize}

However, such challenges also present research 
opportunities, for example, techniques for identification of boosted top quark/Higgs/$W/Z$
have been developed \cite{Altheimer:2012mn}.

The following table summarizes the production cross-sections
for the spin-1 and spin-2 KK particles at the 14 TeV LHC (with KK mass set to 3 TeV) and the dominant decay channels. 
%
%final states
%
Based on the above discussion, note that the polarization of $W/Z$'s in these decay channels is dominantly {\em longitudinal}. The bottom quark is LH, but the top quark can be either LH or RH.
Some more details are in the sections that follow. \\
\vspace{0.1in}
\begin{table}[h]
\label{summarytable}
\begin{center}
\begin{tabular}{|c||c|c|c|}
\hline 
KK particle & total $\sigma$ (fb)
%
%production cross-section 
%
& (SM) final states & references  
\tabularnewline
\hline
\hline 
graviton & $\sim 0.1$ for $k \sim M_{ \rm Pl }$& $t \bar{t}$, $b \bar{b}$, $WW$, $hh$, $ZZ$ & hep-ph/0701186
\tabularnewline
\hline 
gluon & $\sim 100$ & $t \bar{t}$, $b \bar{b}$ & 0706.3960
\tabularnewline
\hline 
$Z$ & a few & $t \bar{t}$, $b \bar{b}$, $Zh$, $WW$ & 0709.0007
\tabularnewline
\hline 
$W$ & ~10 & $WZ$, $Wh$, $t \bar{b}$ & 0810.1497
\tabularnewline
\hline
\end{tabular}
\end{center}
\end{table}

In the following sections, we give 
details of studies of the KK particles in this framework performed as part of the Snowmass 2013 process:
for more details of the specific models (including previous studies), see corresponding references given in each title
and for an overview, see reference \cite{Davoudiasl:2007wf}.

\subsection{A proposal for Two ``Benchmarks" \protect\cite{Agashe:2008jb}}

In order to begin this discussion, we mention the benchmark models studied.
The color
[$SU(3)_c$] structure is standard and is thus is not shown.
On the other hand, the electroweak 
(EW) gauge group is extended: $SU(2)_L \times SU(2)_R \times U(1)_X$ (as motivated by relaxing the 
constraint from the $T$ parameter). The extra gauge bosons (relative to the SM) can be taken (by a suitable
choice of boundary conditions) to have no zero-modes.
The hypercharge is then given by 
$Y = T_{ 3 R } + X$.

In the light of the large
top mass, 
depending on $5D$ Yukawa coupling, it is clear that either (or both) LH and RH top 
have to be localized near the TeV brane.
However if it is the LH top, then so is LH bottom thus creating a 
constraint from 
shift in $Zb \bar{b}$ coupling.
On the other hand, the coupling to $Z$ of $t_R$ as not been precisely
measured thus far and hence such a localization of $t_R$ poses less of a constraint.

In order to alleviate the above constraint, i.e., have the custodial symmetry protection of the $ Zb\bar b $ coupling~\cite{Agashe:2006at}, 
we take the third generation left-handed quarks to be in the representation\footnote{In addition, the two 
$5D$ $SU(2)$ gauge couplings are taken to be equal. The ``canonical" choice would
be that LH fermions are singlets of $SU(2)_R$.}
\beq
Q_L^3 = \bmat q_L^3 & {q_L^\prime}^3 \emat  =  \bmat t_L & \chi_L \\ b_L & T_L \emat \to (2,2)_{2/3} \; \hbox{(for both cases I and II: see below)} \ ,
\label{Q3Ldef.EQ}
\eeq
where $ \chi_L , T_L $ are taken to have no zero-modes\footnote{Only the
$5D$ fermions with SM quantum numbers can have zero modes.}.
We have $ Q(\chi_L) = 5/3 $ and $ Q(T_L) = 2/3 $.
To accommodate the large top and bottom mass difference we take it that $ t_R $ and $ b_R $ do not belong
to the same $SU(2)_R$ multiplet.
With the above choice, one can contemplate 
$Q_L^3$ being localized near the TeV brane.

We consider two cases for the $ t_R $ representations
\bea
{\rm Case \; I} &:&\ t_R \to (1,3)_{2/3} \oplus (3,1)_{2/3} = \bmat \chi_R^{\prime\prime} \\ t_R \\ B_R^{\prime\prime} \emat \oplus \bmat \chi_R^{\prime\prime\prime} \\ T_R^{\prime\prime\prime} \\ B_R^{\prime\prime\prime} \emat  \ ,
\nonumber \\
{\rm Case \; II} &:&\  t_R \to (1,1)_{2/3} \ ,  
\label{tRdef.EQ}
\eea
where (once again) the 
%
%exotic 
%
extra 
fermions are chosen to have no zero-modes. 
%
%and the fermions in the $(3,1)$ representation 
%are not discussed further here. 
%
%since the $Z^{ \prime}$, $W^{ prime }$ decay to a pair of them is kinematically 
%forbidden. 
%

For Case II, $t_R\to (1,1)$, the electroweak precision tests (EWPT) are better 
satisfied for $Q_L^3$ peaked closer to the TeV brane,
while for Case I, $t_R\to (1,3)$, for  $t_R$ peaked closer to the TeV brane. 
So, we choose
\bea
{\rm Case \; I} &:&\ c_{ Q_L^3 } = 0.4 \; \hbox{and} \; c_{ t_R } = 0
 \ ,  \nonumber \\
{\rm Case \;  II} &:&\  c_{ Q_L^3 } = 0 \; \hbox{and} \; c_{ t_R } = 0.4
\eea

After including the charges and the overlap integrals, the largest effective coupling of third generation
fermions to gauge KK modes in Case II would be to $Q_L^3$, being larger than that in Case I,
which would be to $t_R$.   
Consequently, in Case I, the bottom quark is essentially not in the game, where it is  in
Case II. Thus, while on the one hand new gauge KK induced FCNC contributions would be larger in 
Case II (again since $b_L$ and hence down sector flavor violation is involved) and hence more problematic for the simplest constructions, 
on the other hand collider signals would be larger compared to Case I (for example, KK $W$
decays to $ t \bar{b}$ would be suppressed in case I, but significant in case II).

Also, the 
polarization of top quark resulting from the production and decay of KK particles
is different in the two cases and hence this measurement can distinguish them.

Finally, note that 
\begin{itemize}

\item
exact localization of other (light) fermions is basically irrelevant for LHC signals (as long as they are localized 
near the Planck brane, i.e., 
$c \stackrel{>}{\sim} 1/2$, as we assume here).
Hence the values of $c$'s for them are not explicitly shown.

\end{itemize}

\subsection{Sensitivity to narrow $\ttbar$ resonances in the all-hadronic decay channel}
\label{sec:ttbar_allhadronic}

As seen from table \ref{summarytable}, most spin-1 and spin-2 KK particles in the warped extra dimensional framework 
have significant BR to decay into $t \bar{t}$.
Also, in other models, there are $Z^{ \prime }$ which are leptophobic and thus have to be searched for via decays
into quarks, in particular, top.
So, it is very useful to perform a general study of $t \bar{t}$ resonances, 
which is the goal of this section.

In order to predict the expected sensitivity to narrow $\ttbar$ resonances at the 14 TeV LHC, we perform a search in the all-hadronic decay channel.
The expected cross section limits are quoted for resonance masses of 2, 3, 4 and 5 TeV assuming 300~fb$^{-1}$ and 3000~fb$^{-1}$ of integrated luminosity.
Those systematic effects which are expected to dominate are taken into account.
The cross section limits are interpreted as limits on the mass of a leptophobic top-color $Z^\prime$ boson~\cite{Harris:1999ya} in
this section,
as well as limits on the masses
of KK gravitons and KK $Z^\prime$ bosons in Sec.~\ref{sec:KKgraviton} and~\ref{sec:KKZgamma}, respectively\footnote{KK gluon
in the warped extra dimensional framework tends to be rather broad so that it might not be possible to directly 
translate the results of this study into bounds on KK gluon.}.

$Z^\prime$ signal samples are generated using Pythia~\cite{Pythia8} with a narrow resonance width.
The background processes, Standard Model $\ttbar$ production and QCD multijet production, are generated with Herwig++~\cite{HerwigPP},
with a minimum $p_{\rm T}$ requirement of 650~GeV on the two highest-$p_{\rm T}$ partons required at generator level.
All samples are overlaid with simulated minimum-bias events corresponding to three pile-up scenarios with an average number of interactions per
bunch-crossing of $\mu = 0$, 50 and 140.
The samples are then reconstructed using version 3.0.9 of the DELPHES fast detector simulation~\cite{Delphes} using the Snowmass detector~\cite{DelphesSnowMass}.

While the background cross sections are obtained at LO from the Herwig++ generator, the signal cross sections for $Z^\prime$ production calculated in Ref.~\cite{ZprimeXsec}
for a resonance width of 1.2\% are used, and a $k$-factor of 1.3 is applied~\cite{kfactor}.
Tab.~\ref{tab:Xsec} shows an overview of the cross sections considered at leading order (LO).

\begin{table}[h!]
\centering
\caption{LO cross sections considered in the analysis in the $\ttbar$ all-hadronic final state.}
\begin{tabular}{|c||c|}
\hline 
sample & cross section
\tabularnewline
\hline
\hline 
$Z^\prime$ (2 \TeV) & 214 fb
\tabularnewline
\hline 
$Z^\prime$ (3 \TeV) & 23.2 fb
\tabularnewline
\hline 
$Z^\prime$ (4 \TeV) & 3.24 fb
\tabularnewline
\hline 
$Z^\prime$ (5 \TeV) & 0.553 fb
\tabularnewline
\hline 
\ttbar\ ($\pt > 650 \GeV$) & $1.28 \cdot 10^3$ fb
\tabularnewline
\hline 
QCD multijet ($\pt > 650 \GeV$) & $170 \cdot 10^3$ fb
\tabularnewline
\hline 
\end{tabular}
\label{tab:Xsec}
\end{table}

Large-$R$ jets are reconstructed with the Cambridge-Aachen~\cite{ca} algorithm with a radius parameter of 0.8.
They must fulfill $p_{\rm T} > 750$~GeV and $|\eta| < 2.0$.
Top-tagging is implemented requiring the trimmed jet mass, $m$, to be larger than 80~GeV, and the largest di-subjet mass when the jet is de-clustered to
three subjets, $Q_W$, to be larger than 70~GeV.
This top-tagging criterion was optimized using the ratio of the tagging efficiency for top quarks over the square root of the mis-tagging efficiency for non-top jets.
A variety of substructure variables was taken into account, such as $k_t$ splitting scales, $N$-subjettiness variables and the number of subjets.
Adding more variables to the top quark identification does not help to increase the performance significantly.

The decreasing $b$-tagging efficiency, $\varepsilon$, and $b$-tagging rejection of light quark jets, $r$, at high jet $p_{\rm T}$ is taken into account by
applying $b$-tagging weights to jets instead of using the default $b$-tagging as implemented in the Snowmass detector:
\begin{equation*}
\varepsilon(p_{\rm T}) = 0.83 \cdot \exp \left( -0.68 \cdot
p_{\rm T} [{\rm TeV}] \right) \; , \quad
r(p_{\rm T}) = 80 \cdot \exp \left( -0.92 \cdot p_{\rm T} [\TeV] \right) \; .
\end{equation*}
Benchmarks of $\varepsilon(750 \GeV) \approx 50\%$, $\varepsilon(1500 \GeV) \approx 30\%$, $r(750 \GeV) \approx 40$ and $r(1500 \GeV) \approx 20$ were used to
parametrize the curves.
In order to take into account the degradation of $b$-tagging algorithms with increasing pile-up, the fake rate, $f = \frac{1}{r}$, is increased by $\frac{1}{3}$
$\left(\frac{2}{3}\right)$ for $\mu = 50$ (140). 

\begin{figure}[!h]
\centering
  \includegraphics[width=0.47\textwidth]{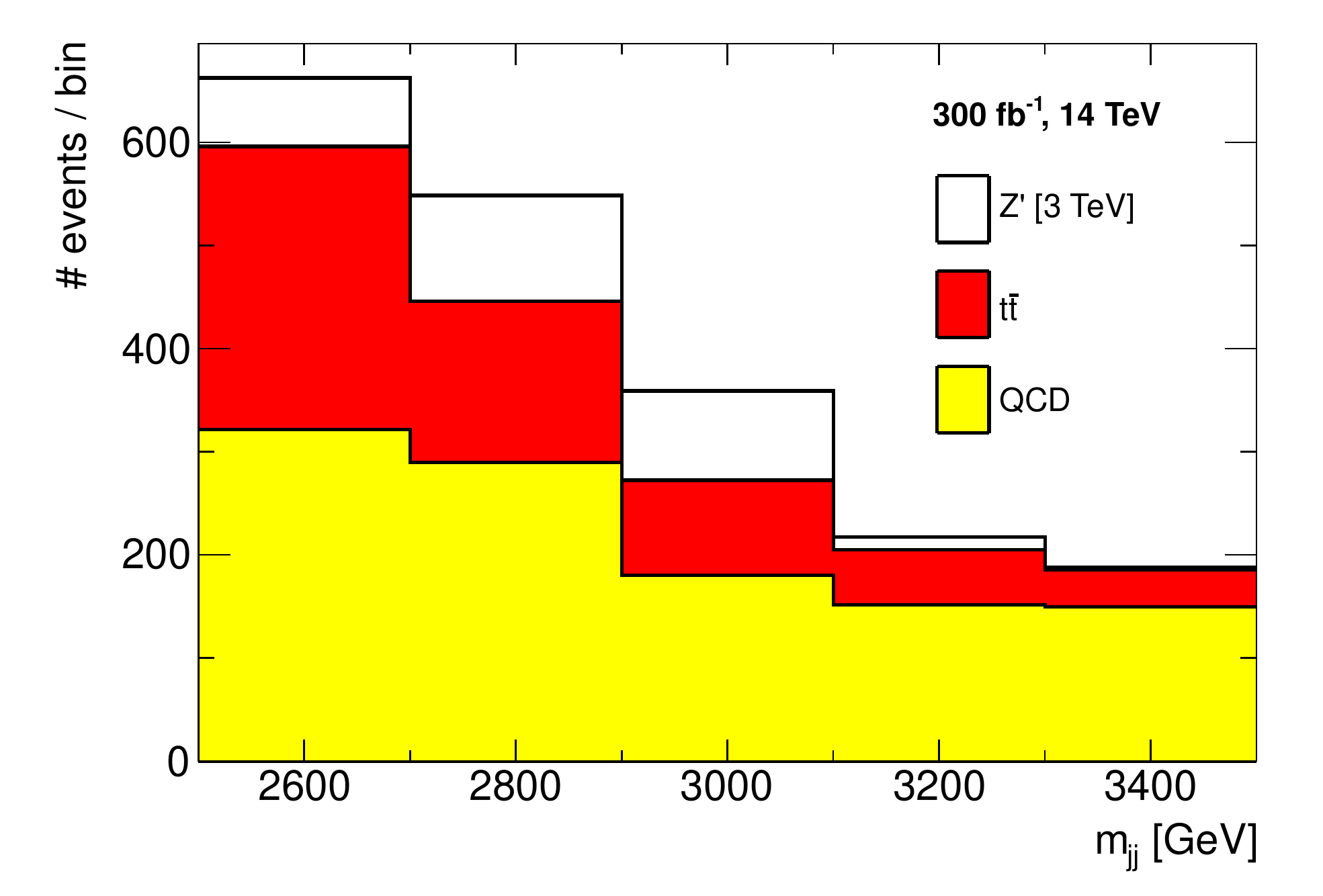}
\caption{
  Reconstructed $Z^\prime$ mass distribution in the all-hadronic channel
  for a resonance mass of 3~TeV, 300~fb$^{-1}$ of 14 TeV data and $\mu = 50$.
  The signal distribution is shown on top of the background contributions.
}
\label{fig:invmass}
\end{figure}

In order to select events for the analysis, two top- and $b$-tagged large-$R$ jets with $p_{\rm T} > 750 \GeV$ are required and
the di-jet invariant mass is calculated.
A signal window of $\pm 500 \GeV$ is chosen around the resonance mass for the statistical analysis, as shown for one
example in Fig.~\ref{fig:invmass}.

Four sources of systematic uncertainties are considered:
\begin{itemize}
\item An overall normalization uncertainty of the Standard Model $\ttbar$ background of 10\% is assumed. With data, it may be constraint by the use of a $\ttbar$
      dominated control region.
\item An overall normalization uncertainty of the QCD multijet background of 50\% is assumed, which reflects the extrapolation uncertainty of the data-driven
      estimate of the QCD multijet background from a control region into the signal region.
\item The jet energy uncertainty is assumed to be 2\% of the jet $p_{\rm T}$. It is evaluated for signal and the Standard Model $\ttbar$ background, because the QCD
      multijet background is assumed to be estimated in a data-driven way.
\item The uncertainty on the $b$-tagging efficiency is assumed to be 10\% and is, analogously to the jet energy uncertainty, evaluated for signal and $\ttbar$ background.
\end{itemize}
All sources of systematic uncertainties are only considered as a change of the yield of the corresponding process.
The effects on the shape of the invariant di-jet mass spectrum are neglected.

The Bayesian Analysis Toolkit~\cite{bat} is used to determine the 95\% CL upper limits on the signal cross section for each mass point
given the SM mass spectrum and that of a given signal model.
Systematic uncertainties are taken into account as nuisance parameters of the fit, hence strongly constraining the mostly unconstrained QCD multijet background with
its prior normalization uncertainty of 50\%.

Fig.~\ref{fig:KKglimit} shows the expected cross section exclusion as a function of the $Z^\prime$ mass for 300~fb$^{-1}$ of 14 TeV data and $\mu = 50$ (left), and for
3000~fb$^{-1}$ of data and $\mu = 140$ (right).
The mass reach for both scenarios is 3.7 and 4.1~TeV, respectively.
Tab.~\ref{tab:Zprimelimits} gives an overview of the cross section limits for the different resonance masses and pile-up scenarios, as well as for the two different
integrated luminosities.

\begin{figure}[!h]
\centering
  \includegraphics[width=0.47\textwidth]{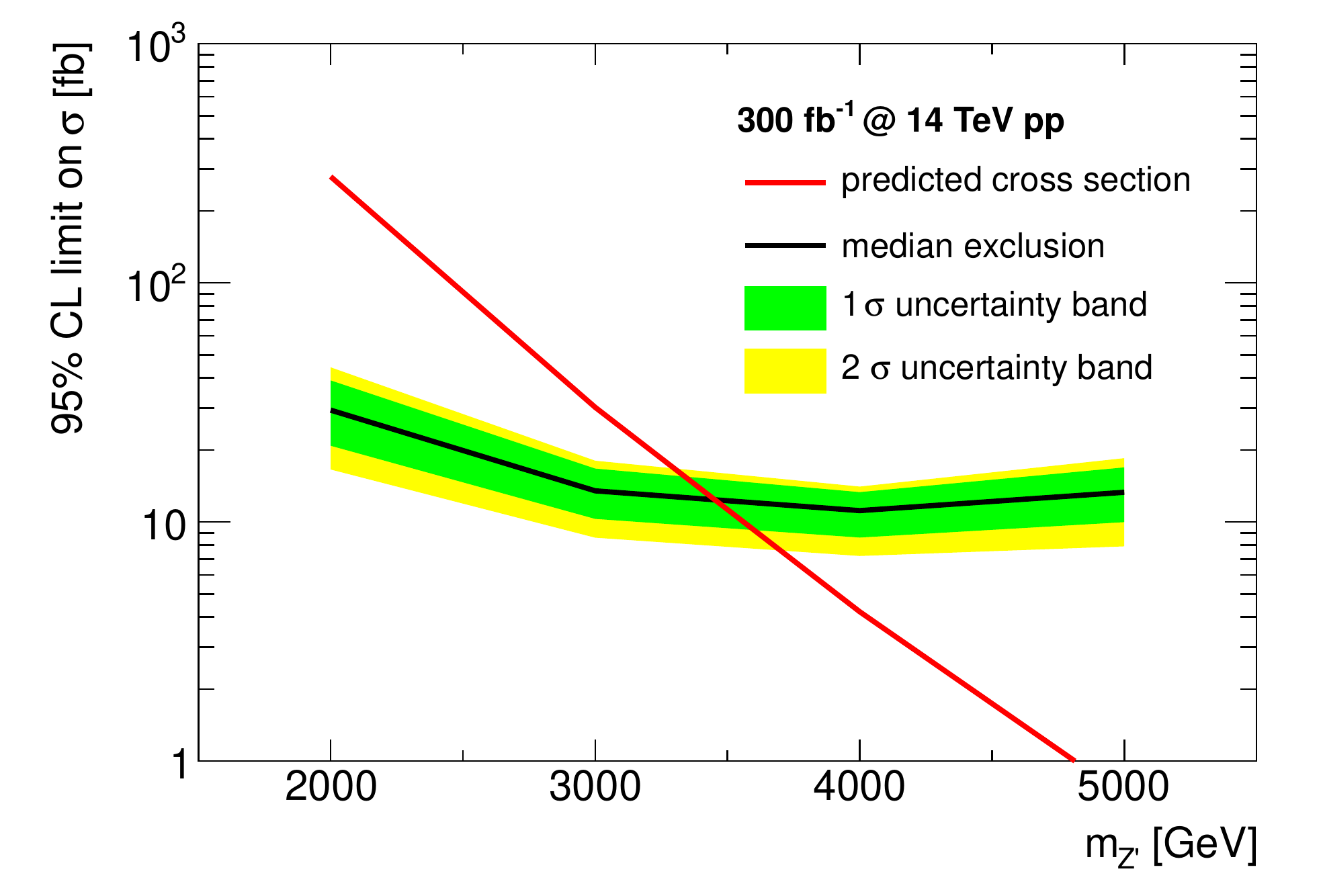}
  \includegraphics[width=0.47\textwidth]{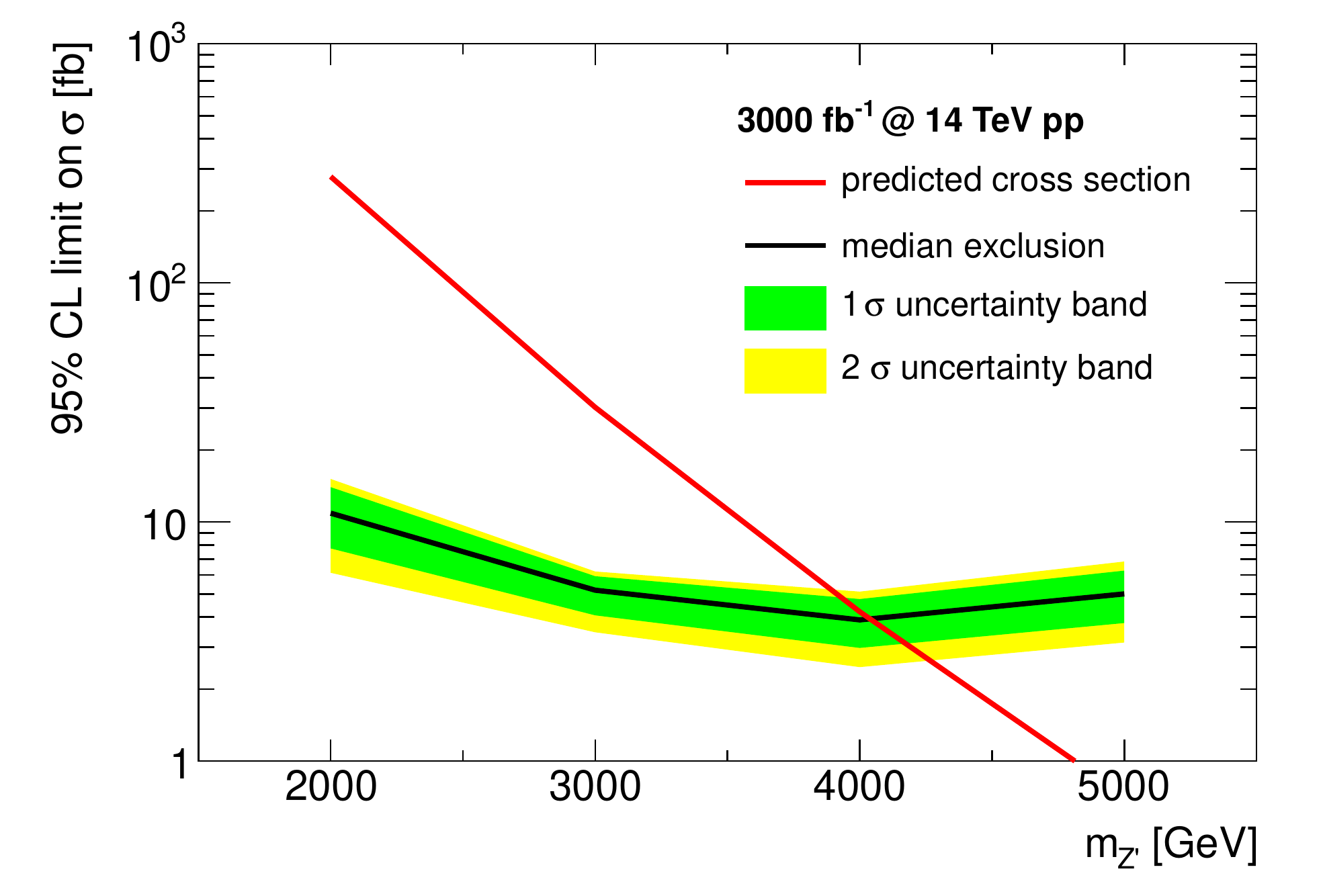}
\caption{
    Expected cross section upper limits at 95\% CL for a leptophobic top-color $Z^\prime$ boson with a width of 1.2\%
    decaying to a top quark pair in the all-hadronic channel. The left plot shows the limits for 300~fb$^{-1}$ of 14 TeV data and $\mu = 50$.
    The right plot shows the limits for 3000~fb$^{-1}$ of data and $\mu = 140$ (right).
    The mass reach is 3.7 (4.1) TeV for 300 (3000) fb$^{-1}$.
}
\label{fig:KKglimit}
\end{figure}

\begin{table}[h!]
\centering
\caption{95\% CL limits for a leptophobic top-color $Z^\prime$ boson with a width of 1.2\% decaying to a top quark pair in the all-hadronic channel.
 Limits are given for different resonance masses, pile-up scenarios and integrated luminosities. The expected mass exclusion is also shown.
}
\begin{tabular}{|c|c|c|c|c|c|c|c|}
\hline
lumi. & pile-up & uncertainties & $m = 2 \TeV$ & $3 \TeV$ & $4 \TeV$ & $5 \TeV$ & mass reach
\tabularnewline
\hline
\hline
300 fb$^{-1}$ & $\mu = 0$ & stat.         &  16 fb & 6.4 fb & 5.3 fb & 6.6 fb & 4.0 TeV \\
300 fb$^{-1}$ & $\mu = 0$ & stat.+syst.   &  26 fb &  12 fb &  10 fb &  11 fb & 3.8 TeV \\
300 fb$^{-1}$ & $\mu = 50$ & stat.        &  16 fb & 6.7 fb & 5.7 fb & 7.6 fb & 3.9 TeV \\
300 fb$^{-1}$ & $\mu = 50$ & stat.+syst.  &  29 fb &  13 fb &  11 fb &  13 fb & 3.7 TeV \\
300 fb$^{-1}$ & $\mu = 140$ & stat.       &  17 fb & 7.2 fb & 6.2 fb & 8.5 fb & 3.9 TeV \\
300 fb$^{-1}$ & $\mu = 140$ & stat.+syst. &  35 fb &  15 fb &  12 fb &  15 fb & 3.7 TeV 
\tabularnewline
\hline
3000 fb$^{-1}$ & $\mu = 0$ & stat.         & 4.9 fb & 2.0 fb & 1.5 fb & 1.9 fb & 4.7 TeV \\
3000 fb$^{-1}$ & $\mu = 0$ & stat.+syst.   & 8.1 fb & 3.6 fb & 3.3 fb & 3.3 fb & 4.3 TeV \\
3000 fb$^{-1}$ & $\mu = 50$ & stat.        & 5.0 fb & 2.1 fb & 1.7 fb & 2.2 fb & 4.6 TeV \\
3000 fb$^{-1}$ & $\mu = 50$ & stat.+syst.  & 8.7 fb & 4.3 fb & 3.6 fb & 4.3 fb & 4.1 TeV \\
3000 fb$^{-1}$ & $\mu = 140$ & stat.       & 5.4 fb & 2.2 fb & 1.9 fb & 2.5 fb & 4.6 TeV \\
3000 fb$^{-1}$ & $\mu = 140$ & stat.+syst. &  11 fb & 5.2 fb & 3.9 fb & 5.0 fb & 4.1 TeV
\tabularnewline
\hline
\end{tabular}
\label{tab:Zprimelimits}
\end{table}

In summary, it can be stated that with 300~fb$^{-1}$ of data at 14 TeV, leptophobic top-color $Z^\prime$ bosons with the above parameters can be excluded
up to masses of 3.7~TeV using this analysis in the all-hadronic $\ttbar$ decay channel.
With ten times more data, taking into account the more challenging pile-up conditions, the mass reach is increased to 4.1~TeV.

\subsection{KK graviton \protect\cite{Fitzpatrick:2007qr}}
\label{sec:KKgraviton}

Armed with the above benchmarks models, we now discuss searches for each type of KK particle in turn.
The dominant production is via gluon fusion and decay channels for KK graviton are
into $t \bar{t}$, $WW$, $ZZ$, $hh$:
\bea
gg \rightarrow & 
\hbox{KK graviton} \rightarrow & ZZ, \; WW, \; h h \; (\hbox{and} \; t \bar{t}, \;
b \bar{b})
\eea
For a $2$ TeV  KK graviton, each of these cross-sections
can be $\sim O(10)$ fb with a total decay width of 
$\sim O(100)$ GeV.

In more detail, the 
production is governed by
\begin{eqnarray}
{\cal L}_{ \rm prod.} & = & 0.053 \frac{c}{ M_{ G^{ (1) } } }
\eta^{ \mu \alpha } \eta^{ \nu \beta } h^{ ( 1 ) }_{ \alpha \beta }
( x ) T_{ \mu \nu }^{ \rm gluon } ( x )
\nonumber 
\end{eqnarray}
where $c = k / M_{ \rm Pl }$, $M_{ G^{ (1) } }$ is KK graviton mass\footnote{As mentioned earlier,
this mass is $\approx 1.5 M_{ \rm KK }$, where $M_{ \rm KK }$ is the {\em gauge} KK mass.}
and $T_{ \mu \nu }^{ \rm gluon }$ is 4D energy-momentum tensor of SM gluon\footnote{we set 
$\sqrt{ k \pi r_c } = 5.83$ here}.

And, 
decay
proceeds via
\begin{eqnarray}
{\cal L}_{ \rm decay } & \ni & 3.83 \frac{ c }{ M_{ G^{ (1) } } }
\eta^{ \mu \alpha } \eta^{ \nu \beta } h^{ ( 1 ) }_{ \alpha \beta }
( x ) T_{ \mu \nu }^{ t,b,H } ( x )
\nonumber 
\end{eqnarray} 
This gives the total decay width: 
\\
\vspace{0.1in} 
\begin{center}
\begin{tabular}{|c||c|c|}
\hline 
& Case I
& Case II 
\tabularnewline
\hline
\hline 
$\Gamma_{ G^{ (1) } }$ &  $0.063 \; c^2 M_{ G^{ (1) } }$ & $0.107 \; c^2 M_{ G^{ (1) } }$
\tabularnewline
\hline 
\end{tabular}
\end{center}
\vspace{0.1in} 
and the branching ratios (BRs):
\\
\vspace{0.1in} 
\begin{center}
\begin{tabular}{|c||c|c|}
\hline 
& Case I
& Case II 
\tabularnewline
\hline
\hline 
$(t,b)_L$ &  $\sim 0$ &  9/22 each
\tabularnewline
\hline 
$t_R$ &  9/13 &  $\sim 0$
\tabularnewline
\hline 
$h$ &  1/13 & 1/22  
\tabularnewline
\hline 
$W_L$ &  2/13 & 2/22  
\tabularnewline
\hline 
$Z_L$ &  1/13 &  1/22
\tabularnewline
\hline 
\end{tabular}
\end{center}

After the above overview of KK graviton couplings, we discuss studies of two specific decay channels.

\subsubsection{KK graviton $\rightarrow ZZ$}

We estimate the KK graviton discovery potential at the LHC using the ZZ decay channel where one Z bosons decays leptonically and the other hadronically. For high graviton masses, the produced Z bosons are highly boosted and the two quarks from the hadronically decaying Z boson are expected to be inside the same jet. Thus the final state topology consists of two isolated leptons and one highly energetic massive jet. The events are selected using the following criteria: 
\begin{equation}
p_\perp^l > 25\, \text{GeV} \quad p_\perp^j > 25\, \text{GeV}, \quad |\eta^{j,l}|<2.4, \quad 66\, \text{GeV} < m_{ll} <116\, \text{GeV}.
\end{equation}¥
To reduce the background, the signal region is defined by requiring:
\begin{equation}
p_\perp^{ll}\, >\, 400\, \text{GeV},\quad p_\perp^{j}\, > \, 400 \text{GeV}, \quad m_j > 40 \, \text{GeV}
\end{equation}¥

The $Z$+Jets events form the dominant SM background. The background and the signal are both simulated with \texttt{MadGraph 5}. The model is implemented into \texttt{Madgraph 5} \cite{Alwall:2011uj} using the \texttt{FeynRules} \cite{Christensen:2008py} package. The events are then passed to \texttt{Pythia 6} \cite{Sjostrand:2006za} for showering and hadronization.  The \texttt{Delphes 3} program is used for fast detector simulation and for object reconstruction. The jets are reconstructed using the anti-$k_t$ algorithm~\cite{anti_kt} with a radius parameter 0.6. 

In Figs.\ref{fig:1a} and \ref{fig:1b} we present the 95$ \%$ confidence level exclusion plots for KK graviton mass in the decay channel $pp\rightarrow \hbox{KK graviton} \rightarrow  ZZ \rightarrow \mu^+\mu^- jj$ at the $\sqrt{s}=14$ TeV and $\sqrt{s}=33$ TeV respectively. On both plots, the blue dashed curve corresponds to the Case I (the $t_R$ near TeV brane) while the solid green curve to the Case II ($Q_L^3$ near the TeV brane). The limits on the signal rate are determined using the CL$_s$  method \cite{Junk:1999kv,Read:2002hq}.

\begin{figure}[bt]

\subfloat[ The 95$ \%$ confidence level exclusion for the KK graviton mass at $\sqrt{s}=14$ TeV
in the decay channel $pp\rightarrow G^{(1)} \rightarrow  ZZ \rightarrow \mu^+\mu^- jj$.
] 
{\includegraphics[width=.52\textwidth]{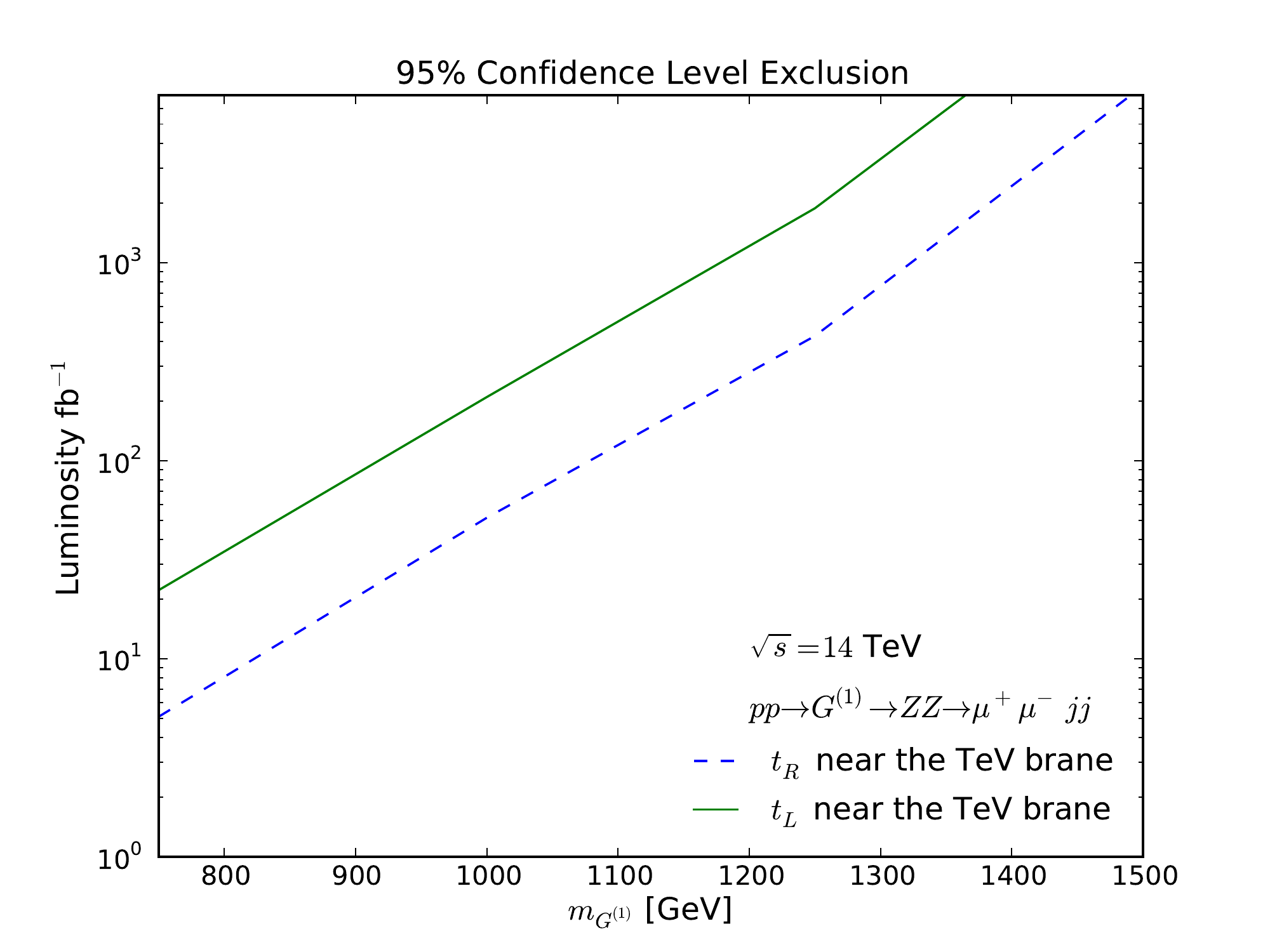}\label{fig:1a}}
  \hfill
\subfloat[The 95$ \%$ confidence level exclusion for the KK graviton mass at $\sqrt{s}=33$ TeV
in the decay channel $pp\rightarrow G^{(1)} \rightarrow  ZZ \rightarrow \mu^+\mu^- jj$.]{\includegraphics[width=.52\textwidth]{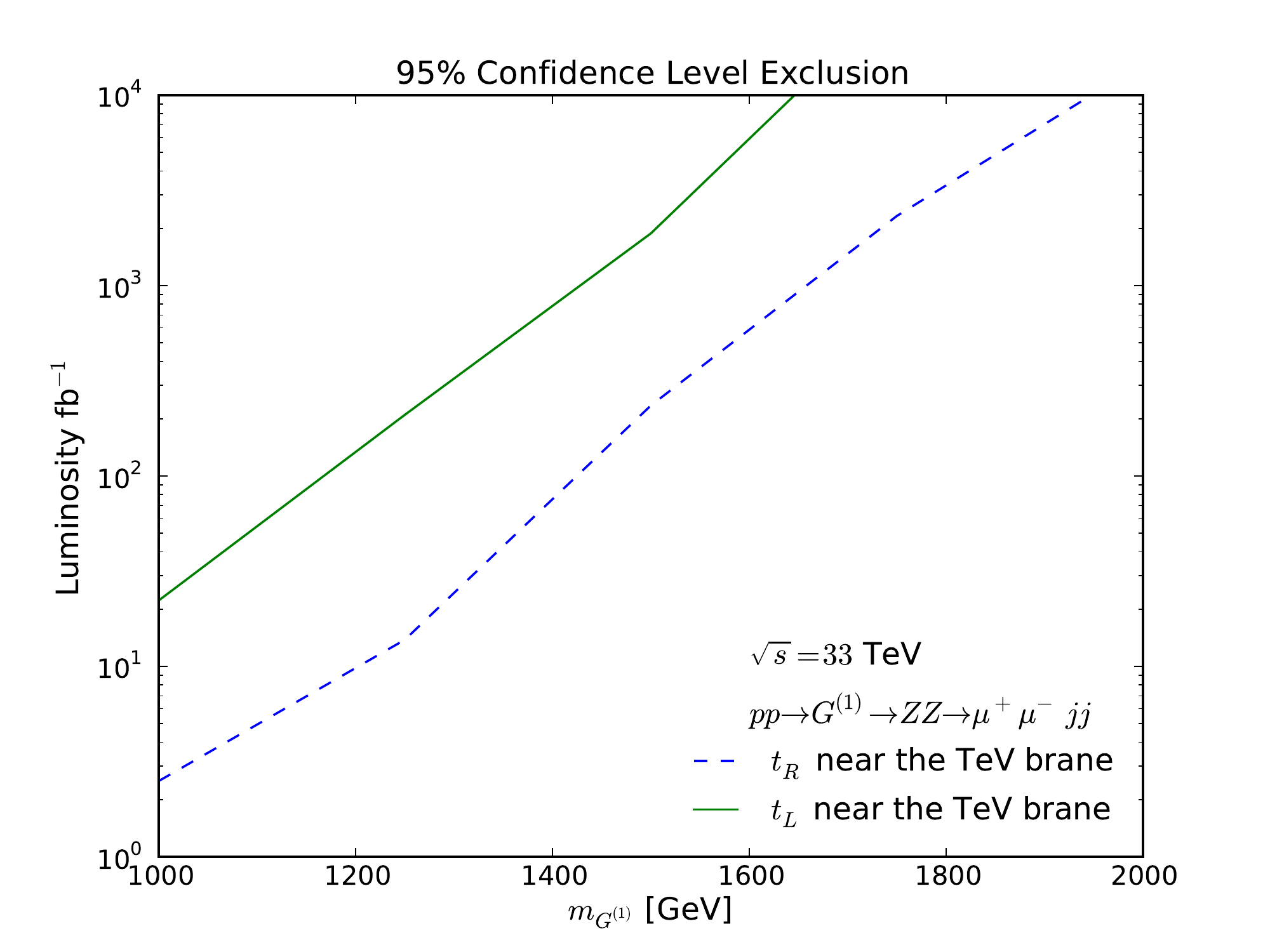}\label{fig:1b}}
\addtocounter{figure}{1}
\end{figure} 

%
%Case I: total decay width, $\Gamma_{ G^{ (1) } } = 0.063 \; c^2 M_{ G^{ (1) } }$, 
%with BR's to $t_R$, $h$, $W_L$ and $Z_L$: $9/13$, $1/13$, $2/13$ and
%$1/13$. 
%
%Similarly, for Case II, we get $\Gamma_{ G^{ (1) } } = 0.107 \; c^2 M_{ G^{ (1) } }$, 
%with BR's to $t_L$, $b_L$, $h$, $W_L$ and $Z_L$: $9/22$, $9/22$, $1/22$, $2/22$ and
%$1/22$.
%

\subsubsection{Reinterpretation of the search for narrow $\ttbar$ resonances}

The search for narrow $\ttbar$ resonances in the all-hadronic final state described in Sec.~\ref{sec:ttbar_allhadronic}
is reinterpreted in terms of limits on KK gravitons as discussed in Ref.~\cite{Fitzpatrick:2007qr}
decaying to $\ttbar$.
Four scenarios are investigated following the notation convention as in the reference:
\begin{itemize}
\item $t_R$ near the TeV brane with $c = 0.5$,
\item $t_R$ near the TeV brane with $c = 1.0$,
\item $t_L$ near the TeV brane with $c = 0.5$, and
\item $t_L$ near the TeV brane with $c = 1.0$.
\end{itemize}
Tab.~\ref{tab:Xsec_grav} shows the signal cross sections used in the reinterpretation.
The $k$-factor of 1.3 used in the leptophobic top-color $Z^\prime$ analysis is also used here, although it has only limited
validity in this scenario.

\begin{table}[h!]
\centering
\caption{Cross sections at 14 TeV considered in the analysis of the search for narrow $\ttbar$ resonances.}
\begin{tabular}{|c||c|c|}
\hline
scenario & KK graviton mass & cross section
\tabularnewline
\hline
\hline
R 0.5 & 2 \TeV & 4.7 fb \\
R 0.5 & 3 \TeV & 0.22 fb \\
R 0.5 & 4 \TeV & 0.016 fb \\
R 0.5 & 5 \TeV & 0.0014 fb
\tabularnewline
\hline
R 1.0 & 2 \TeV & 18 fb \\
R 1.0 & 3 \TeV & 0.87 fb \\
R 1.0 & 4 \TeV & 0.066 fb \\
R 1.0 & 5 \TeV & 0.0065 fb
\tabularnewline
\hline
L 0.5 & 2 \TeV & 2.7 fb \\
L 0.5 & 3 \TeV & 0.13 fb \\
L 0.5 & 4 \TeV & 0.0096 fb \\
L 0.5 & 5 \TeV & 0.00088 fb
\tabularnewline
\hline
L 1.0 & 2 \TeV & 10 fb \\
L 1.0 & 3 \TeV & 0.50 fb \\
L 1.0 & 4 \TeV & 0.040 fb \\
L 1.0 & 5 \TeV & 0.0042 fb
\tabularnewline
\hline
\end{tabular}
\label{tab:Xsec_grav}
\end{table}

\begin{figure}[p]
\centering
  \includegraphics[width=0.4\textwidth]{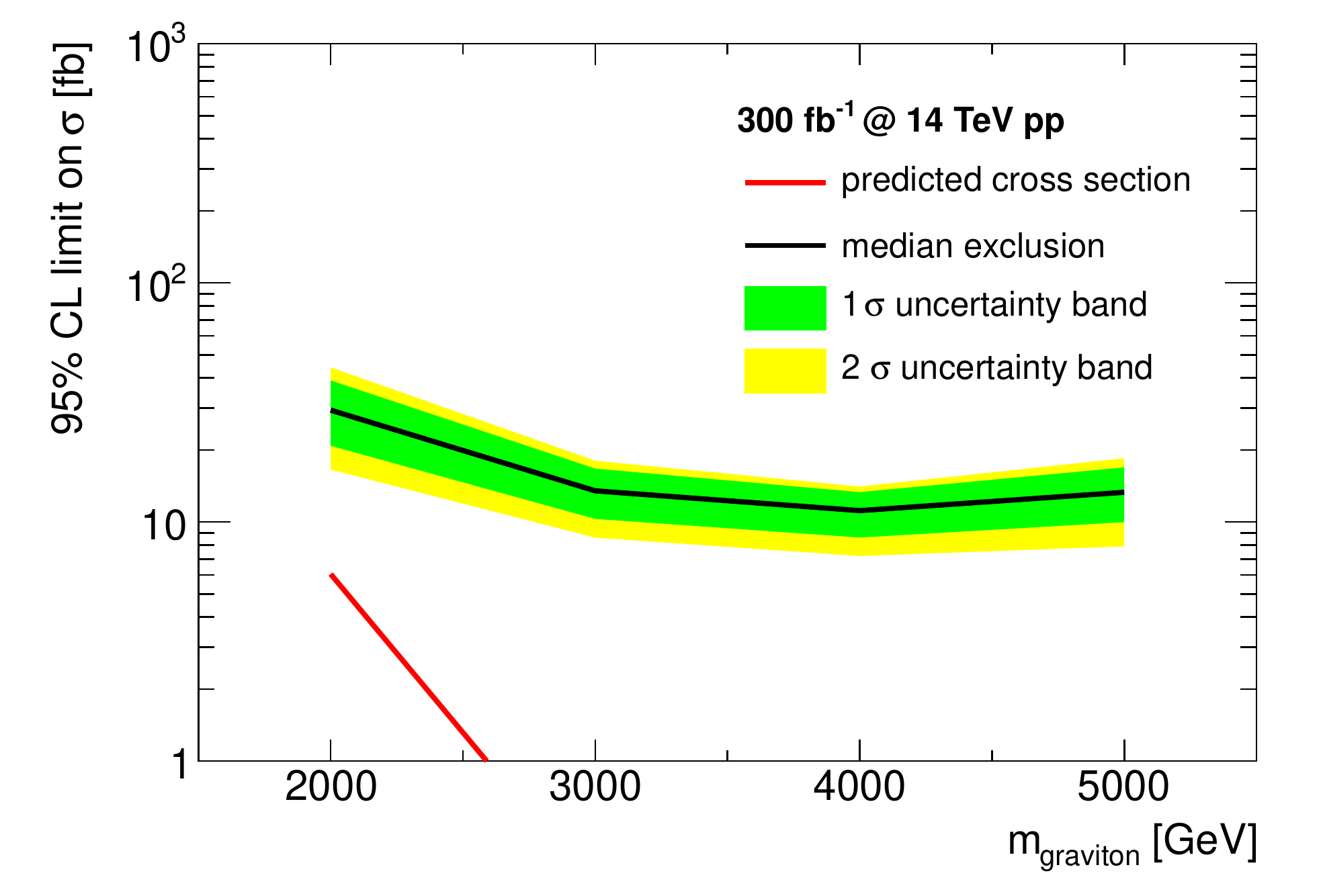}
  \includegraphics[width=0.4\textwidth]{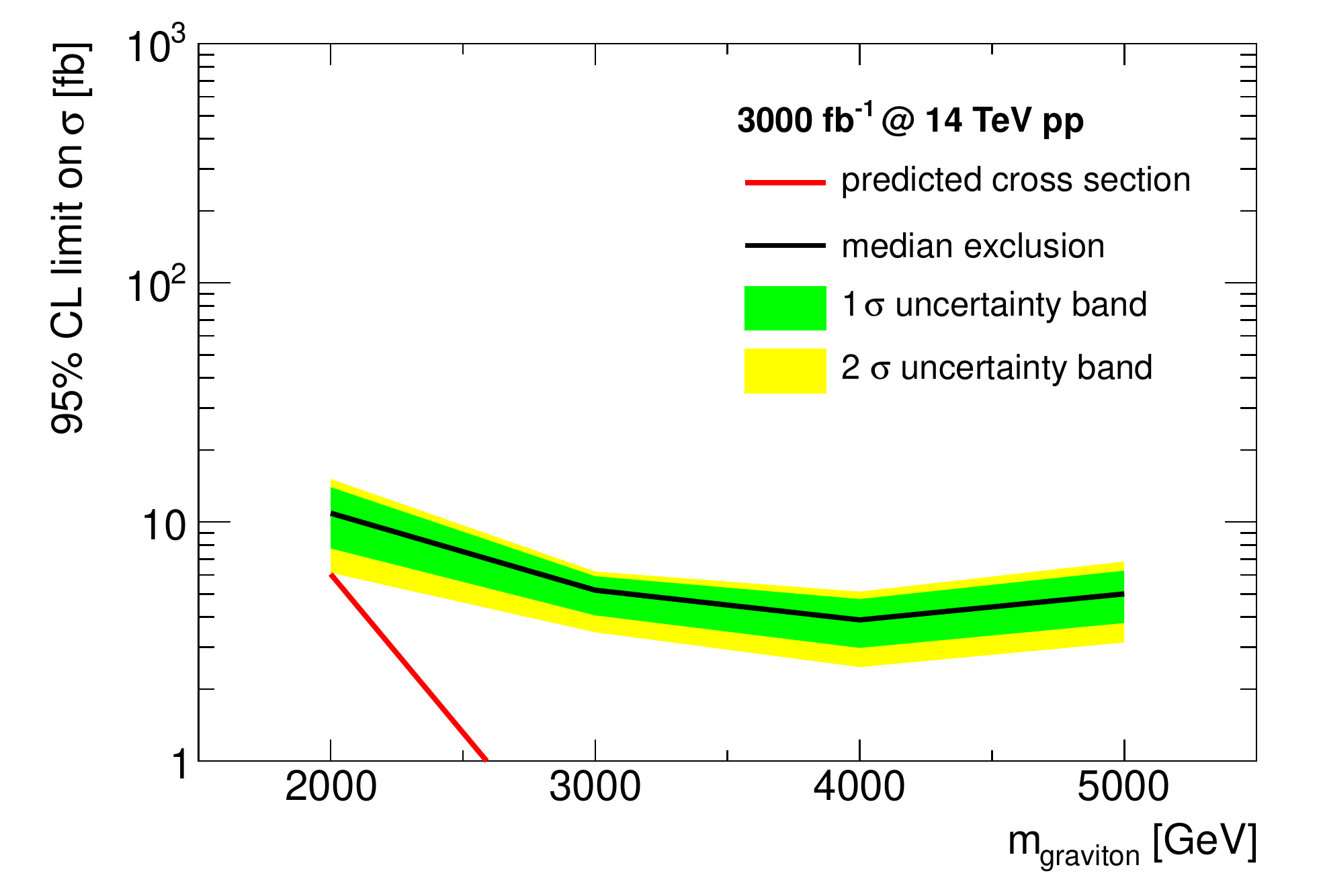}
  \includegraphics[width=0.4\textwidth]{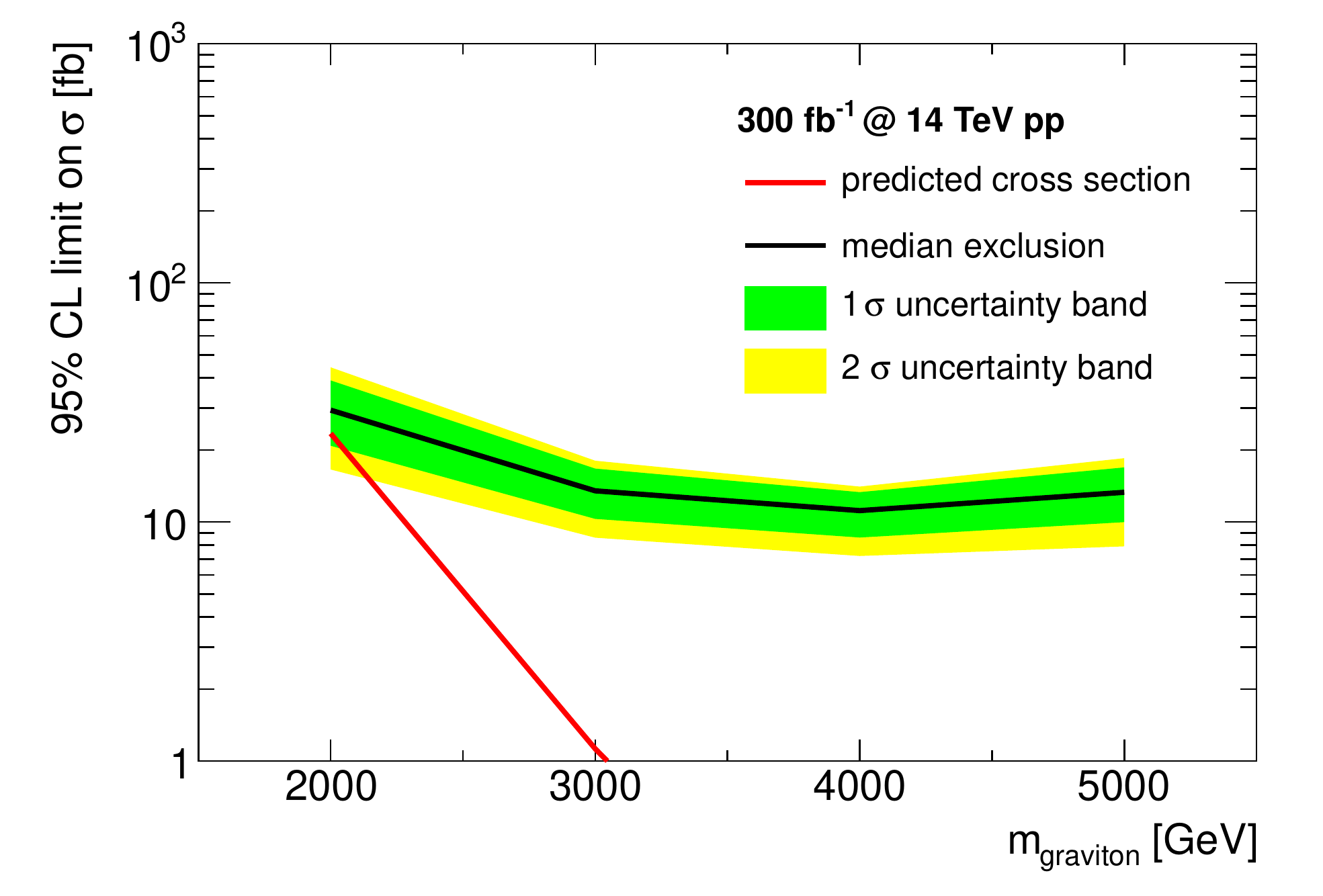}
  \includegraphics[width=0.4\textwidth]{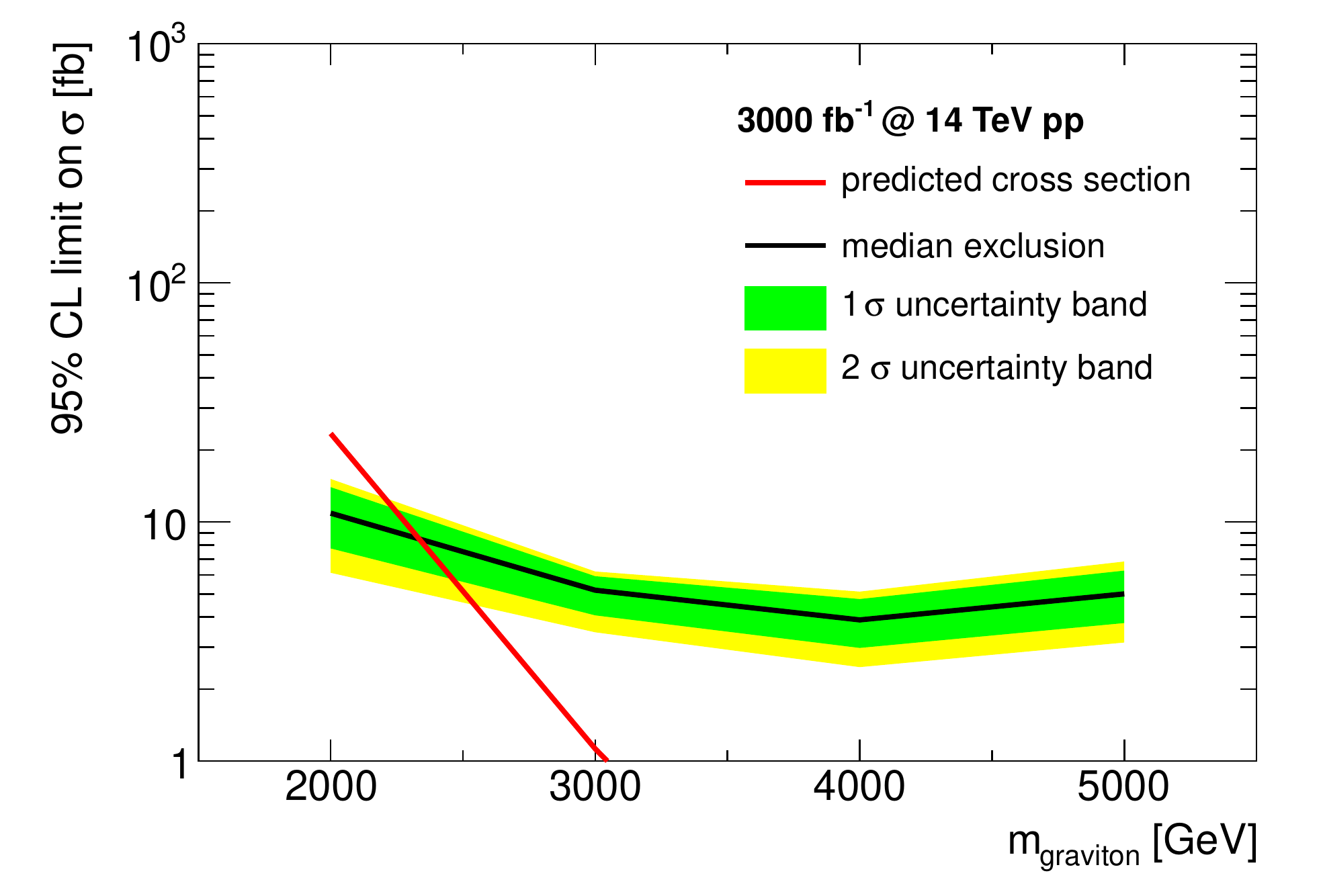}
\caption{
    Expected cross section upper limits at 95\% CL for the $t_R$ KK graviton model with parameter $c = 0.5$ and
    $c = 1.0$ (upper and lower plots)
    decaying to a top quark pair in the all-hadronic channel. The left plots show the limits for 300~fb$^{-1}$ of 14 TeV data and $\mu = 50$.
    The right plots show the limits for 3000~fb$^{-1}$ of data and $\mu = 140$ (right).
}
\vspace{0.5cm}
\label{fig:KKglimit_grav1}
  \includegraphics[width=0.4\textwidth]{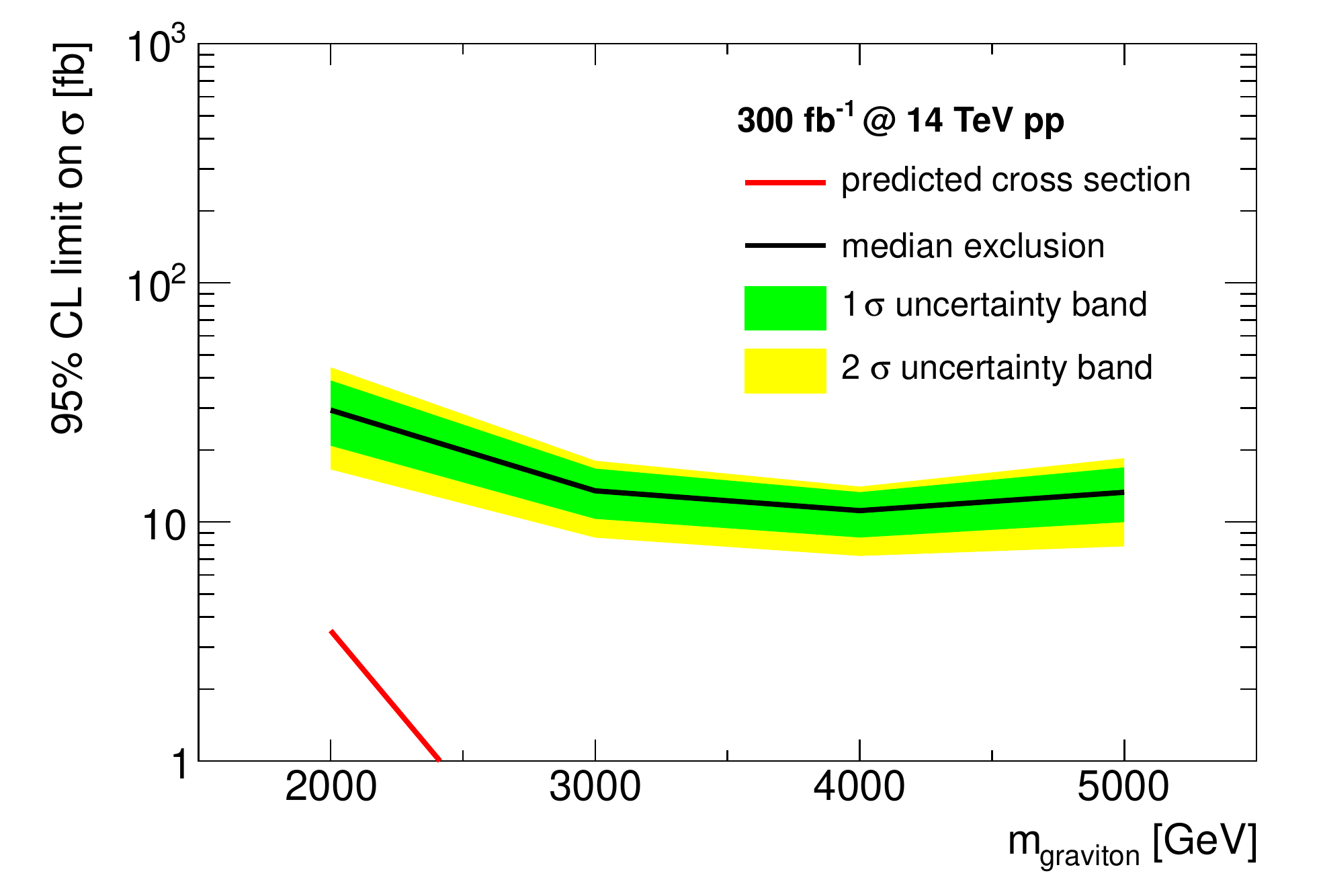}
  \includegraphics[width=0.4\textwidth]{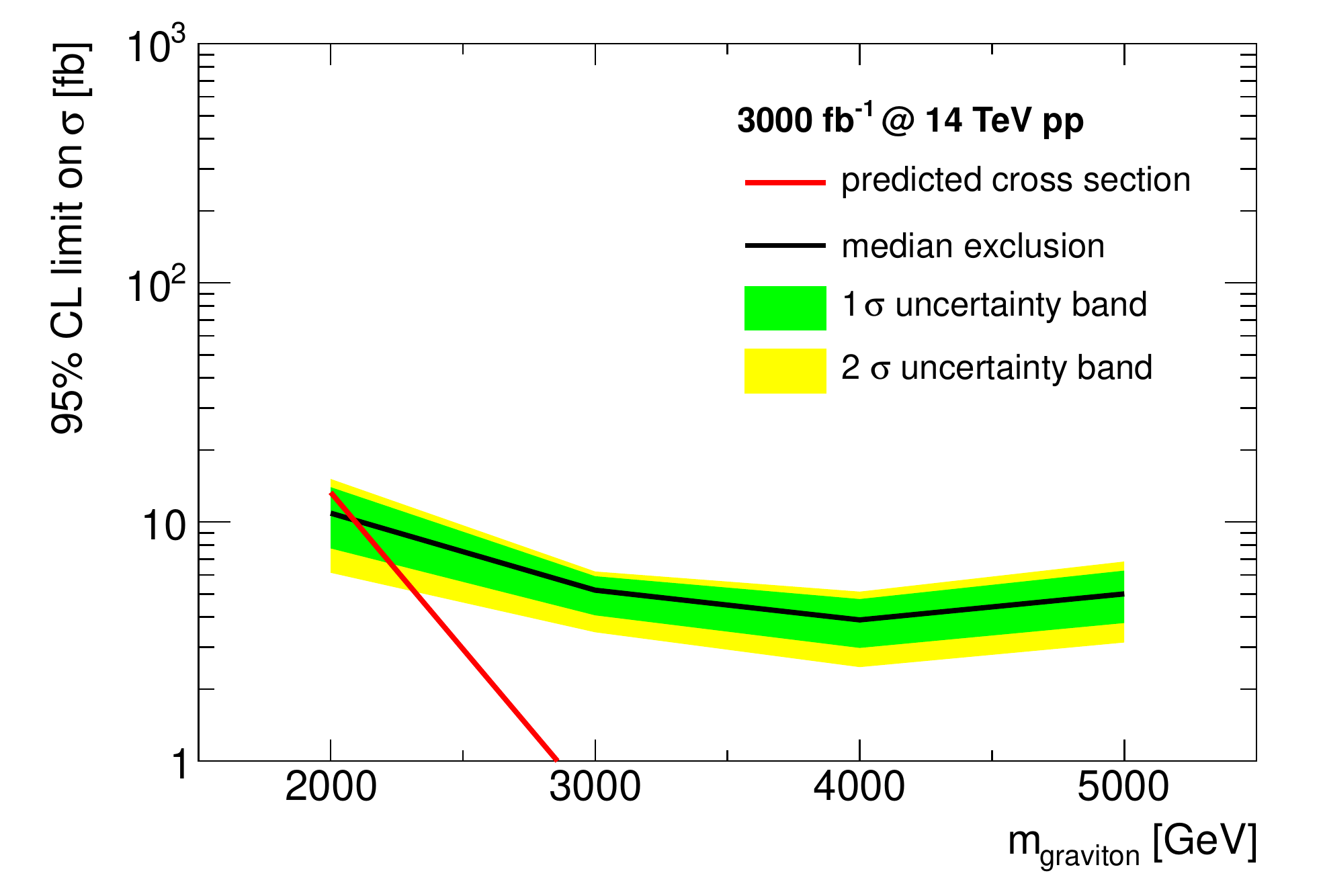}
  \includegraphics[width=0.4\textwidth]{figures/limit_GRAV_L_05_50_300_syst.pdf}
  \includegraphics[width=0.4\textwidth]{figures/limit_GRAV_L_1_140_3000_syst.pdf}
\caption{
    Expected cross section upper limits at 95\% CL for the $t_L$ KK graviton model with parameter $c = 0.5$ and
    $c = 1.0$ (upper and lower plots)
    decaying to a top quark pair in the all-hadronic channel. The left plots show the limits for 300~fb$^{-1}$ of 14 TeV data and $\mu = 50$.
    The right plots show the limits for 3000~fb$^{-1}$ of data and $\mu = 140$ (right).
}
\label{fig:KKglimit_grav2}
\end{figure}

Figs.~\ref{fig:KKglimit_grav1} and~\ref{fig:KKglimit_grav2}
show the 95\% CL limits on the cross section of a narrow resonance overlaid with the expected cross sections for the KK graviton models.
The mass limits for the KK gravitons are given in Tab.~\ref{tab:Zprimelimits_grav} for the different pile-up and luminosity scenarios for statistical
uncertainties only and with systematic uncertainties included.
If the mass limits is denoted $< 2 \TeV$, the analysis is not sensitive to resonance mass above 2~TeV.
With 3000~fb$^{-1}$ of data at 14 TeV, KK gravitons can be excluded with masses up to 2.8 (2.3)~TeV in the $t_R$ ($t_L$) scenarios with $c = 1.0$.
For $c = 0.5$, however, the sensitivity does not reach 2~TeV for the resonance mass when systematic uncertainties are taken into account.

%Tab.~\ref{tab:Zprimelimits_grav}

\begin{table}[p]
\centering
\caption{95\% CL limits on the resonance mass for the two KK graviton models with $c = 0.5$ and $c = 1.0$
  decaying to a top quark pair in the all-hadronic channel.
  Limits are given for pile-up scenarios and integrated luminosities.
}
{
\small
\begin{tabular}{|c|c|c|c|c|}
\hline
scenario & lumi. & pile-up & uncertainties & mass reach
\tabularnewline
\hline
\hline
R 0.5 & 300 fb$^{-1}$ & $\mu = 0$ & stat.          & $<$ 2 TeV \\
R 0.5 & 300 fb$^{-1}$ & $\mu = 0$ & stat.+syst.    & $<$ 2 TeV \\
R 0.5 & 300 fb$^{-1}$ & $\mu = 50$ & stat.         & $<$ 2 TeV \\
R 0.5 & 300 fb$^{-1}$ & $\mu = 50$ & stat.+syst.   & $<$ 2 TeV \\
R 0.5 & 300 fb$^{-1}$ & $\mu = 140$ & stat.        & $<$ 2 TeV \\
R 0.5 & 300 fb$^{-1}$ & $\mu = 140$ & stat.+syst.  & $<$ 2 TeV
\tabularnewline
\hline
R 0.5 & 3000 fb$^{-1}$ & $\mu = 0$ & stat.         & 2.4 TeV \\
R 0.5 & 3000 fb$^{-1}$ & $\mu = 0$ & stat.+syst.   & $<$ 2 TeV \\
R 0.5 & 3000 fb$^{-1}$ & $\mu = 50$ & stat.        & 2.3 TeV \\
R 0.5 & 3000 fb$^{-1}$ & $\mu = 50$ & stat.+syst.  & $<$ 2 TeV \\
R 0.5 & 3000 fb$^{-1}$ & $\mu = 140$ & stat.       & 2.3 TeV \\
R 0.5 & 3000 fb$^{-1}$ & $\mu = 140$ & stat.+syst. & $<$ 2 TeV
\tabularnewline
\hline
R 1.0 & 300 fb$^{-1}$ & $\mu = 0$ & stat.          & 2.6 TeV \\
R 1.0 & 300 fb$^{-1}$ & $\mu = 0$ & stat.+syst.    & $<$ 2 TeV \\
R 1.0 & 300 fb$^{-1}$ & $\mu = 50$ & stat.         & 2.6 TeV \\
R 1.0 & 300 fb$^{-1}$ & $\mu = 50$ & stat.+syst.   & $<$ 2 TeV \\
R 1.0 & 300 fb$^{-1}$ & $\mu = 140$ & stat.        & 2.5 TeV \\
R 1.0 & 300 fb$^{-1}$ & $\mu = 140$ & stat.+syst.  & $<$ 2 TeV
\tabularnewline
\hline
R 1.0 & 3000 fb$^{-1}$ & $\mu = 0$ & stat.         & 3.0 TeV \\
R 1.0 & 3000 fb$^{-1}$ & $\mu = 0$ & stat.+syst.   & 2.9 TeV \\
R 1.0 & 3000 fb$^{-1}$ & $\mu = 50$ & stat.        & 2.9 TeV \\
R 1.0 & 3000 fb$^{-1}$ & $\mu = 50$ & stat.+syst.  & 2.8 TeV \\
R 1.0 & 3000 fb$^{-1}$ & $\mu = 140$ & stat.       & 2.9 TeV \\
R 1.0 & 3000 fb$^{-1}$ & $\mu = 140$ & stat.+syst. & 2.8 TeV
\tabularnewline
\hline
L 0.5 & 300 fb$^{-1}$ & $\mu = 0$ & stat.          & $<$ 2 TeV \\
L 0.5 & 300 fb$^{-1}$ & $\mu = 0$ & stat.+syst.    & $<$ 2 TeV \\
L 0.5 & 300 fb$^{-1}$ & $\mu = 50$ & stat.         & $<$ 2 TeV \\
L 0.5 & 300 fb$^{-1}$ & $\mu = 50$ & stat.+syst.   & $<$ 2 TeV \\
L 0.5 & 300 fb$^{-1}$ & $\mu = 140$ & stat.        & $<$ 2 TeV \\
L 0.5 & 300 fb$^{-1}$ & $\mu = 140$ & stat.+syst.  & $<$ 2 TeV
\tabularnewline
\hline
L 0.5 & 3000 fb$^{-1}$ & $\mu = 0$ & stat.         & $<$ 2 TeV \\
L 0.5 & 3000 fb$^{-1}$ & $\mu = 0$ & stat.+syst.   & $<$ 2 TeV \\
L 0.5 & 3000 fb$^{-1}$ & $\mu = 50$ & stat.        & $<$ 2 TeV \\
L 0.5 & 3000 fb$^{-1}$ & $\mu = 50$ & stat.+syst.  & $<$ 2 TeV \\
L 0.5 & 3000 fb$^{-1}$ & $\mu = 140$ & stat.       & $<$ 2 TeV \\
L 0.5 & 3000 fb$^{-1}$ & $\mu = 140$ & stat.+syst. & $<$ 2 TeV
\tabularnewline
\hline
L 1.0 & 300 fb$^{-1}$ & $\mu = 0$ & stat.          & $<$ 2 TeV \\
L 1.0 & 300 fb$^{-1}$ & $\mu = 0$ & stat.+syst.    & $<$ 2 TeV \\
L 1.0 & 300 fb$^{-1}$ & $\mu = 50$ & stat.         & $<$ 2 TeV \\
L 1.0 & 300 fb$^{-1}$ & $\mu = 50$ & stat.+syst.   & $<$ 2 TeV \\
L 1.0 & 300 fb$^{-1}$ & $\mu = 140$ & stat.        & $<$ 2 TeV \\
L 1.0 & 300 fb$^{-1}$ & $\mu = 140$ & stat.+syst.  & $<$ 2 TeV
\tabularnewline
\hline
L 1.0 & 3000 fb$^{-1}$ & $\mu = 0$ & stat.         & 2.9 TeV \\
L 1.0 & 3000 fb$^{-1}$ & $\mu = 0$ & stat.+syst.   & 2.6 TeV \\
L 1.0 & 3000 fb$^{-1}$ & $\mu = 50$ & stat.        & 2.9 TeV \\
L 1.0 & 3000 fb$^{-1}$ & $\mu = 50$ & stat.+syst.  & 2.6 TeV \\
L 1.0 & 3000 fb$^{-1}$ & $\mu = 140$ & stat.       & 2.8 TeV \\
L 1.0 & 3000 fb$^{-1}$ & $\mu = 140$ & stat.+syst. & 2.3 TeV
\tabularnewline
\hline
\end{tabular}
}
\label{tab:Zprimelimits_grav}
\end{table}

\newpage
\subsection{KK gluon \protect\cite{Agashe:2006hk}}

Kinematics of heavy $s$-channel resonance decays are characterized by decay products with a significant boost. The energy frontier searches for heavy di-top resonances thus hinge on our ability to efficiently tag highly energetic top quarks. Identifying boosted tops with $p_T < 1 \TeV$ has become possible in the light of numerous recent developments in the field of jet substructure \cite{Ellis:2007ib,Abdesselam:2010pt,Salam:2009jx,Nath:2010zj,Almeida:2011ud,Soper:2012pb, Soper:2011cr,Almeida:2010pa, Plehn:2011tg}, while efficiently tagging the tops in the ultra-high boosted regime (e.g. $p_T > 1 \TeV$) remains a challenge.  In this section, using RS KK gluon as an example, we estimate the discovery potential for a top pair resonance in the future 14  $\TeV$ LHC experiment using jet substructure techniques (more precisely, the Template Overlap Method (TOM) \cite{Almeida:2010pa} ).
TOM has been extensively studied in the past in the context of theoretical studies of boosted top and boosted Higgs decays \cite{ Almeida:2011ud, Almeida:2010pa,Backovic:2012jj}, as well as used by the ATLAS collaboration for a boosted resonance search \cite{Aad:2012raa} at $\sqrt{s} = 7 \TeV$. The method is designed to match the energy distribution of a boosted jet to the parton-level configuration of a boosted top decay, with all kinematic constraints taken into account. Because our analysis focuses on events in which one top decays hadronically and the other semi-leptonically, we employ two different formulations of TOM. The hadronic TOM \, which is designed to tag the fully hadronic decays of the top, and the leptonic TOM \cite{LeptonicTemp}, designed to take into account decays with missing energy \footnote{We use the TemplateTagger v.2.0 \cite{Backovic:2012jk} implementation of TOM for both the hadronic and the leptonic formulations. }. The latter is particularly useful in compensating for the loss of rejection power due to the absence of efficient $b$-tagging methods at high $p_T$. 

Low susceptibility to intermediate levels of pileup (i.e. 20-50 interactions per bunch crossing), makes TOM particularly attractive for boosted top analyses at the LHC. Refs. \cite{  Backovic:2012jj, LeptonicTemp} showed that the ability of TOM to tag the top jets is nearly unaffected even at 20-50 interactions per bunch crossing, with the $W$+jets mistag rate changing by as little as $10 \%$ at, thus nearly eliminating the need for pileup subtraction / correction at intermediate pileup levels.

\subsubsection{Template Overlap Method}

We begin with the definition of the hadronic peak template overlap \cite{ Almeida:2011ud,Almeida:2010pa, Backovic:2012jj}:
\begin{equation}
 Ov_3^{h}(j,f[j]) = \max_{\{f\}}\,\left[ \exp\left[ -\sum_{a=1}^3 \frac{1}{\sigma_a^2}\left( \epsilon \,  p_{T,a} -\sum_{i\in j} p_{T,i} \,F(\hat n_i,\hat n_a) \right)^2 \right] \right], \label{eq:ovHad}
\end{equation}

where $p_{T,a}$ is the transverse momentum of the $a^{th}$ template parton, $p_{T,i}$ is the transverse momentum of the $i^{th}$ jet constituent and $f$ is the set of all possible parton level configurations which satisfy the kinematic constraints of a boosted top decay (called ``templates''): 
\be
	\sum_a p_{T,a} = P, \,\,\,\,\,\,\, P^2 = m_t^2, \,\,\,\,\,\,\,\, (p_1 + p_2)^2 = m_W^2,
\ee
where $p_{1,2}$ are two of the three template momenta and $P$ is the four momentum of the boosted top. 

The function $ \,F(\hat n_i,\hat n_a) $ restricts the sum over $i$ and $a$ to non-intersecting angular regions centered around each template momentum $p_a$:
\begin{equation}
 F(\hat n_i,\hat n_a) = \left\{ \begin{array}{rl}
 1 &\mbox{ if $ \Delta R < r_a$} \\
  0 &\mbox{otherwise}
       \end{array} \right. , \label{eq:kernel}
\end{equation}
where $\Delta R = \sqrt{\delta \phi^2 + \delta \eta^2}$ is the distance between the template parton and the jet constituent.

 The coefficient $\epsilon$ in Eq. \ref{eq:ovHad} compensates for the energy deposited outside the template sub-cone, while the weight $\sigma_a$ defines the resolution of the observable $Ov_3^h$. Here we use $\sigma_a = p_{T,a} / 3$. For simplicity, we will keep all $\epsilon = 1,$ without the loss of generality or significant effects on the mass reach of resonance searches.

Similar to Eq. \ref{eq:ovHad} we use the definition of leptonic peak overlap of Ref. \cite{LeptonicTemp}: 
\footnotesize
\begin{equation}
 Ov_3^l =\max_{\{f\}}\, \left[ \underbrace{ \exp \frac{-1}{\sigma_b^2}\left( \epsilon \,  k_{T,b} -\sum_{i\in j} p_{T,i} \,F(\hat n_i,\hat n_a) \right)^2  }_{\text{b quark}}\,\, \underbrace{\exp \frac{-1}{\sigma_l^2}\left( \epsilon_l \,  k_{T,l} -p_{T,l}  \right)^2  }_{\text{lepton}} 
 		 \,\, \underbrace{\exp \frac{-1}{\sigma_\nu^2}\left( \epsilon_\nu \,  k_{T,\nu} - \sl{E_T} \,\,\,F'(\phi_\nu, \phi_{\sl{E_T}}\,\,\,) \right)^2}_{\text{neutrino}}   \right], \label{eq:lep_overlap}
\end{equation}
\normalsize
where $F'$ is defined as
\begin{equation}
F'(\phi_\nu, \phi_{\sl{E_T}}\,\,\,) = \left\{ \begin{array}{rl}
 1 &\mbox{ if $ \Delta \phi_{\nu, \sl{E_T}} < r_\nu$} \\
  0 &\mbox{otherwise}
       \end{array} \right. ,
\end{equation}
 and $\Delta \phi$ is the azimuthal distance between the template parton and the total $\sl{E_T}\,\,\,$. 

The $Ov_3^l$ definition differs from Eq. \ref{eq:ovHad} in two major ways. First, Eq. \ref{eq:lep_overlap} explicitly keeps track of the particle species and second, only the transverse component of the missing energy is taken into account. We choose to rotate each leptonic top template into the frame in which the first template parton is aligned with the lepton, due to the absence of a well defined ``jet axis''. Note that the two formulations of template overlap are compatible enough so that the same set of template states can be used in both. 

For the purpose of our analysis, we generate the template sets in the boosted top frame at fixed transverse momentum with a sequential scan in $50$ steps in $\eta, \phi$ for the first two template momenta, while the third parton is determined by conservation of energy. We consider a range of $p_T = 500 - 1600 \GeV$ in steps of $100 \GeV,$ resulting in 12 template $p_T$ bins. 

For hadronic top templates we use fixed template sub-cones at fixed template $p_T$, with dependence of the sub-cone size with the transverse momentum of the boosted top described by the scaling rule
\be
	r_a(p_{T,a}) =\frac{70 \GeV}{p_{T,a}}, \label{eq:scaling}
\ee
where we limit the range of $r_a$ to the interval of $[0.07, 0.3]$. The lower limit on the template sub cone size serves to take into account the detector resolution, as well as match the optimal value for the template sub cone at $p_T = 1 \TeV$. 

The leptonic top template for the $b$ quark uses the same scaling rule of Eq. \ref{eq:scaling}, while for the neutrino we use a fixed $r_\nu = 0.2$. 

The output of TOM is not limited to the peak overlap score, $Ov_3^{h,l}. $ Additional information about the boosted top decay is contained in the peak template momenta. An additional correlation, useful in discriminating boosted tops from light jets is Template Planar Flow \cite{Almeida:2010pa}:
\begin{equation}
tPf \equiv \frac{4 \,\det(I)}{{\rm tr}(I)^2}\,, \label{eq:pf}
\end{equation}
where $$ I ^{kl} \equiv \frac{1}{m_t} \, \sum_{i=1}^3 \frac{p_k^i p^i_l}{ p_T^i}\, , $$ and $ p^i_k \equiv p_T^i (\eta^i, \phi^i)$ are the two-vectors constructed from $p_T, \eta, \phi$ of the three peak template momenta. Template planar flow is useful in distinguishing decay configurations which lie on a straight line in the plane perpendicular to the top boost (low $tPf \approx 0$) from the decays which are not aligned on a line ($tPf \approx 1$). The formulation of $tPf$ is similar to Jet Planar Flow, with an important difference that it is intrinsically infra-red safe and weakly sensitive to pileup.

\subsubsection{Results}

We consider two Snowmass benchmark points: the Case 1 RS KK gluon  (Right handed, top near the brane) of $M_{G'_{\rm KK}} = 3, \, 5 \TeV,$ and a Case 2 RS KK gluon (Left handed, bottom near the brane) with $M_{G'_{\rm KK}} = 3, \, 5 \TeV$ as well.  The coupling strengths to left and right heavy quarks in both cases are given in Table \ref{tab:KKg_couplings}, while for production, we take the coupling to light quarks to be $-0.2$ of SM QCD (for both Cases 1 and 2). Note that the SM QCD coupling here should be evaluated at a few TeV.

We consider a signal of the form
\bea
q \bar{q} \rightarrow & \hbox{KK gluon} \rightarrow & t \bar{t},
\eea

where we require one top to decay hadronically and the other leptonically. The main background channels consist of SM $t\bar{t}$ and $Wjj$.

\begin{table}
\begin{center}
\begin{tabular}{|c||c|c|}
\hline 
& Case 1
& Case 2 
\tabularnewline
\hline
\hline 
$(t,b)_L$ & 1 & 3.9
\tabularnewline
\hline 
$t_R$ & 3.9 & 1 
\tabularnewline
\hline 
\end{tabular}
\end{center}
\caption{Coupling strengths of the KK gluon to the left/right heavy flavors for the Case 1 and 2 benchmark points.}
\label{tab:KKg_couplings}
\end{table}

We generate the data using MadGraph \cite{Alwall:2011uj} + Pythia \cite{Pythia8}, with CTEQ6M \cite{Nadolsky:2008zw} parton distribution functions at $\sqrt{s} = 14 \TeV$, while jet clustering is performed using the FastJet \cite{Cacciari:2008gp} implementation of the anti-$k_T$ algorithm with a jet cone of $R=1.0$ for the fat jet. All SM events are matched to an extra jet using the MLM matching \cite{Mangano:2006rw} prescription. Note that we do not impose an implicit mass window on the top jets, and instead rely on the intrinsic mass filtering ability of TOM as a pileup insensitive alternative to a mass window. 

We separately analyze events with no pileup and with $\langle N_{vtx} \rangle = 50$ interactions per bunch crossing in order to illustrate the effects of pileup on the analysis. Note that we do not employ any pileup subtraction / correction to the events, but instead focus on observables which are weakly sensitive to pileup. For instance, we determine the transverse momentum of the hadronically decaying top based on the scalar sum of the transverse momenta of the decay products of the leptonic top which recoils against it. The method serves both as a pileup-insensitive estimate of the hadronic top $p_T$ as well as a discriminant against SM di-top events in which significant next-to-leading order effects of hard gluon emissions are non-negligible.

%Fig. \ref{fig:ov3gkk} show the results of the overlap analysis on the signal events. In both cases the method is able to tag the signal tops with a high efficiency, whereas the peak at low overlap values can be attributed to the events  in which one of the top decay products is outside the jet cone, thus producing a fat-jet with lower mass (background will be shown later). 

We pre-select the events which satisfy the following ``Basic Cuts'':
\begin{eqnarray}
	p_T^{j\, R=1.0} > 500 \GeV,   \,\, \,\,\,&     \sl{E_T}\,\,\, > 40 \GeV,  \,\,& N_l^{out} (p_T^l > 25 \GeV) =1,      \nn \\      
	N_j^{out} (p_T^j > 25 \GeV) \ge 1, \,\, & \Delta \phi_{jl} > 2.3, \,\, & \eta_{j, \,l} < 2.5,
	\label{eq:bc} 
\end{eqnarray}

where $p_T^{j\, R=1.0}$ is the transverse momentum of the hardest fat jet, $N_{l,j}^{out}$ is the number of leptons with $p_T > 25 \GeV$ and mini-ISO $>0.95 \,(l)$ \cite{Rehermann:2010vq}  or anti-$k_T$ $r=0.4$ jets outside the fat jet $(j)$, and $\Delta \phi_{jl}$ is the transverse plane angle between the fat jet and the lepton. 

Our ``leptonic top''  candidate consists of the mini-isolated lepton, missing energy and a $b$-jet candidate, which we select to be the hardest $r=0.4$ anti-$k_T$ jet within the distance of $R = 1.5$ from the lepton.

%We then consider the other Snowmass benchmark point we consider is the Case 1 RS KK gluon  (Right handed, near the brane) of $M_{G'_{\rm KK}} = 5$, 6 and $8 \TeV$ at $\sqrt{s} = 14,\, 33 \TeV$, and show how far can $14 \TeV$ with 300 fb$^{-1}$ vs 3000 fb$^{-1}$ and 3000 fb$^{-1}$ $33 \TeV$ LHC reach for higher resonance searches (to be done).

% Mkk = 3TeV 50 p

We define our signal region with the following substructure cuts (in addition to Basic Cuts):
\begin{eqnarray}
 Ov_3^l > 0.5, \,\,\,\,\,\,\,\, Ov_3^h > 0.5, \,\,\,\,\,\,\,\,\, Ov_3^h + tPf > 1.0.
\label{eq:ovcuts}
\end{eqnarray}
In addition, we only consider events with $m_{T\bar{T}} > 2.8, 3.5 \TeV$ for $M_{G'_{\rm KK}}  = 3, 5 \TeV$ respectively, where $m_{T\bar{T}}$ is the invariant mass of the leptonic and hadronic \textit{peak templates}.

Table \ref{tab:temp_results} summarizes our results for Case 1. TOM is able to improve $S/B$ relative to the Basic Cuts by a factor of 10, while the signal significance improves by three-fold. The effect of 50 average pileup events is mild, with the apparent improvement in significance being due to the requirement on the lower $p_T$ of the fat jet in the Basic Cuts. More background events pass the Basic Cuts in the presence of pileup but are efficiently rejected by TOM, thus causing a virtual improvement in the signal significance. The efficiency of the overlap cut on SM $t\bar{t}$ events is significantly lower than the signal $t\bar{t}$ events, due to the higher order effects becoming more prominent in high energy SM $t\bar{t}$ events. 

We achieve significantly better sensitivity for higher masses at $L = 3000 {\rm \, fb}^{-1}$, with $S/\sqrt{B} \approx 4-5$  for $M_{G'_{\rm KK}}  = 5 \TeV$. Compared to $300 {\rm \,fb}^{-1},$ and $S/\sqrt{B}\approx 1.5,$ we find that increasing luminosity in Case 1 could significantly improve the reach of the search.

\begin{figure}[htb]
\begin{center}
\includegraphics[width=3in]{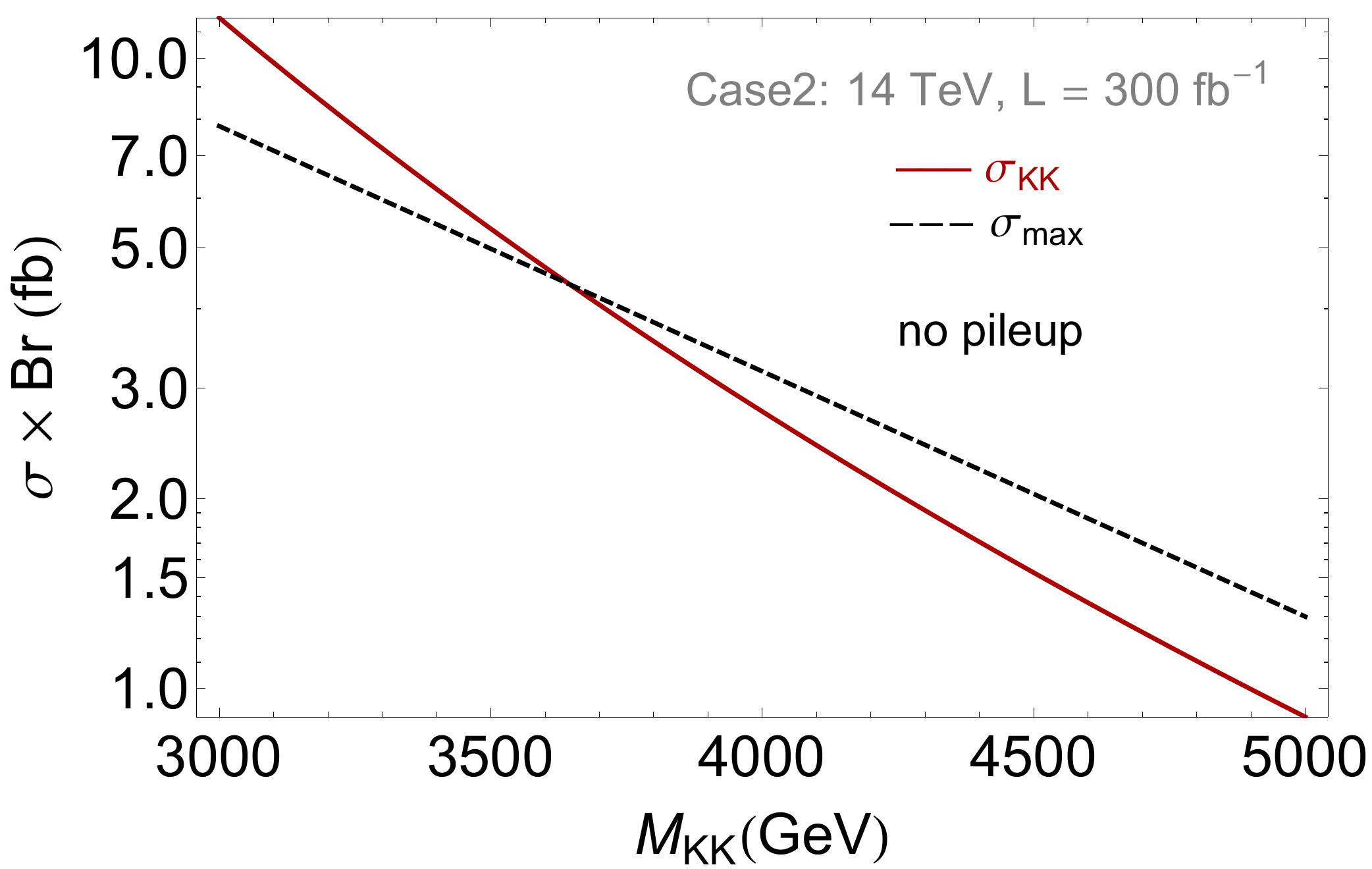}\includegraphics[width=3in]{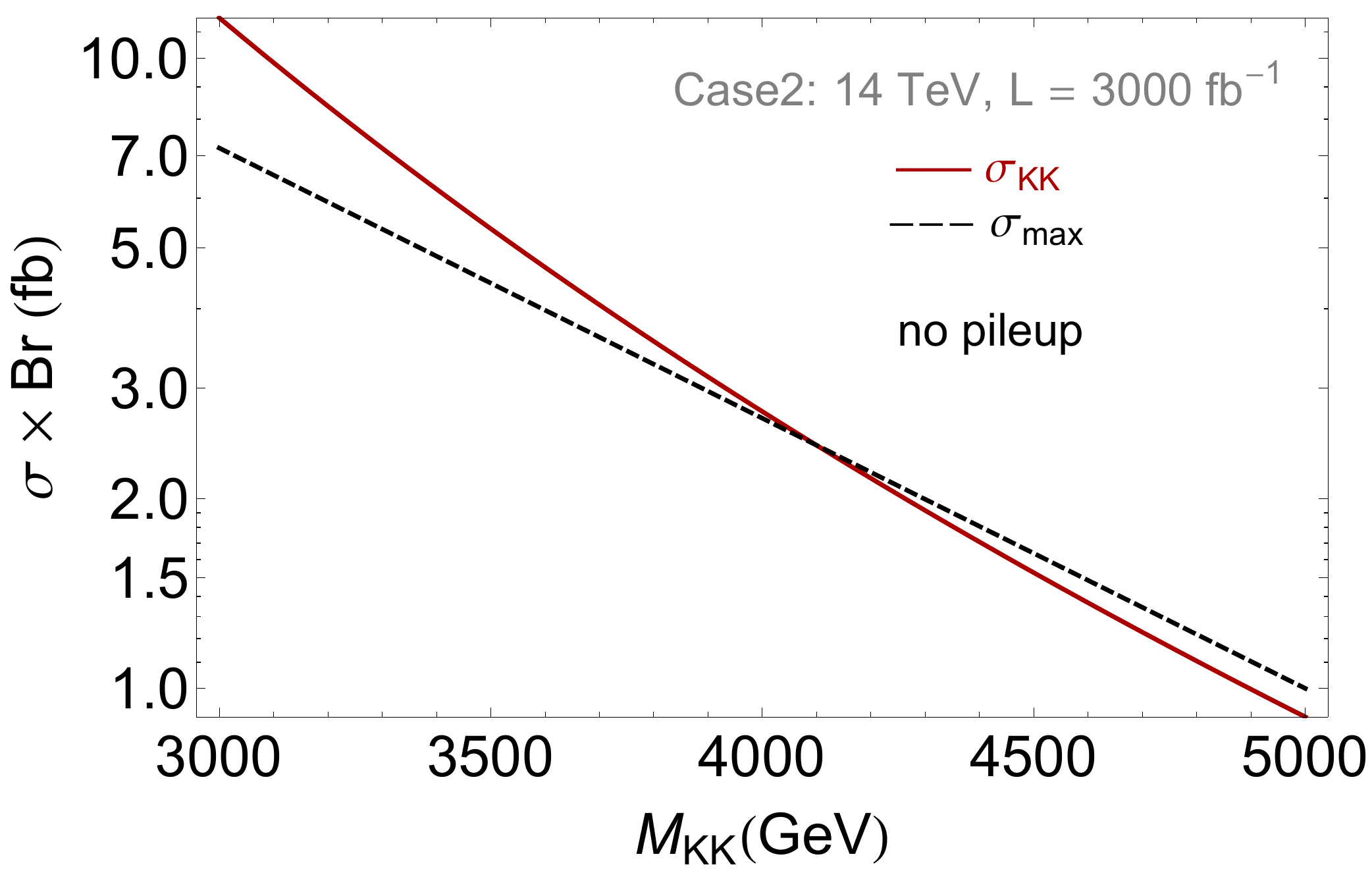}
\caption{95 $\%$ confidence level reach for the Case 2 KK gluon search at $\sqrt{s} = 14 \TeV$ with $L = 300, 3000 \, {\rm fb}^{-1}$. The dashed line represents the upper limit on the cross section whereas the solid line is the leading order KK gluon cross section.  }
\label{fig:reaches2}
\end{center}
\end{figure}

\begin{figure}[htb]
\begin{center}
\includegraphics[width=3in]{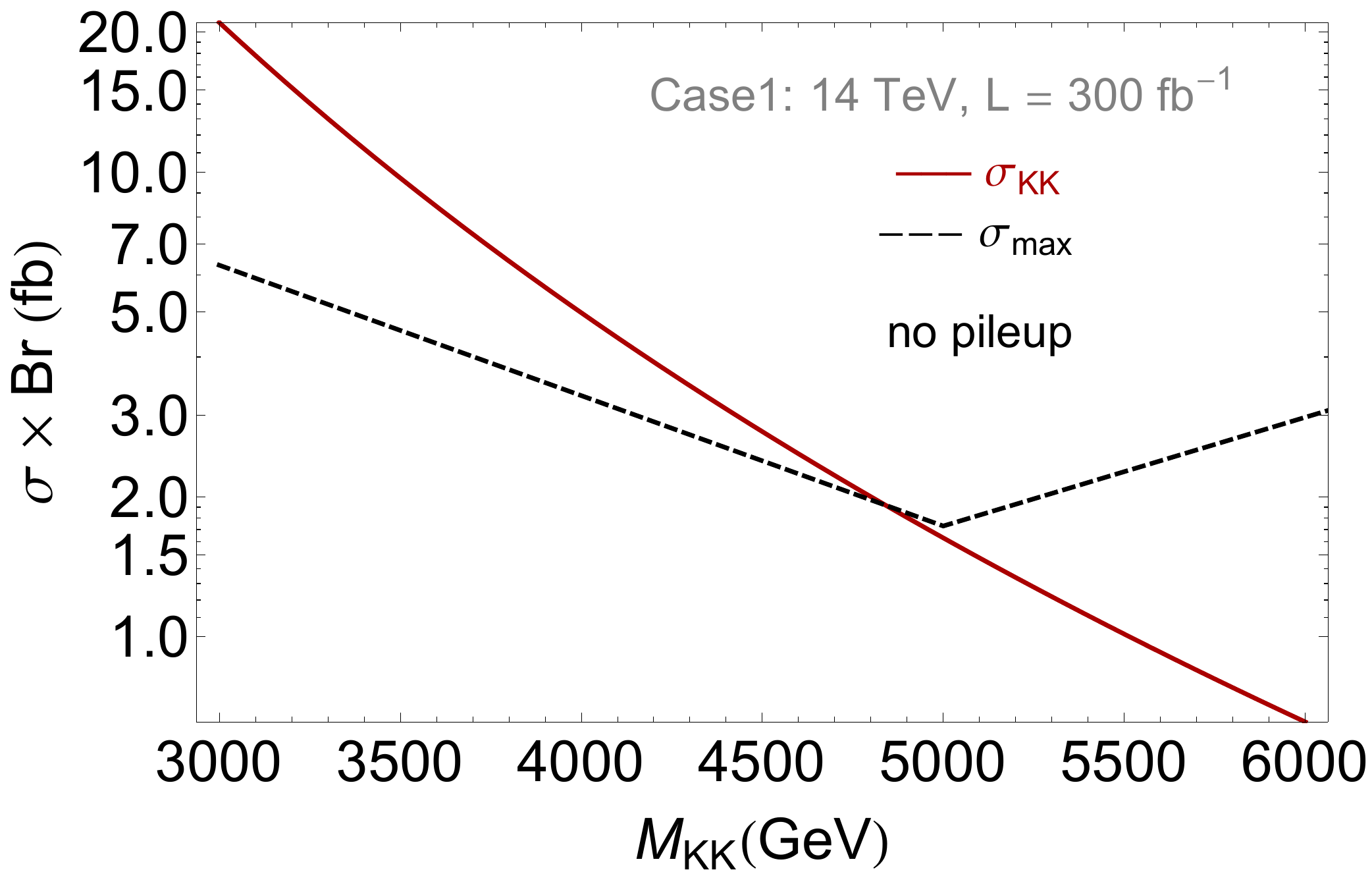}\includegraphics[width=3in]{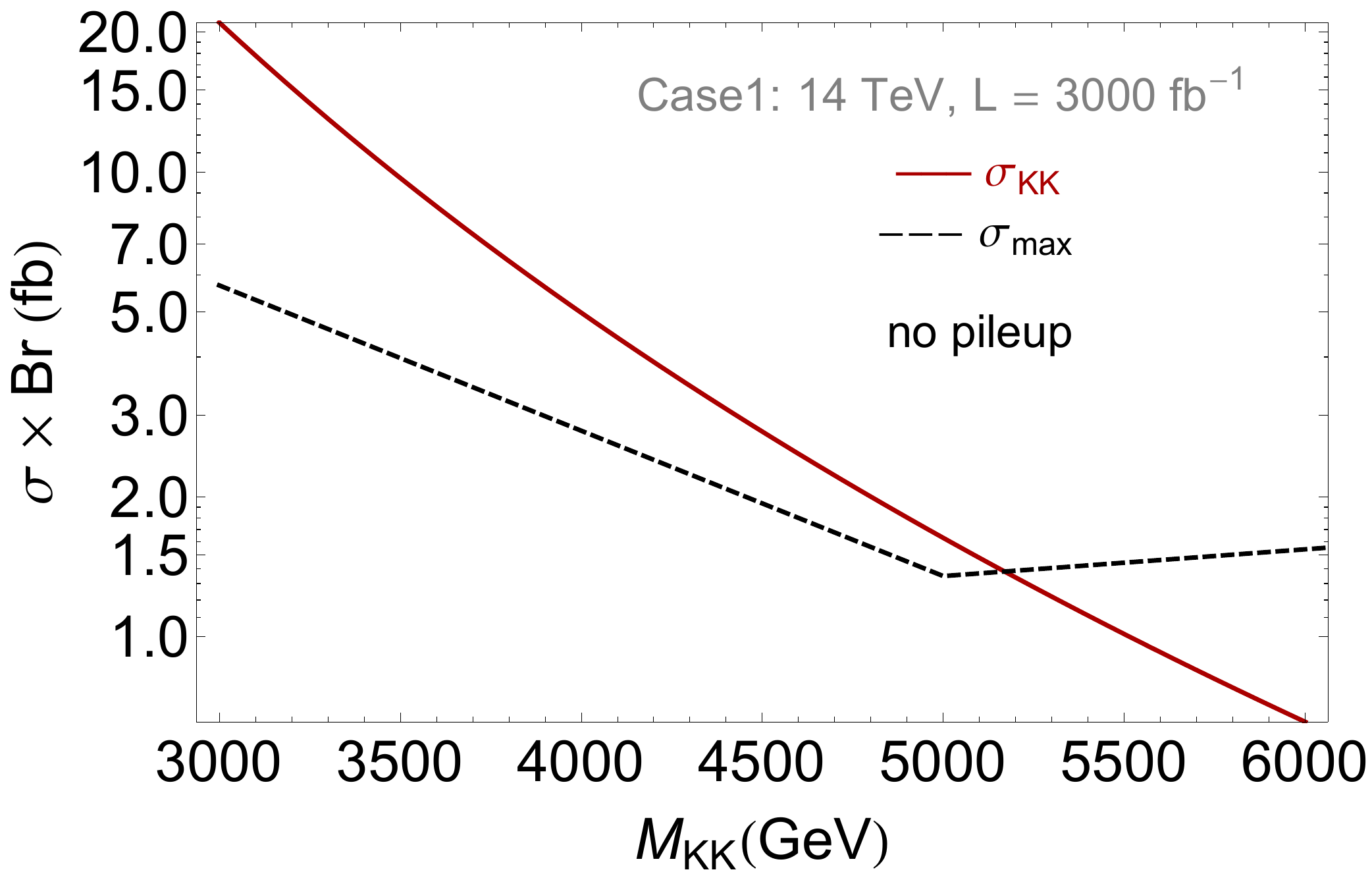}
\caption{ 95 $\%$ confidence level reach for the Case 1 KK gluon search at $\sqrt{s} = 14 \TeV$ with $L = 300, 3000 \, {\rm fb}^{-1}$. The dashed line represents the upper limit on the cross section whereas the solid line is the leading order KK gluon cross section.  }
\label{fig:reaches1}
\end{center}
\end{figure}

Similarly, Table \ref{tab:temp_results2} summarizes our results for Case 2. We find that TOM again improves the signal significance by three-fold while $S/B$ is improved by a factor of roughly 10. However, due to the lower signal cross section, neither the increase in luminosity to $L = 3000 {\rm \, fb}^{-1}$  nor TOM sufficiently improves the significance for $M_{G'_{\rm KK}}  = 5 \TeV$. Effects of pileup at $50$ average interactions per bunch crossing are mild, with the results in $S/B$ and $S/\sqrt{B}$ being within $10 \%$. We find that an increase in center of mass energy alone does not suffice to extend the experimental reach for the Case 2 KK gluon with mass $M_{G'_{\rm KK}}  = 5 \TeV$. However, a combination of increase in $\sqrt{s}$ to $33 \TeV$ and the integrated luminosity of $3000 \,{\rm fb^{-1}}$ can yield a $7 \sigma$ signal as shown in Table~\ref{tab:33TeVCase2}.

Fig. \ref{fig:reaches2} shows the results for the $95 \%$ CL sensitivity of the Case 2 KK gluon search at $\sqrt{s} = 14 \TeV$ and $L = 300, 3000 \,{\rm fb}^{-1}.$ We include a $5 \%$ systematic uncertainty in the background into the calculation via a convolution with a log-normal distribution centered around the expected number of background events.  We find that KK gluon masses up to $\approx 3.6 \TeV$ can be excluded at $L = 300 {\rm \, fb}^{-1},$ and masses up to $\approx 4.1 \TeV$ at $L = 3000 {\rm \, fb}^{-1}.$ 

Fig.  \ref{fig:reaches1} shows the corresponding sensitivity for the Case 1 KK gluon. We find that masses up to $M_{G'_{\rm KK}}  \approx 4.8 \TeV$ can be excluded at $14 \TeV$ with $300 {\rm \, fb}^{-1}$, while the increase in luminosity to $3000  {\rm \, fb}^{-1}$ only modestly improves  the exclusion limit on  $M_{G'_{\rm KK}} $. Table \ref{tab:mass_sensitivity} summarizes the results for $2 \sigma, 3\sigma$ and $5 \sigma$ reach.

\begin{table}
\begin{center}
\begin{tabular}{|c|c|c|c|}
\hline
Luminosity & CL & Case 1 & Case 2 \\
\hline
\hline
&$2\sigma$ & 4.8 TeV & 3.6 TeV \\
$300  {\rm \, fb}^{-1}$ &$3\sigma$ & 3.8 TeV & $<$ 3 TeV\\
&$5\sigma$ &3.2 TeV& $ <$ 3 TeV\\
\hline

\hline
&$2\sigma$ & 5.1 TeV & 4.1 TeV \\
$3000  {\rm \, fb}^{-1}$ &$3\sigma$ &4.4 TeV & 3.5 TeV\\
&$5\sigma$ &3.5 TeV& $ <$ 3 TeV\\
\hline

\end{tabular}

\end{center}
\caption{Mass reach of the $\sqrt{s} =14 \TeV$ run search for RS KK gluons for $2, 3$ and $5 \sigma$ confidence intervals. All results include $5\%$ systematics. }
\label{tab:mass_sensitivity}
\end{table}

\begin{figure}[htb]
\begin{center}
\includegraphics[width=3.5in]{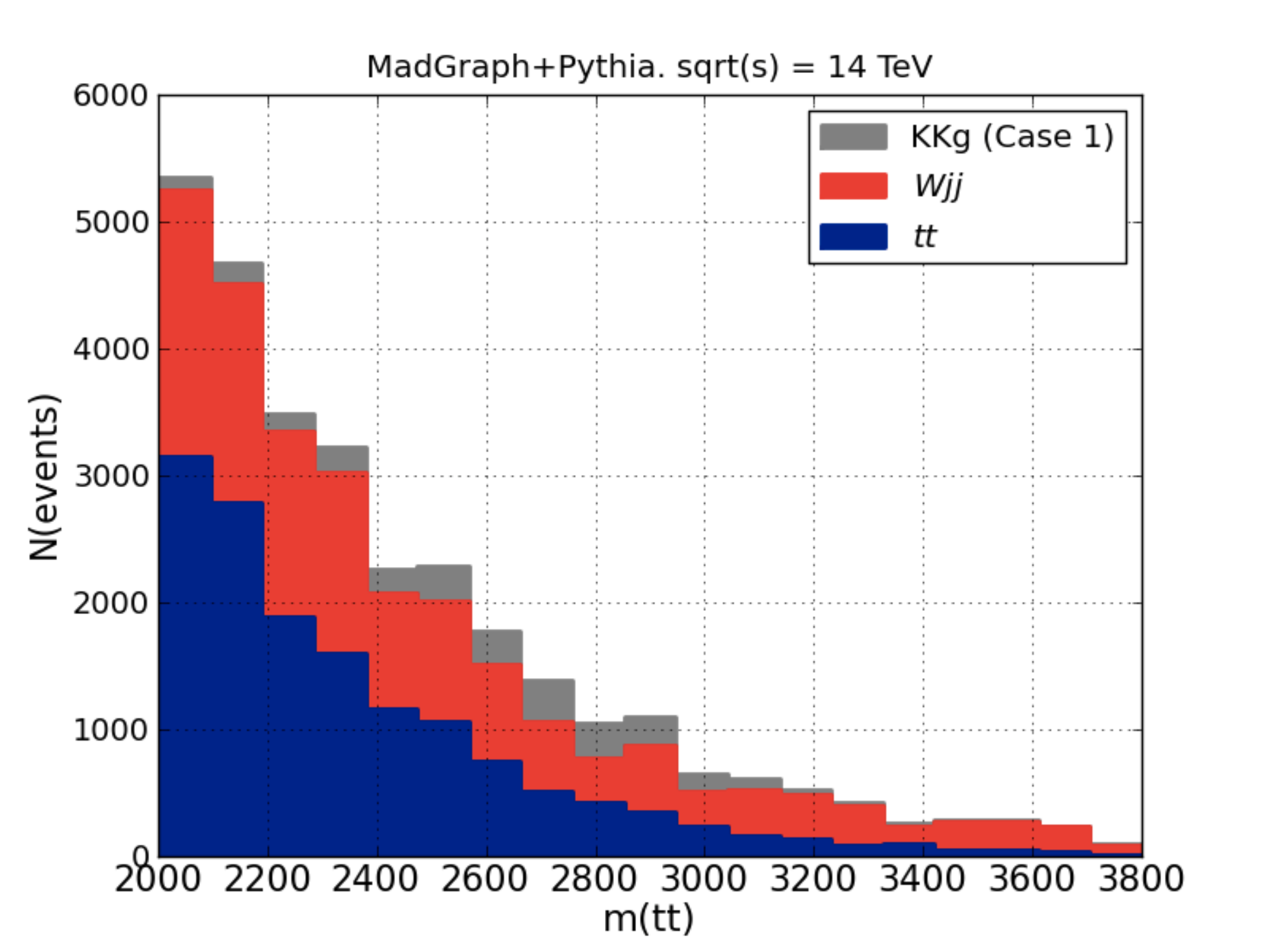}\includegraphics[width=3.5in]{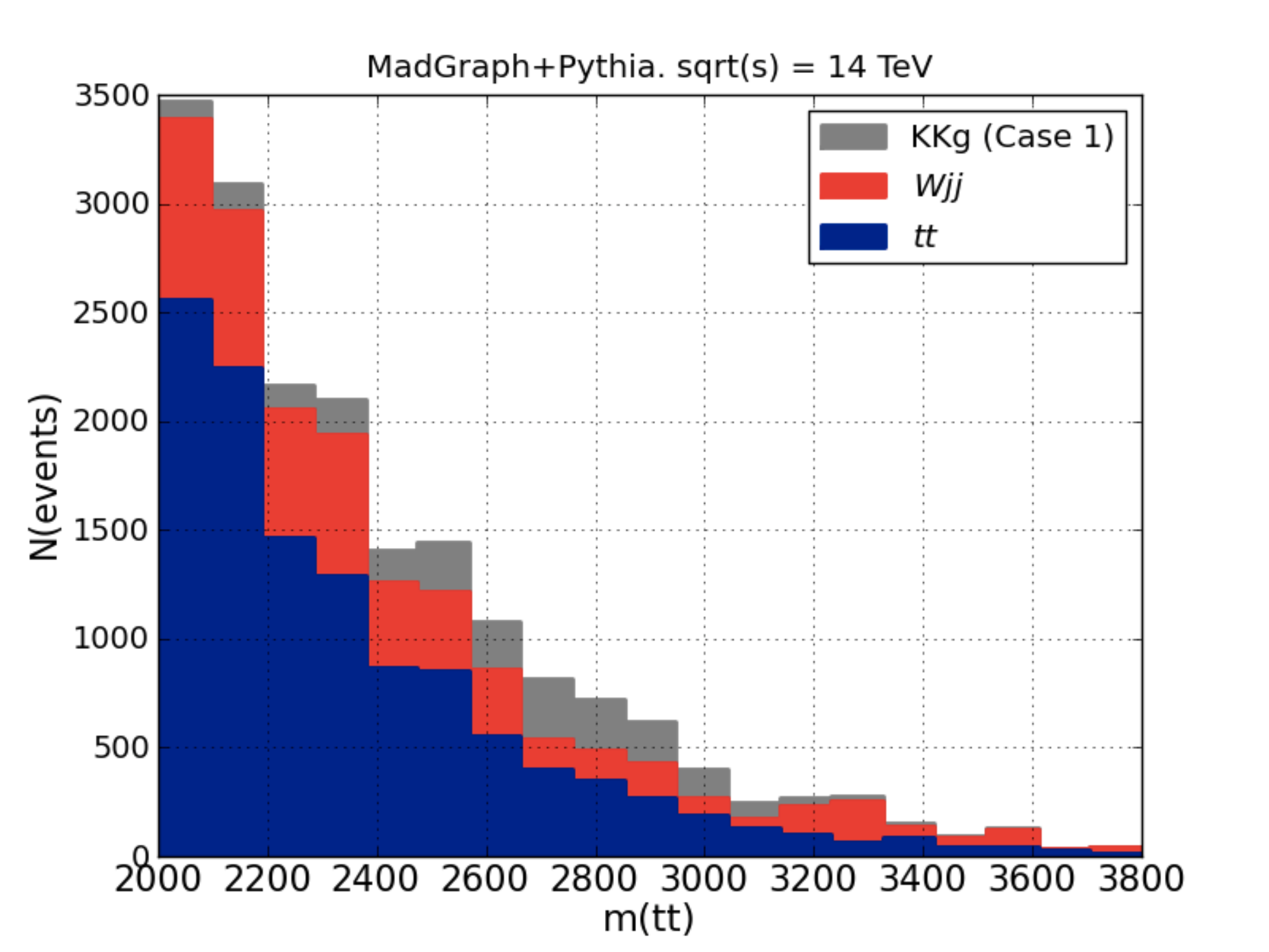}
\caption{Invariant mass of the peak hadronic template and the peak leptonic template with basic cuts (left) and additional overlap cuts (right). }
\label{fig:ov3gkk}
\end{center}
\end{figure}

Although the prospects for discovery of a KK gluon with mass of several TeV using jet substructure are good, the mass measurement is a more challenging task. Fig. \ref{fig:ov3gkk} shows an illustration. Strongly coupled resonances are typically characterized by large total decay widths, which in combination with the effects of PDF broadening result in a signal, which is smeared over a wide range of $m_{T\bar{T}}$. The mass measurement using a ``bump fitting'' technique will thus be challenging. Without the use of TOM, the measurement of $m_{T\bar{T}}$ is further complicated by the fact the Basic Cuts are sensitive to higher order effects (i.e. a hard gluon emission from a top quark) which further contribute to the smearing of the mass peak.  

\begin{table}[htb]

\begin{tabular}{c}
Case 1, $\langle N_{vtx} \rangle = 0$, $M_{G'_{\rm KK}}  = 3 \TeV$ \\
\small
\begin{tabular}{|c|c|c|c|c|c|c|c|c|c|}
\hline
Cuts & $\sigma_{t\bar{t} }({\rm fb})$ &$ \epsilon_{t\bar{t}}$& $\sigma_{Wjj}({\rm fb}) $& $\epsilon_{Wjj}$ &$ \sigma_{\rm KK}$ & $\epsilon_{\rm KK}$ & $S/B$ & $S/\sqrt{B} (300 {\rm \, fb}^{-1}) $& $S/\sqrt{B}(3000 {\rm \, fb}^{-1})$ \\
\hline

Basic Cuts &12.2 & 1.00 & 121.0 & 1.00 & 3.3 & 1.00 & 0.02 & 4.9 & 15.5 \\
%\& $Ov_3^h > 0.5 $ & 1231.0 & 0.72 & 791.0 & 0.16 & 14.6 & 0.85 & 0.006 & 4.7 & 15.0 \\
%\& $Ov_3^h + tPf > 1.0$ & 1067.0 & 0.63 & 433.0 & 0.09 & 11.1 & 0.76 & 0.007 & 4.9 & 15.6 \\
\& $Ov$ cuts  & 4.1 & 0.34 & 3.9 & 0.03 & 2.3 & 0.70& 0.30 & 14.3 & 45.4 \\
\hline

\end{tabular}
\end{tabular}\\\\

\begin{tabular}{c}
Case 1, $\langle N_{vtx} \rangle = 50$, $M_{G'_{\rm KK}}  = 3 \TeV$ \\
\small
\begin{tabular}{|c|c|c|c|c|c|c|c|c|c|}
\hline
Cuts & $\sigma_{t\bar{t} }({\rm fb})$ &$ \epsilon_{t\bar{t}}$& $\sigma_{Wjj}({\rm fb}) $& $\epsilon_{Wjj}$ &$ \sigma_{\rm KK}$ & $\epsilon_{\rm KK}$ & $S/B$ & $S/\sqrt{B} (300 {\rm \, fb}^{-1}) $& $S/\sqrt{B}(3000 {\rm \, fb}^{-1})$ \\
\hline
Basic Cuts  & 18.7 & 1.00 & 208.5  & 1.00 & 4.1 & 1.00  & 0.02 & 4.7 & 14.9 \\
%\& $Ov_3^h > 0.5$ &852 & 0.76 & 337.0 & 0.16 &11.6 & 0.85 & 0.01 & 5.8 & 18.4  \\
%\& $Ov_3^h + tPf > 1.0$ & 752.0 & 0.67 & 196.0 & 0.09 & 10.4 & 0.76 & 0.01 & 5.9 & 18.5 \\
%\& $tPf > 0.7$ & 73.0 &0.20 &15.0 & 0.009 & 1.5 & 0.22 \\
\& $Ov$ cuts & 5.2 & 0.25 & 5.2  &0.025 &  2.9 & 0.70 & 0.30 & 15.7 & 49.0 \\
\hline 
\end{tabular}
\end{tabular} \\\\ 

\begin{tabular}{c}
Case 1, $\langle N_{vtx} \rangle = 0$, $M_{G'_{\rm KK}}  = 5 \TeV$ \\
\small
\begin{tabular}{|c|c|c|c|c|c|c|c|c|c|}
\hline
Cuts & $\sigma_{t\bar{t} }({\rm fb})$ &$ \epsilon_{t\bar{t}}$& $\sigma_{Wjj}({\rm fb}) $& $\epsilon_{Wjj}$ &$ \sigma_{\rm KK}$ & $\epsilon_{\rm KK}$ & $S/B$ & $S/\sqrt{B} (300 {\rm \, fb}^{-1}) $& $S/\sqrt{B}(3000 {\rm \, fb}^{-1})$ \\
\hline

Basic Cuts &2.5 & 1.00 & 44.6 & 1.00 & 0.18 & 1.00 & 0.004 & 0.4 & 1.4\\
%\& $Ov_3^h > 0.5 $ & 1231.0 & 0.72 & 791.0 & 0.16 & 0.53 & 0.85 & $2\times 10^{-4}$ & 0.2 & 0.7\\
%\& $Ov_3^h + tPf > 1.0$ & 1067.0 & 0.63 & 433.0 & 0.09 & 0.50 & 0.77 & $2 \times 10^{-4}$ & 0.2 & 0.7 \\
\& $Ov$ cuts  & 0.6 & 0.33 & 1.3 & 0.03 & 0.12 & 0.70 & 0.07 & 1.5 & 4.8  \\
\hline

\end{tabular}
\end{tabular} \\\\

\begin{tabular}{c}
Case 1, $\langle N_{vtx} \rangle = 50$, $M_{G'_{\rm KK}}  = 5 \TeV$ \\
\small
\begin{tabular}{|c|c|c|c|c|c|c|c|c|c|}
\hline
Cuts & $\sigma_{t\bar{t} }({\rm fb})$ &$ \epsilon_{t\bar{t}}$& $\sigma_{Wjj}({\rm fb}) $& $\epsilon_{Wjj}$ &$ \sigma_{\rm KK}$ & $\epsilon_{\rm KK}$ & $S/B$ & $S/\sqrt{B} (300 {\rm \, fb}^{-1}) $& $S/\sqrt{B}(3000 {\rm \, fb}^{-1})$ \\
\hline

Basic Cuts  & 4.2 & 1.00 & 73.0  & 1.00  & 0.18 & 1.00  & $0.002$ &  0.4 & 1.1  \\
%\& $Ov_3^h > 0.5$ &852 & 0.76 & 337.0 & 0.16 &0.50 & 0.86 & $2\times  10^{-4}$& 0.2 & 0.7  \\
%\& $Ov_3^h + tPf > 1.0$ & 752.0 & 0.67 & 196.0 & 0.09  & 0.48 & 0.78 &$3 \times 10^{-4} $ & 0.2 & 0.7\\
%\& $tPf > 0.7$ & 73.0 &0.20 &15.0 & 0.009 & 1.5 & 0.22 \\
\& $Ov$ cuts & 0.9 & 0.27 & 2.0  &0.027 &  0.12 & 0.69 & $0.041$ &  1.3 &  4.1 \\
\hline 

\end{tabular}
\end{tabular}

\caption{Results for Case 1 KK gluons with  $M_{G'_{\rm KK}}  = 3, 5 \TeV$, without and with $N_{vtx} = 50$ of pileup. The tables show the signal and background cross-sections at $\sqrt{s} = 14 \TeV$ with the corresponding signal significance at $300 {\, \rm fb}^{-1}$ and $3000 {\, \rm fb}^{-1}$ of integrated luminosity. The basic cuts include the pre-selections of Eq. \ref{eq:bc} and a $m_{T \bar{T} }> 2.8 \TeV$ for $M_{G'_{\rm KK}}  = 3 \TeV$, and   $m_{T \bar{T} }> 3.5 \TeV$ for $M_{G'_{\rm KK}}  = 5 \TeV$ . The $Ov$ cuts include additional cuts of Eq. \ref{eq:ovcuts}. The apparent improved significance in the presence of pileup is due to the pileup sensitive lower $p_T$ cut on the fat jet. }
\label{tab:temp_results}
\end{table}

%CASE 2
\begin{table}[!]

%C2: Mkk = 3 TeV 0 p
\begin{tabular}{c}
Case 2: $\langle N_{vtx} \rangle = 0$, $M_{G'_{\rm KK}}  = 3 \TeV$ \\
\small
\begin{tabular}{|c|c|c|c|c|c|c|c|c|c|}
\hline
Cuts & $\sigma_{t\bar{t} }({\rm fb})$ &$ \epsilon_{t\bar{t}}$& $\sigma_{Wjj}({\rm fb}) $& $\epsilon_{Wjj}$ &$ \sigma_{\rm KK}$ & $\epsilon_{\rm KK}$ & $S/B$ & $S/\sqrt{B} (300 {\rm \, fb}^{-1}) $& $S/\sqrt{B}(3000 {\rm \, fb}^{-1})$ \\
\hline

Basic Cuts &12.2 & 1.00 & 121.0 & 1.00&  1.38 & 1.00  & $0.01$ &  2.1 & 6.6   \\
%\& $Ov_3^h > 0.5$ &852 & 0.76 & 337.0 & 0.16 &0.50 & 0.86 & $2\times  10^{-4}$& 0.2 & 0.7  \\
%\& $Ov_3^h + tPf > 1.0$ & 752.0 & 0.67 & 196.0 & 0.09  & 0.48 & 0.78 &$3 \times 10^{-4} $ & 0.2 & 0.7\\
%\& $tPf > 0.7$ & 73.0 &0.20 &15.0 & 0.009 & 1.5 & 0.22 \\
\& $Ov$ cuts  & 4.1 & 0.34 & 3.9 & 0.03&  1.02 & 0.74 & $ $0.13 &  6.3 & 19.8  \\
\hline

\end{tabular}
\end{tabular} \\\\

%C2: Mkk = 5 TeV 0 p
\begin{tabular}{c}
Case 2: $\langle N_{vtx} \rangle = 0$, $M_{G'_{\rm KK}}  = 5 \TeV$ \\
\small
\begin{tabular}{|c|c|c|c|c|c|c|c|c|c|}
\hline
Cuts & $\sigma_{t\bar{t} }({\rm fb})$ &$ \epsilon_{t\bar{t}}$& $\sigma_{Wjj}({\rm fb}) $& $\epsilon_{Wjj}$ &$ \sigma_{\rm KK}$ & $\epsilon_{\rm KK}$ & $S/B$ & $S/\sqrt{B} (300 {\rm \, fb}^{-1}) $& $S/\sqrt{B}(3000 {\rm \, fb}^{-1})$ \\
\hline

Basic Cuts &2.5 & 1.00 & 44.6 & 1.00   &  0.17 & 1.00  & $0.001$ & 0.3 &0.8  \\
%\& $Ov_3^h > 0.5$ &852 & 0.76 & 337.0 & 0.16 &0.50 & 0.86 & $2\times  10^{-4}$& 0.2 & 0.7  \\
%\& $Ov_3^h + tPf > 1.0$ & 752.0 & 0.67 & 196.0 & 0.09  & 0.48 & 0.78 &$3 \times 10^{-4} $ & 0.2 & 0.7\\
%\& $tPf > 0.7$ & 73.0 &0.20 &15.0 & 0.009 & 1.5 & 0.22 \\
\& $Ov$ cuts  & 0.6 & 0.33 & 1.3 & 0.03& 0.12  & 0.72 & $0.01$ &0.7  &2.3  \\
\hline

\end{tabular}
\end{tabular} \\\\

\begin{tabular}{c}
Case 2:  $\langle N_{vtx} \rangle = 50$, $M_{G'_{\rm KK}}  = 3 \TeV$ \\
\small
\begin{tabular}{|c|c|c|c|c|c|c|c|c|c|}
\hline
Cuts & $\sigma_{t\bar{t} }({\rm fb})$ &$ \epsilon_{t\bar{t}}$& $\sigma_{Wjj}({\rm fb}) $& $\epsilon_{Wjj}$ &$ \sigma_{\rm KK}$ & $\epsilon_{\rm KK}$ & $S/B$ & $S/\sqrt{B} (300 {\rm \, fb}^{-1}) $& $S/\sqrt{B}(3000 {\rm \, fb}^{-1})$ \\
\hline

Basic Cuts  & 18.7 & 1.00 & 208.5  & 1.00 &  1.6 & 1.00  &0.007 &  1.8 &  5.8 \\
%\& $Ov_3^h > 0.5$ &852 & 0.76 & 337.0 & 0.16 &11.6 & 0.85 & 0.01 & 5.8 & 18.4  \\
%\& $Ov_3^h + tPf > 1.0$ & 752.0 & 0.67 & 196.0 & 0.09 & 10.4 & 0.76 & 0.01 & 5.9 & 18.5 \\
%\& $tPf > 0.7$ & 73.0 &0.20 &15.0 & 0.009 & 1.5 & 0.22 \\
\& $Ov$ cuts & 5.2 & 0.25 & 5.2  &0.025 & 1.2 & 0.74 &0.11 &  6.3 &  20.0   \\
\hline
\end{tabular}
\end{tabular}\\\\

% C2: Mkk = 5 TeV 50 p
\begin{tabular}{c}
Case 2: $\langle N_{vtx} \rangle = 50$, $M_{G'_{\rm KK}}  = 5 \TeV$ \\
\small
\begin{tabular}{|c|c|c|c|c|c|c|c|c|c|}
\hline
Cuts & $\sigma_{t\bar{t} }({\rm fb})$ &$ \epsilon_{t\bar{t}}$& $\sigma_{Wjj}({\rm fb}) $& $\epsilon_{Wjj}$ &$ \sigma_{\rm KK}$ & $\epsilon_{\rm KK}$ & $S/B$ & $S/\sqrt{B} (300 {\rm \, fb}^{-1}) $& $S/\sqrt{B}(3000 {\rm \, fb}^{-1})$ \\
\hline

Basic Cuts  & 4.2 & 1.00 & 73.0  & 1.00  &  0.17 & 1.00  & $7 \times 10^{-4}$ &0.2   & 0.6  \\
%\& $Ov_3^h > 0.5$ &852 & 0.76 & 337.0 & 0.16 &0.50 & 0.86 & $2\times  10^{-4}$& 0.2 & 0.7  \\
%\& $Ov_3^h + tPf > 1.0$ & 752.0 & 0.67 & 196.0 & 0.09  & 0.48 & 0.78 &$3 \times 10^{-4} $ & 0.2 & 0.7\\
%\& $tPf > 0.7$ & 73.0 &0.20 &15.0 & 0.009 & 1.5 & 0.22 \\
\& $Ov$ cuts & 0.9 & 0.27 & 2.0  &0.027 & 0.12  &0.72  & $0.01$ &  0.7 & 2.0\\
\hline

\end{tabular}
\end{tabular}

\caption{Results for Case 2 KK gluons with  $M_{G'_{\rm KK}}  = 3, 5 \TeV$, without and with $N_{vtx} = 50$ of pileup. The tables show the signal and background cross-sections at $\sqrt{s} = 14 \TeV$ with the corresponding signal significance at $300 {\, \rm fb}^{-1}$ and $3000 {\, \rm fb}^{-1}$ of integrated luminosity. The basic cuts include the pre-selections of Eq. \ref{eq:bc} and a $m_{T \bar{T} }> 2.8 \TeV$ for $M_{G'_{\rm KK}}  = 3 \TeV$, and   $m_{T \bar{T} }> 3.5 \TeV$ for $M_{G'_{\rm KK}}  = 5 \TeV$ . The $Ov$ cuts include additional cuts of Eq. \ref{eq:ovcuts}. The apparent improved significance in the presence of pileup is due to the pileup sensitive lower $p_T$ cut requirement on the fat jet. }
\label{tab:temp_results2}
\end{table}

\begin{table}
\begin{tabular}{c}
Case 2: $\langle N_{vtx} \rangle = 0 $, $M_{G'_{\rm KK}}  = 5 \TeV$ \\
\small
\begin{tabular}{|c|c|c|c|c|c|c|c|c|c|}
\hline
Cuts & $\sigma_{t\bar{t} }({\rm fb})$ &$ \epsilon_{t\bar{t}}$& $\sigma_{Wjj}({\rm fb}) $& $\epsilon_{Wjj}$ &$ \sigma_{\rm KK}$ & $\epsilon_{\rm KK}$ & $S/B$ & $S/\sqrt{B} (300 {\rm \, fb}^{-1}) $& $S/\sqrt{B}(3000 {\rm \, fb}^{-1})$ \\
\hline

Basic Cuts  &173.0 & 1.00 & 800.0  & 1.00  &  2.16 & 1.00  & $0.002$ &1.2  & 3.7  \\
%\& $Ov_3^h > 0.5$ &852 & 0.76 & 337.0 & 0.16 &0.50 & 0.86 & $2\times  10^{-4}$& 0.2 & 0.7  \\
%\& $Ov_3^h + tPf > 1.0$ & 752.0 & 0.67 & 196.0 & 0.09  & 0.48 & 0.78 &$3 \times 10^{-4} $ & 0.2 & 0.7\\
%\& $tPf > 0.7$ & 73.0 &0.20 &15.0 & 0.009 & 1.5 & 0.22 \\
\& $Ov$ cuts & 48.0 & 0.28 & 39.0  &0.04 & 1.26 &0.58  & $0.01$ &  2.4 & 7.4\\
\hline

\end{tabular}
\end{tabular}

\caption{Results for Case 2 KK gluons with  $M_{G'_{\rm KK}}  = 5 \TeV$. The table shows the signal and background cross-sections at $\sqrt{s} = 33 \TeV$ with the corresponding signal significance at $300 {\, \rm fb}^{-1}$ and $3000 {\, \rm fb}^{-1}$ of integrated luminosity. The basic cuts include the pre-selections of Eq. \ref{eq:bc} and and   $m_{T \bar{T} }> 3.5 \TeV$ for $M_{G'_{\rm KK}}  = 5 \TeV$ . The $Ov$ cuts include additional cuts of Eq. \ref{eq:ovcuts}.}
\label{tab:33TeVCase2}
\end{table}

\subsection{KK $Z$, $\gamma$  \protect\cite{Agashe:2007ki}}
\label{sec:KKZgamma}

Schematically, we get 
\bea
q \bar{q} \rightarrow & \hbox{KK} \; Z, \; \gamma \rightarrow & WW, \; Z h \; (\hbox{and} \; t \bar{t}, \; b \bar{b})
\eea

There are actually three neutral KK states (generically denoted by
$Z^{ \prime }$):
KK photon (denoted by $A_1$), KK $Z$, and a KK mode of an extra $U(1)$ (again, with no 
corresponding zero-mode).
The latter states mix after EWSB\footnote{In obtaining the values of couplings shown below, this mixing is evaluated 
using $2$ TeV KK mass: for more details of the dependence on KK mass, see references above.} and the mass eigenstates are 
denoted by $\tilde{Z}_1$ and $\tilde{Z}_{ X_1 }$, where $\tilde{Z}_1$ is ``mostly'' KK $Z$ and $\tilde{Z}_{ X_1 }$ is mostly the extra $U(1)$.

The dominant decay modes are to $t \bar{t}$, $WW$ and $Zh$, each with a 
cross-section of 
$\sim O(10)$ fb for a
$2$ TeV $Z^{ \prime }$ with a total decay width of $\sim 100$ GeV.
However, the $t \bar{t}$ channel can be swamped by KK gluon
$\rightarrow t \bar{t}$ if the $Z^{ \prime }$ and KK gluon have similar mass.

The couplings relevant for production (for both cases I, II) are  \\
\vspace{0.1in} 
\begin{center}
\begin{tabular}{|c||c|c|c|}
\hline 
& $A_1$
& $\tilde{Z}_{ X_1} $ & $\tilde{Z}_1$
\tabularnewline
\hline
\hline 
$u_L \overline{ u_L }$ & $-0.04$ & $-0.025$ & $-0.046$
\tabularnewline
\hline
$u_R \overline{ u_R }$ & $-0.04$ & $0.01$ & $0.018$
\tabularnewline
\hline
$d_L \overline{ d_L }$ & $0.02$ & $0.03$ & $0.055$
\tabularnewline
\hline
$d_R \overline{ d_R }$ & $0.02$ & $-0.005$ & $0.0092$
\tabularnewline
\hline 
\end{tabular}
\end{center}
\vspace{0.1in} 
The 
notation used is that the coupling $Z u_L \overline{ u_L }$ in SM is $g_Z \left( +1/2 - 2/3 \sin^2 \theta_W \right)$, 
with $g_Z \approx 0.74$, $\sin^2 \theta_W \approx 0.23$.

The decay into gauge bosons (for both cases I, II) is determined by  \\
\vspace{0.1in} 
\begin{center}
\begin{tabular}{|c||c|c|c|}
\hline 
& $A_1$
& $\tilde{Z}_{ X_1} $ & $\tilde{Z}_1$
\tabularnewline
\hline
\hline 
$WW$ & $-0.0062$ & $-0.0069$ & $-0.0078$
\tabularnewline
\hline 
$Zh$ & $0$ & $-67.3$ GeV & $482$ GeV
\tabularnewline
\hline
\end{tabular}
\end{center}
\vspace{0.1in} 
where the notation used is that the couplings $\gamma
WW$ and $ZWW$in SM are $-g \sin \theta_W$ and $-g \cos \theta_W$ (respectively),
with $g \approx 0.65$.

Finally, the decays to SM fermions are given by \\
\vspace{0.1in} 
\begin{center}
\begin{tabular}{|c||c|c|c|}
\hline 
& $A_1$
& $\tilde{Z}_{ X_1} $ & $\tilde{Z}_1$
\tabularnewline
\hline
\hline 
$t_L \overline{ t_L }$ & $0.21$ & $-0.32$ & $0.47$
\tabularnewline
\hline 
$b_L \overline{ b_L }$ & $-0.1$ & $-0.6$ & $-0.03$
\tabularnewline
\hline
$t_R \overline{ t_R }$ & $0.81$ & $-0.77$ & $-0.085$
\tabularnewline
\hline
\end{tabular}
\end{center}
\vspace{0.1in} 
and \\
\vspace{0.1in}
\begin{center}
\begin{tabular}{|c||c|c|c|}
\hline 
& $A_1$
& $\tilde{Z}_{ X_1} $ & $\tilde{Z}_1$
\tabularnewline
\hline
\hline 
$t_L \overline{ t_L }$ & $0.81$ & $-1.05$ & $1.72$
\tabularnewline
\hline 
$b_L \overline{ b_L }$ & $-0.4$ & $-2.1$ & $-0.24$
\tabularnewline
\hline
$t_R \overline{ t_R }$ & $0.21$ & $-0.22$ & $-0.008$
\tabularnewline
\hline
\end{tabular}
\end{center}

\vspace{0.1in}
\noindent
for both cases I, II, respectively.

Having summarized the couplings of KK $Z$, $\gamma$, we now move on to two specific searches that were studied
as part of the Snowmass 2013 process.

\subsubsection{KK $Z \to ZH$}

Samples of 10,000 signal events are generated 
using our implementation of the model in CalcHEP~\cite{CalcHEP}
with the first KK excitation $Z_{KK}$ forced to decay to the $ZH$
final state. $Z_{KK}$ masses of 2 and 3 \TeV\ are chosen and the
corresponding cross sections are shown in Table~\ref{signal_xsec}.
For background processes, $Z$+jets events are generated with
MadGraph 5.11~\cite{Alwall:2011uj} and Pythia 6.420~\cite{Sjostrand:2006za} for
parton shower and hadronization. Those events are produced separately for
one- and two-jet topologies, and light vs. heavy flavors,
see Table~\ref{bkg_xsec}. A minimum $p_T$ requirement is imposed on
the dilepton pair to enhance the available integrated luminosity
in the final signal region with a lower (higher) minimum value for the
study of 2 (3) \TeV\ $Z_{KK}$, corresponding to the upper (lower)
portions of Table~\ref{bkg_xsec}.

\begin{table}[h]
\caption{Cross sections for signal processes in $pp$ collisions
         at $\sqrt{s} = 14$ \TeV.}
\label{signal_xsec}
\centering
\vspace{0.2cm}
\begin{tabular}{|c||c|c|c|}
\hline
Process & mass [\TeV] & cross section [fb] & \# events
\tabularnewline
\hline
\hline
$pp \to Z_{KK} \to ZH$ & 2.0 & 19.1 & 10,000
\tabularnewline
\hline 
$pp \to Z_{KK} \to ZH$ & 3.0 & 2.40 & 10,000
\tabularnewline
\hline
\end{tabular}
\end{table}

\begin{table}[tb]
\caption{Cross sections for background processes in $pp$ collisions
         at $\sqrt{s} = 14$ \TeV. Jets $j$ include the following particles:
         $g, u, \bar{u}, d, \bar{d}, s, \bar{s}, c, \bar{c}$.
         The $pp \to Zb$ process includes both $Zb$
         and $Z\bar{b}$ final states.}
\label{bkg_xsec}
\centering
\vspace{0.2cm}
\begin{tabular}{|c||c|c|c|}
\hline
Process & min $p_T^{\ell\ell}$ [\TeV] & cross section [fb] & \# events
\tabularnewline
\hline
\hline 
$pp \to Zj$  & 0.5 & 72.6 & 100,000 \\
$pp \to Zjj$ & 0.5 & 132 & 100,000 \\
$pp \to Zb$  & 0.5 & 1.40 & 50,000 \\
$pp \to Z\bbbar$ & 0.4 & 4.51 & 50,000
\tabularnewline
\hline 
$pp \to Zj$  & 0.7 & 13.0 & 100,000 \\
$pp \to Zjj$ & 0.7 & 25.1 & 80,000 \\
$pp \to Zb$  & 0.7 & 0.172 & 50,000 \\
$pp \to Z\bbbar$ & 0.7 & 0.222 & 50,000
\tabularnewline
\hline 
\end{tabular}
\end{table}

  Signal samples are processed through Pythia 8.176~\cite{Pythia8}
to produce parton showers and hadronize partons. 
Final decays $Z \to e^+e^-$, $Z \to \mu^+\mu^-$ and
$H \to \bbbar$ are forced at this stage with a
Higgs mass set to 125~\GeV.
For the data analysis described below the following
branching ratios are assumed:
$BR(Z \to \ell^+\ell^-) = 0.067$ and
$BR(H \to \bbbar) = 0.588$.
Detector simulation is performed with Delphes 3.0.9~\cite{Delphes}
including the effects of 50 pileup collisions.

Signal event candidates are selected based on the following object
requirements:
\begin{itemize}
\item Electrons and muons have transverse momentum $p_T > 25$ \GeV and
      pseudorapidity $|\eta| < 2.5$;
\item Both are required to be isolated from other tracks in the event,
      i.e. the ratio between the sum of transverse 
      momenta of track within a cone of
      $\Delta R = \sqrt{(\Delta \eta)^2 + (\Delta \phi)^2} = 0.3$ and
      the lepton $p_T$ and is less than 10\%: $\Sigma p_T^{\rm tracks} / p_T < 0.1$,
      where the sum excludes all leptons of the same flavor as the candidate
      lepton; this is done to avoid removing $Z \to \ell\ell$ decays
      with a highly boosted lepton pair;
\item Jets reconstructed with the anti-$k_t$ algorithm with a
      distance parameter of 0.5 must have $p_T > 100$ \GeV\ and $|\eta| < 2.5$;
\item Jets must pass the loose $b$-tag requirements.
\end{itemize}

$Z$ candidates are formed from pairs of same-flavor leptons without
any requirement on their electric charge. In case there are more than
one candidate in each event, a single candidate is chosen by selecting
the lepton pair with mass closest to the $Z$ boson mass. Finally, this
pair is required to have a mass within 15 \GeV\ of the $Z$ boson mass.
The above selection has an efficiency of about 40\% for signal events.

\begin{figure}[htb]
\begin{center}
\includegraphics[width=0.49\linewidth]{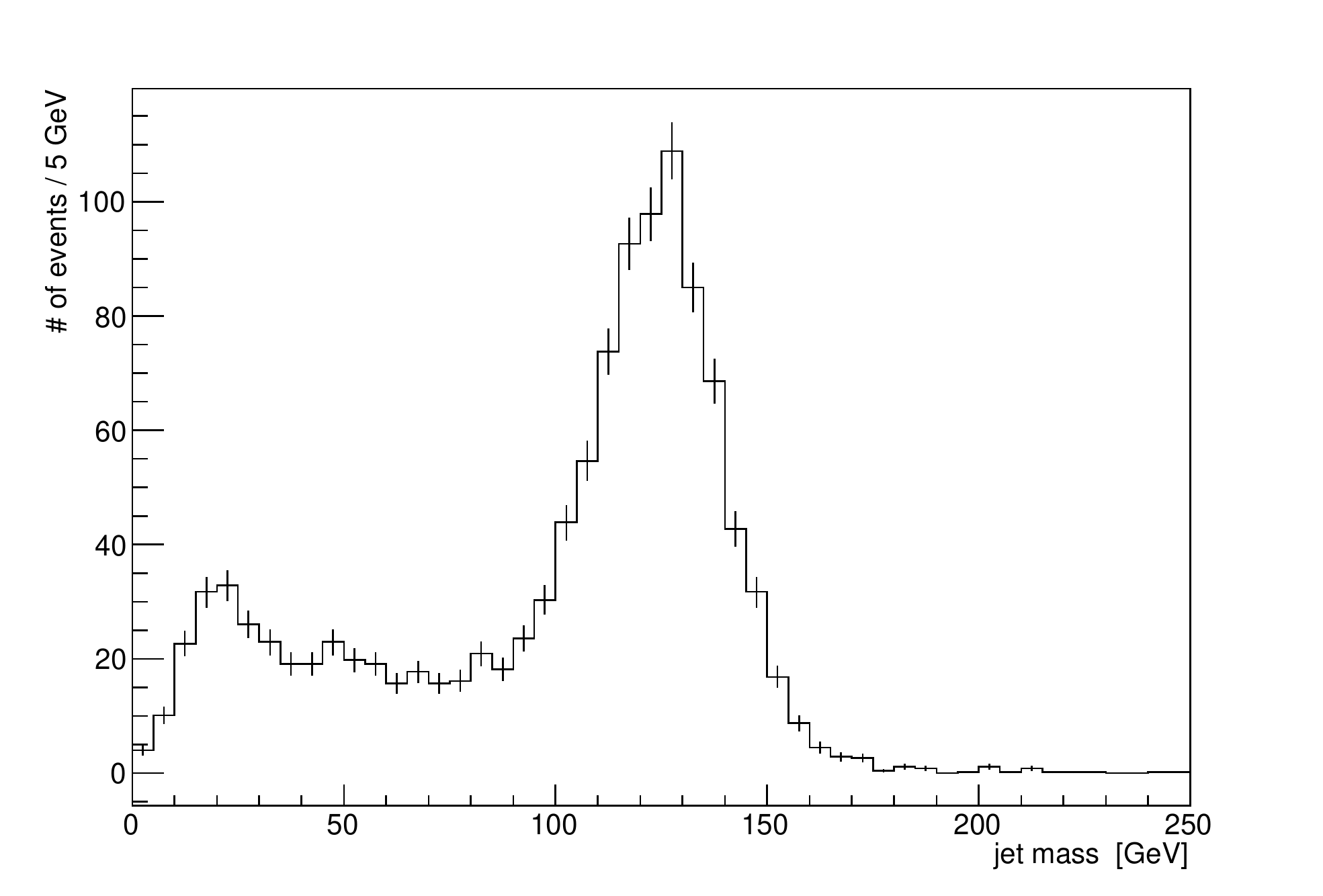}
\includegraphics[width=0.49\linewidth]{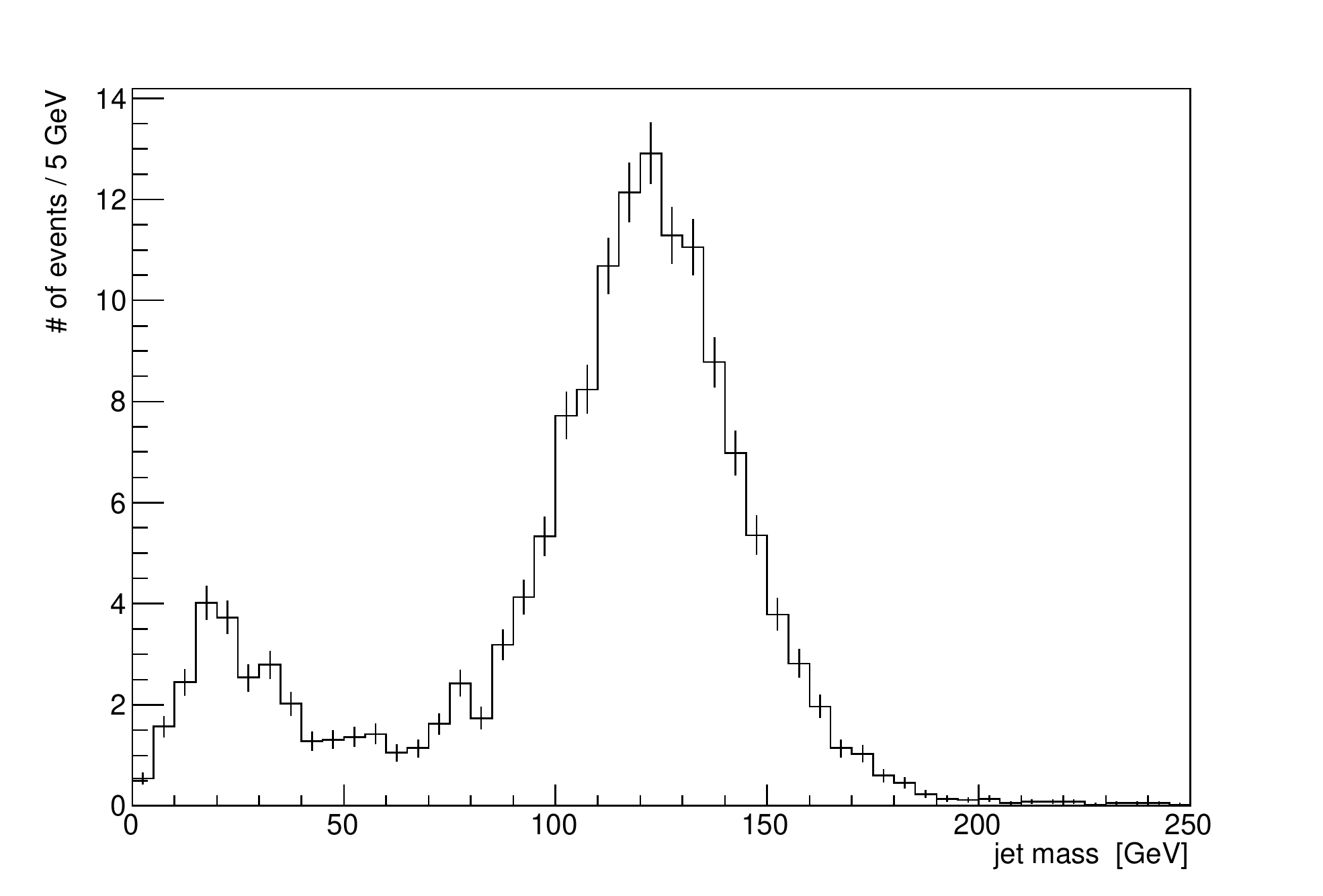}
\end{center}
\vspace{-.4cm}
\caption{Mass distribution for jets passing the initial selection:
         $p_T > 100$ \GeV, $|\eta| < 2.5$, and loose $b$ tag, for
         signal events with $Z_{KK}$ mass of 2 \TeV\ (left) and 3 \TeV\ (right).}
\label{fig_jetmass}
\end{figure}

$H$ candidates are identified from the set of $b$-tagged jets with a
jet mass within 20 (25) \GeV\ of the Higgs boson mass for
a $Z_{KK}$ mass of 2 (3) \TeV, respectively.
With a distance parameter $R = 0.5$, $H \to \bbbar$ final state
particless are expected to be merged into a single jet. 
For example, a Higgs boson
with $p_T = 1 \TeV$ would produce two $b$ quarks with an angular
separation $\Delta R = 2 m_H / p_T = 0.25$.
Jet mass distributions for signal decays of 2 and 3 \TeV\ masses
are presented in Fig.~\ref{fig_jetmass}.
The jet selection is about 31\% (33\%) efficient for signal events.

Finally, $ZH$ candidates must pass the following requirements:
\begin{itemize}
\item $Z$ candidate $p_T > 500$ \GeV;
\item $H$ candidate $p_T > 500$ \GeV;
\item cosine of the angle between the $Z$ and $H$ candidate momentum
      vectors is less than $-0.5$;
\item $ZH$ mass is within 200 (300) \GeV\ of the target $Z_{KK}$ mass
      for the 2 (3) \TeV\ $Z_{KK}$ cases.
\end{itemize}
The overall acceptance times efficiency for the full selection
is about 12\% (14\%) for $Z_{KK}$ mass of 2 (3) \TeV.

\begin{figure}[htb]
\begin{center}
\includegraphics[width=0.7\linewidth]{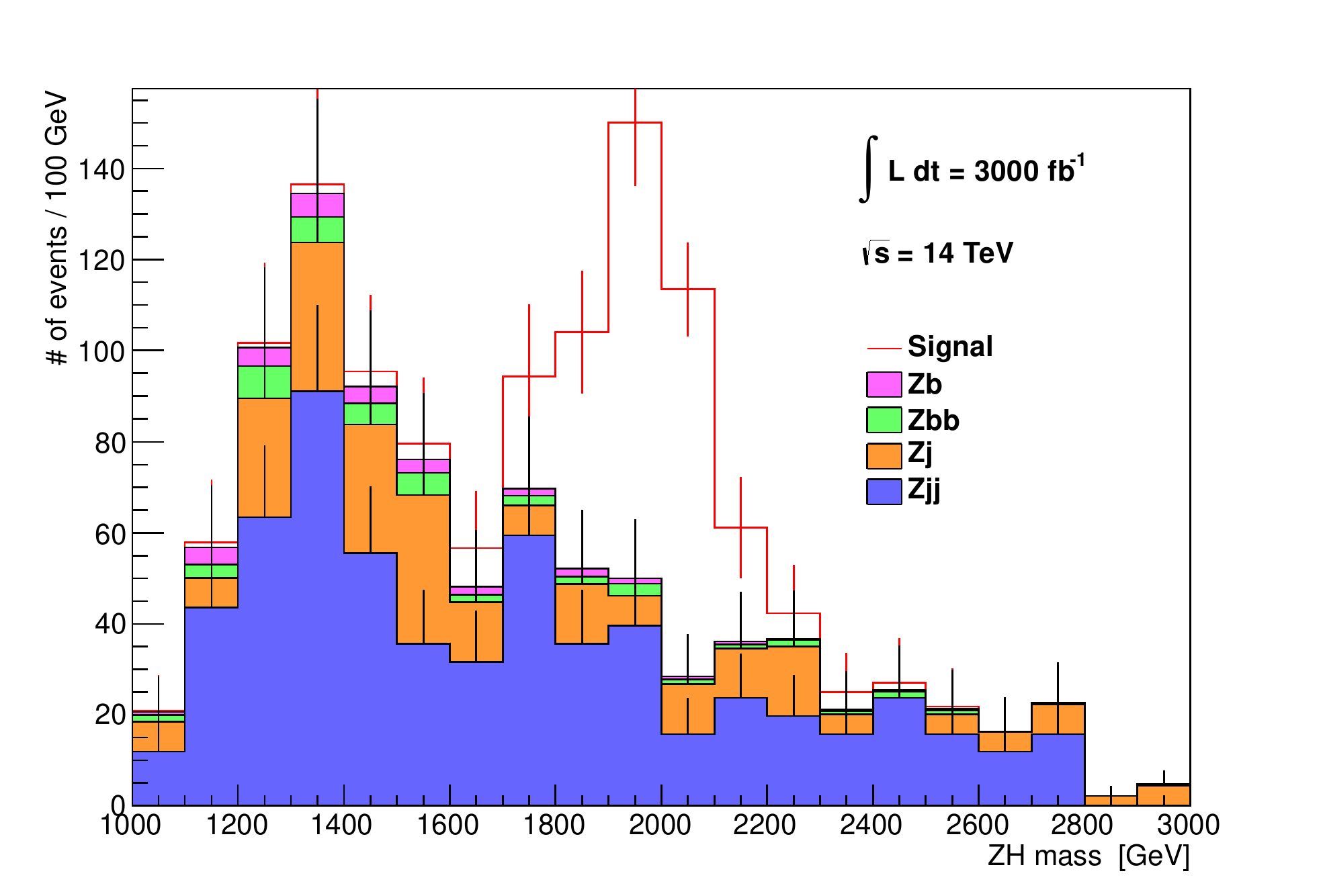}
\end{center}
\vspace{-.4cm}
\caption{Mass distribution for $ZH$ candidates passing the
         selection criteria for the $Z_{KK}$ mass of 2 \TeV, except
         for the mass requirement.}
\label{fig_ZHmass_2TeV}
\end{figure}

\begin{figure}[htb]
\begin{center}
\includegraphics[width=0.7\linewidth]{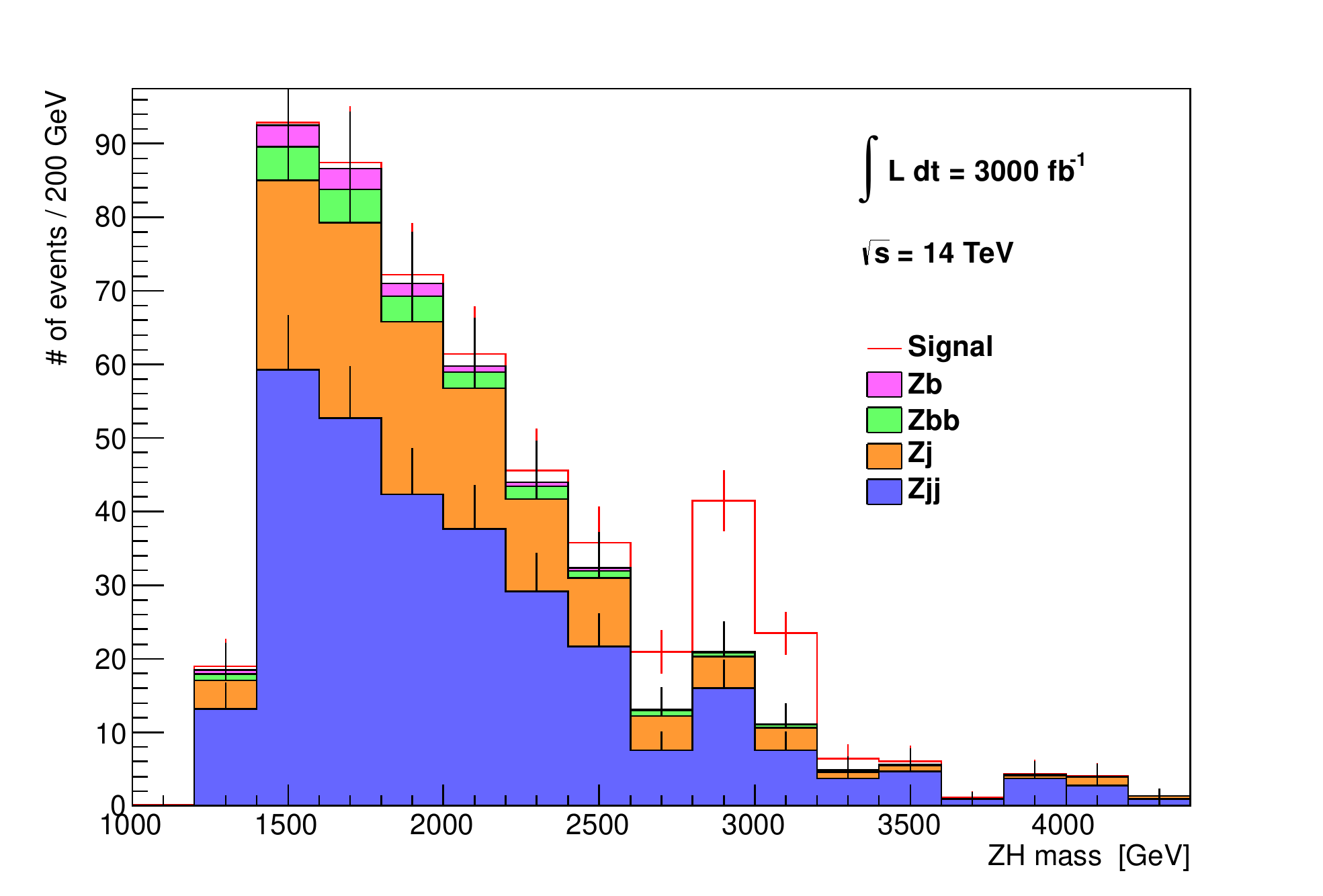}
\end{center}
\vspace{-.4cm}
\caption{Mass distribution for $ZH$ candidates passing the
         selection criteria for the $Z_{KK}$ mass of 3 \TeV, except
         for the mass requirement.}
\label{fig_ZHmass_3TeV}
\end{figure}

Figs.~\ref{fig_ZHmass_2TeV} and \ref{fig_ZHmass_3TeV} show 
the mass distributions for the $ZH$ candidates passing all selection criteria,
except the final mass window cut, in the $Z_{KK}$ mass = 2 (3) \TeV\ cases,
respectively.

Table~\ref{ZHyields} presents the signal and background event yields
after all selection requirements listed above are applied. Both
2 and 3 \TeV\ mass points can be observed with significance ($S/\sqrt{B}$) above
$5\sigma$ with an integrated luminosity of 3000~fb$^{-1}$. For the
2 \TeV\ case, 300~fb$^{-1}$ would be sufficient for a discovery with
a significance of 6.1.

\begin{table}[tb]
\caption{Event yields for signal and background processes in $pp$ collisions
         at $\sqrt{s} = 14$ \TeV\ with 3000~fb$^{-1}$ of integrated luminosity.}
\label{ZHyields}
\centering
\vspace{0.2cm}
\begin{tabular}{|c||c|c|}
\hline
Process & $Z_{KK} {\rm ~mass} = 2 \TeV$ &  $Z_{KK} {\rm ~mass} = 3 \TeV$
\tabularnewline
\hline
\hline
$pp \to Z_{KK} \to ZH$ & $264.2 \pm 7.7$ & $39.3 \pm 1.1$
\tabularnewline
\hline
$pp \to Zj$  &  $ 50 \pm 10 $ &  $10.5 \pm 2.0$ \\
$pp \to Zjj$ &  $ 123 \pm 22 $ &  $28 \pm 5$ \\
$pp \to Zb$  &  $ 4.3 \pm 0.6$ &  $0.32 \pm 0.06$ \\
$pp \to Z\bbbar$ &  $ 6.2 \pm 1.3$ & $1.49 \pm 0.14$
\tabularnewline
\hline
Total background & $183 \pm 24$ & $40.6 \pm 5.5$
\tabularnewline
\hline
Signal significance & 19.5 & 6.2
\tabularnewline
\hline
\end{tabular}
\end{table}

%*********************************************

\subsubsection{Reinterpretation of the search for narrow $\ttbar$ resonances}

The search for narrow $\ttbar$ resonances in the all-hadronic final state described in Sec.~\ref{sec:ttbar_allhadronic}
is reinterpreted in terms of limits on KK $Z^\prime$ and $Z_1$ bosons decaying to $\ttbar$.
Tab.~\ref{tab:Xsec_ZKK} shows the signal cross sections used in the reinterpretation.
The $k$-factor of 1.3 used in the leptophobic top-color $Z^\prime$ analysis is also used here, although it has only limited
validity in this scenario.

\begin{table}[h!]
\centering
\caption{Cross sections at 14 TeV considered in the analysis of the search for narrow $\ttbar$ resonances.}
\begin{tabular}{|c||c|c|}
\hline
scenario & $Z_{KK}$ mass & cross section
\tabularnewline
\hline
\hline
KK $Z^\prime$ & 2 \TeV & 17.2 fb \\
KK $Z^\prime$ & 3 \TeV & 1.8 fb \\
KK $Z^\prime$ & 4 \TeV & 0.31 fb \\
KK $Z^\prime$ & 5 \TeV & 0.075 fb
\tabularnewline
\hline
KK $Z_1$ & 2 \TeV & 3.43 fb \\
KK $Z_1$ & 3 \TeV & 0.42 fb \\
KK $Z_1$ & 4 \TeV & 0.074 fb \\
KK $Z_1$ & 5 \TeV & 0.017 fb
\tabularnewline
\hline
\end{tabular}
\label{tab:Xsec_ZKK}
\end{table}

\begin{figure}[!h]
\centering
  \includegraphics[width=0.47\textwidth]{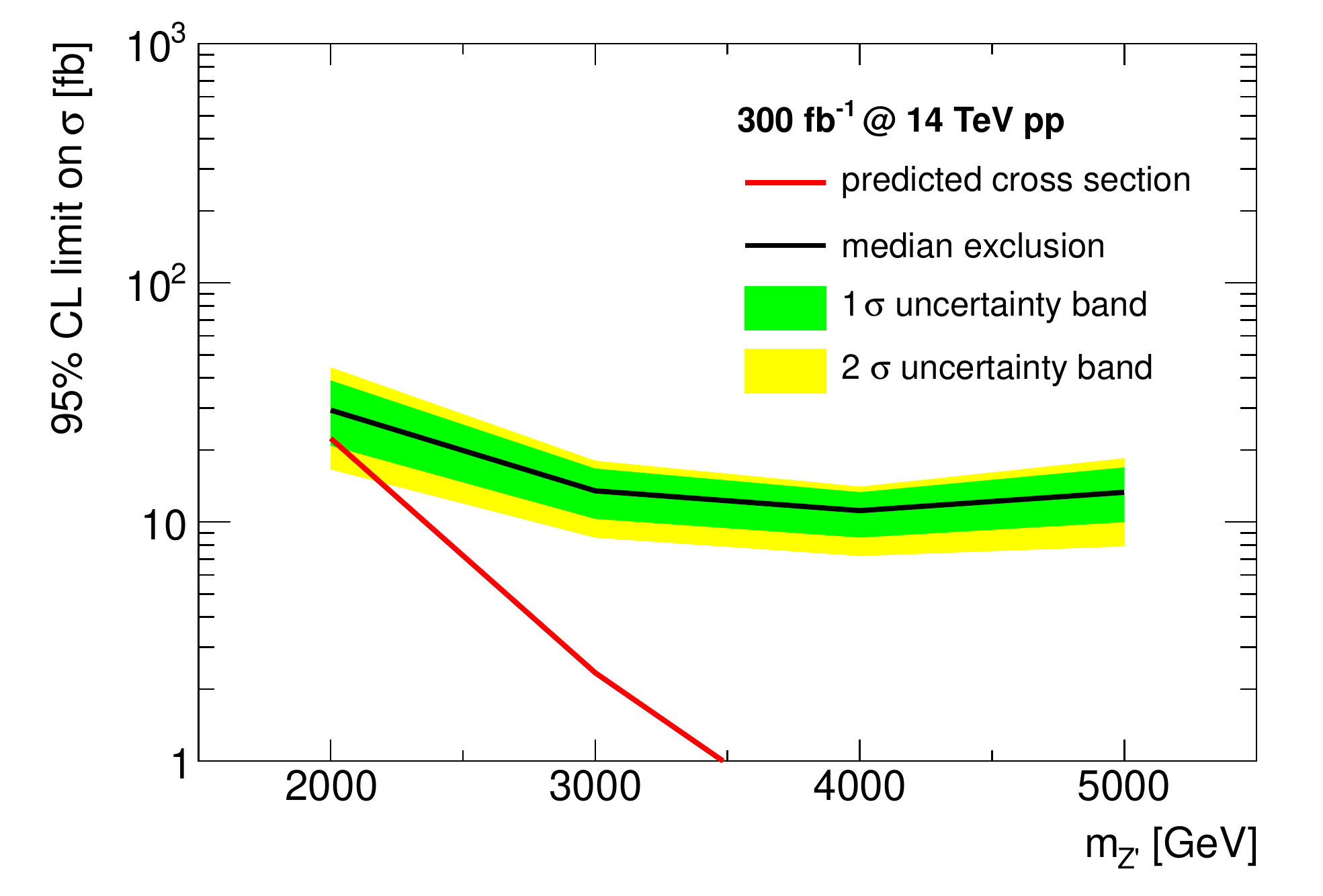}
  \includegraphics[width=0.47\textwidth]{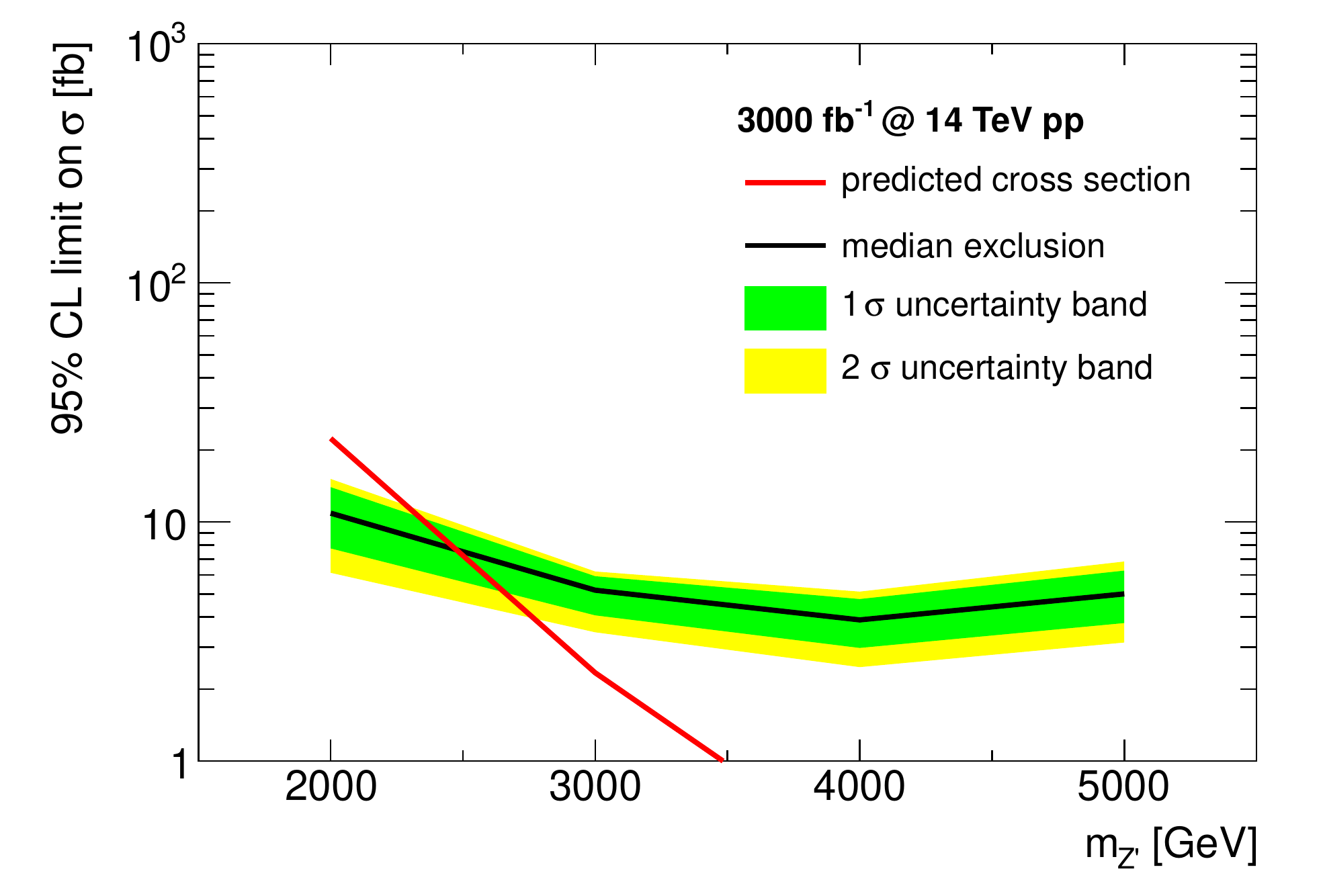} \\
  \includegraphics[width=0.47\textwidth]{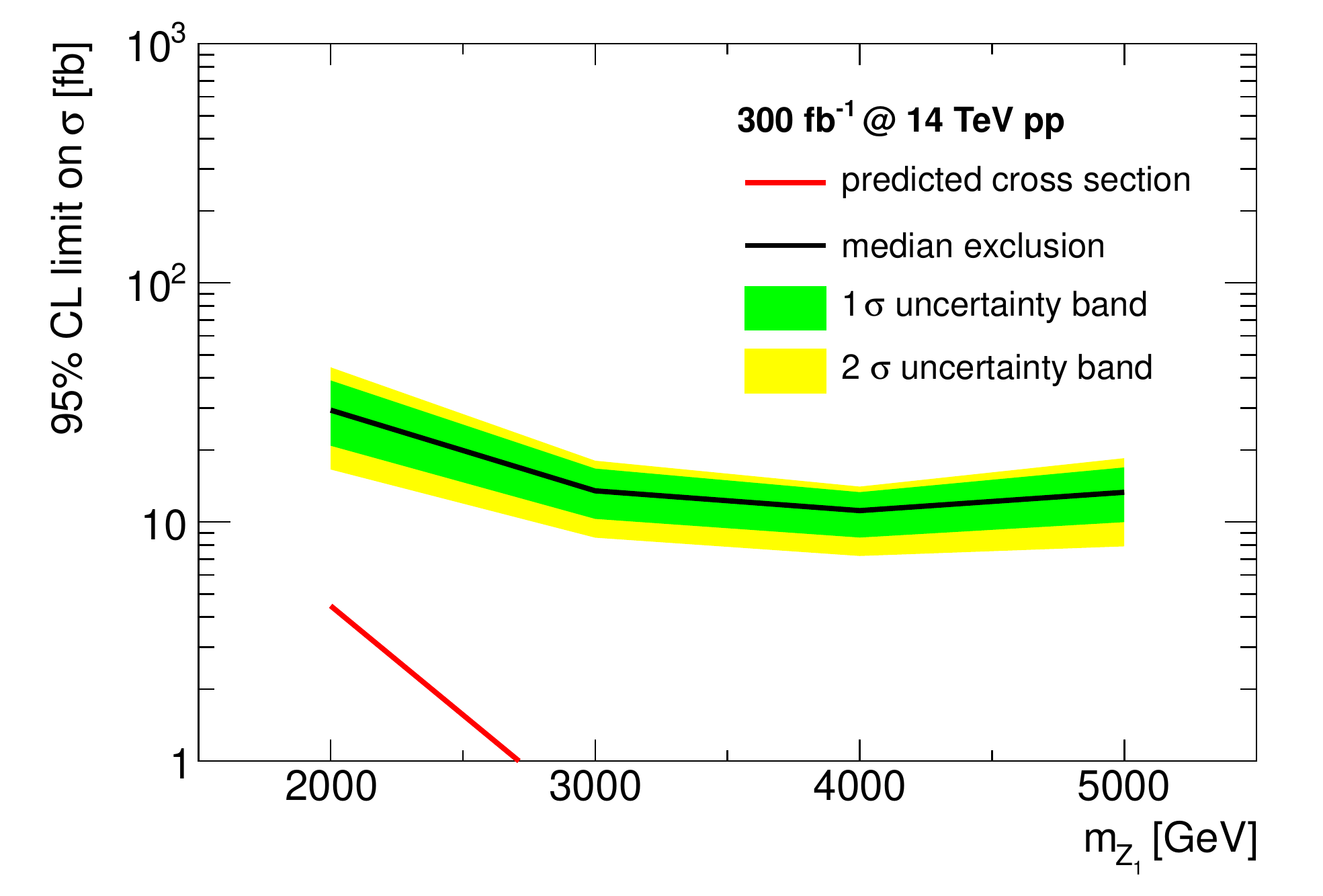}
  \includegraphics[width=0.47\textwidth]{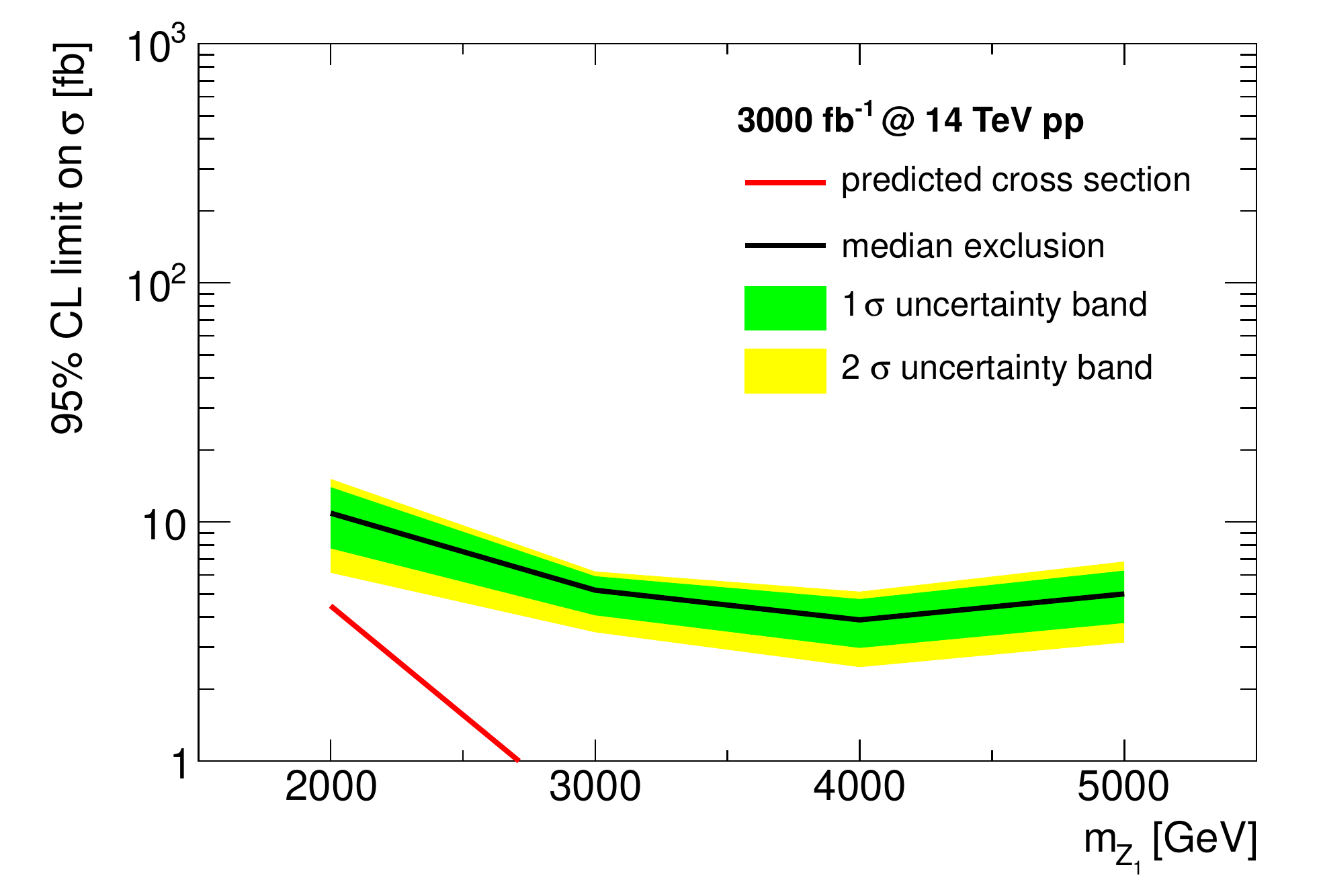}
\caption{
    Expected cross section upper limits at 95\% CL for the KK $Z^\prime$ and $Z_1$ models (upper and lower plots)
    decaying to a top quark pair in the all-hadronic channel. The left plots show the limits for 300~fb$^{-1}$ of 14 TeV data and $\mu = 50$.
    The right plots show the limits for 3000~fb$^{-1}$ of data and $\mu = 140$ (right).
}
\label{fig:KKglimit_ZKK}
\end{figure}

Fig.~\ref{fig:KKglimit_ZKK} shows the 95\% CL limits on the cross section of a narrow resonance overlaid with the expected cross sections for the KK models.
The mass limits for the KK $Z^\prime$ boson are given in Tab.~\ref{tab:Zprimelimits_ZKK} for the different pile-up and luminosity scenarios for statistical
uncertainties only and with systematic uncertainties included.
If the mass limit is denoted $< 2 \TeV$, the analysis is not sensitive to resonance mass above 2~TeV.
No mass limits are given for the $Z_1$ boson given the low predicted cross section for this process.
With 3000~fb$^{-1}$ of data at 14 TeV, KK $Z^\prime$ bosons can be excluded with masses up to 2.8~TeV using this analysis in the all-hadronic $\ttbar$ decay channel.

\begin{table}[h!]
\centering
\caption{95\% CL limits on the resonance mass for the KK $Z^\prime$ model decaying to a top quark pair in the all-hadronic channel.
 Limits are given for pile-up scenarios and integrated luminosities.
}
\begin{tabular}{|c||c|c|c|c|}
\hline
scenario & lumi. & pile-up & uncertainties & mass reach
\tabularnewline
\hline
\hline
KK $Z^\prime$ & 300 fb$^{-1}$ & $\mu = 0$ & stat.          & 2.6 TeV \\
KK $Z^\prime$ & 300 fb$^{-1}$ & $\mu = 0$ & stat.+syst.    & $<$ 2 TeV \\
KK $Z^\prime$ & 300 fb$^{-1}$ & $\mu = 50$ & stat.         & 2.6 TeV \\
KK $Z^\prime$ & 300 fb$^{-1}$ & $\mu = 50$ & stat.+syst.   & $<$ 2 TeV \\
KK $Z^\prime$ & 300 fb$^{-1}$ & $\mu = 140$ & stat.        & 2.5 TeV \\
KK $Z^\prime$ & 300 fb$^{-1}$ & $\mu = 140$ & stat.+syst.  & $<$ 2 TeV
\tabularnewline
\hline
KK $Z^\prime$ & 3000 fb$^{-1}$ & $\mu = 0$ & stat.         & 3.3 TeV \\
KK $Z^\prime$ & 3000 fb$^{-1}$ & $\mu = 0$ & stat.+syst.   & 2.9 TeV \\
KK $Z^\prime$ & 3000 fb$^{-1}$ & $\mu = 50$ & stat.        & 3.2 TeV \\
KK $Z^\prime$ & 3000 fb$^{-1}$ & $\mu = 50$ & stat.+syst.  & 2.9 TeV \\
KK $Z^\prime$ & 3000 fb$^{-1}$ & $\mu = 140$ & stat.       & 3.2 TeV \\
KK $Z^\prime$ & 3000 fb$^{-1}$ & $\mu = 140$ & stat.+syst. & 2.8 TeV
\tabularnewline
\hline
\end{tabular}
\label{tab:Zprimelimits_ZKK}
\end{table}

\subsection{KK $W$ \protect\cite{Agashe:2008jb}}

Here, we get
\bea
q^{ \prime } \bar{q} \rightarrow & 
\hbox{KK} \; W \rightarrow & WZ, \; W h (\hbox{and} \; t \bar{b})
\eea

It turns out that in addition to KK $W_L^+$, these models also have a KK $W_R^+$
(with no corresponding zero-mode), due to the custodial (i.e., extended $5D$ gauge) symmetry.
These two KK states mix after EWSB and the mass eigenstates
are denoted by 
$\tilde{W}^+_{ L 1 }$ and $\tilde{W}^+_{ R 1 }$, with former being mostly KK $W_L^+$.
The dominant decay modes for
$W^{ \prime }$ (generic notation) are into $WZ$ and $Wh$.
For a 
2 TeV $W^{ \prime }$, the cross-section $\sim O(10)$ fb each
with a total decay width of $\sim  O(100)$ GeV.
In some models, 
$W^{ \prime }$ decays to $t \bar{b}$ -- giving boosted top and bottom --
can also have similar cross-section. Interestingly, the process
KK gluon $\rightarrow t \bar{t}$ -- with KK
gluon mass being similar to $W^{ \prime }$ -- can be a significant background to this channel since a highly boosted top quark can fake a bottom quark: 
techniques similar to the ones used to identify highly boosted tops can now be applied to {\em veto} this possibility!

The production (for both cases I and II) goes via \\
\vspace{0.1in}
\begin{center}
\begin{tabular}{|c||c|c|}
\hline 
& $\tilde{W}^+_{ L_1 }$
& $\tilde{W}^+_{ R_1 }$
\tabularnewline
\hline
\hline 
$d_L \overline{ u_L }$ & $-0.074$ & $-0.055$
\tabularnewline
\hline 
\end{tabular}
\end{center}
\vspace{0.1in} 
where the notation used is that the coupling $d_L \overline{ u_L } W$ in SM is $g / \sqrt{2}$.

The decay to gauge bosons (for both cases I and II)
depend on \\
\vspace{0.1in} 
\begin{center}
\begin{tabular}{|c||c|c|}
\hline 
& $\tilde{W}^+_{ L_1 }$
& $\tilde{W}^+_{ R_1 }$
\tabularnewline
\hline
\hline 
$WZ$ & $-0.0026$ & $-0.0182$
\tabularnewline
\hline 
$Wh$ & $429$ GeV & $-61.3$ GeV
\tabularnewline
\hline
\end{tabular}
\end{center}
\vspace{0.1in} 
while 
that to SM fermions 
are given by \\
\vspace{0.1in}
\begin{center}
\begin{tabular}{|c||c|c|}
\hline 
& $\tilde{W}^+_{ L_1 }$
& $\tilde{W}^+_{ R_1 }$
\tabularnewline
\hline
\hline 
$b_L \overline{ t_L} $ & $0.37$ & $0.28$
\tabularnewline
\hline 
\end{tabular}
\end{center}
\vspace{0.1in} 
for case I and \\
\vspace{0.1in} 
\begin{center}
\begin{tabular}{|c||c|c|}
\hline 
& $\tilde{W}^+_{ L_1 }$
& $\tilde{W}^+_{ R_1 }$
\tabularnewline
\hline
\hline 
$b_L \overline{ t_L} $ & $1.44$ & $1.08$
\tabularnewline
\hline 
\end{tabular}
\end{center}
\vspace{0.1in} 
for case II.

Next, we 
describe a study of a specific decay channel for KK $W$.

\subsubsection{KK $W \rightarrow Wh$}

The discovery potential of the KK $W^+_{L}$ at the 3000~fb$^{-1}$ LHC is studied using the Wh decay channel where W bosons decays leptonically and H bosons decaying to $b\bar{b}$. 
The signal samples are generated using 
our implementation of the model in \texttt{CalcHep 3.4.2}~\cite{CalcHEP}.
The $W$+jets and $t\bar{t}$ events, the dominant SM background processes, are simulated with \texttt{Madgraph 5.11} \cite{Alwall:2011uj}.
Both signal and background events are then passed to \texttt{Pythia 6.420} \cite{Sjostrand:2006za} for showering and hadronization.  
The \texttt{Delphes 3.0.9}~\cite{Delphes} program is used with Snowmass fast detector simulation settings  for object reconstruction. Note that this study was done without pile-up. However, because of the high kinematic cuts we applied we expect that it should be relatively insensitive to pile-up effects.

For high KK $W^+_{L}$ masses, the produced Higgs bosons are highly boosted and the two bottom quarks from the hadronically decaying Higgs boson are expected to be reconstructed as one jet due to limited hadronic calorimeter resolution. Thus the final state topology consists of one isolated lepton ($e$ or $\mu$), large transverse missing energy and one highly energetic massive jet. The events are selected using the following criteria:
\begin{itemize}
 \item one and only one lepton: $p_{T}^l > 25\, \text{GeV}, |\eta^l|< 2.5, \Sigma p_{T}^{tracks}/p_{T}<0.1$
 \item one and only one b-jet: $p_{T}^j > 30\, \text{GeV}, |\eta^j|< 2.5$, loose b-tag
 \item transverse missing energy $> 200 \text{GeV}$.
\end{itemize}
The lepton is either electron or muon and the track isolation criteria is calculated within distance parameter 0.3. 
The jets are reconstructed using the anti-$k_t$ algorithm with a  distance parameter 0.5.
The DELPHES loose b-tag efficieincy is applied; it corresponds to 80\% efficiency for b-jets and 6\% efficiency for light jets.
The one and only one b-jet selection can significantly reduce the top background which usually produces multiple b-jets.

To further improve the signal purity, the additional Higgs candidate selection is applied to thet jet:
\begin{itemize}
\item $p_{T}^j > 1000+(M_{\text{KK} W}-3000)/2\, \text{GeV}$,
\item $100~\text{GeV} < m_j < 150~\text{GeV}$.
\end{itemize}
This selection is based on the fact that the Higgs decaying from the KK $W_{L}^+$ is highly boosted with large transverse momentum and so the jet, with distance parameter R=0.5, is expected to contain both decayed b-quarks.

Finally, the KK W candidate must pass the following additional criteria:
\begin{itemize}
\item W candidate $p_{T} > 500~\text{GeV}$,
\item $\lvert\phi_{l} - \phi_{missing~E_{T}}\rvert < 1$,
\item $\lvert\phi_{l} - \phi_{j}\rvert > 2.5$.
\item $2.4~\text{TeV} < m_{W,H} < 3.2~\text{TeV}~(2.9~\text{TeV} < m_{W,H} < 4.1~\text{TeV})$ for the 3 (4) TeV $W_{KK}$ cases.
\end{itemize}
The $\phi_{l,\nu}$ selection is based on the fact that the W is highly boosted and so the lepton and neutrino are expected to be highly collimated. The angle between the lepton and the missing energy in the xy-plane should be small. Furthermore, since the KK W has little transverse momentum, we expect the angle between the W and Higgs, and therefore the lepton and jet, in the xy-plane to be very large.

\begin{figure}[h]
  \centering
  \includegraphics[width=0.49\textwidth]{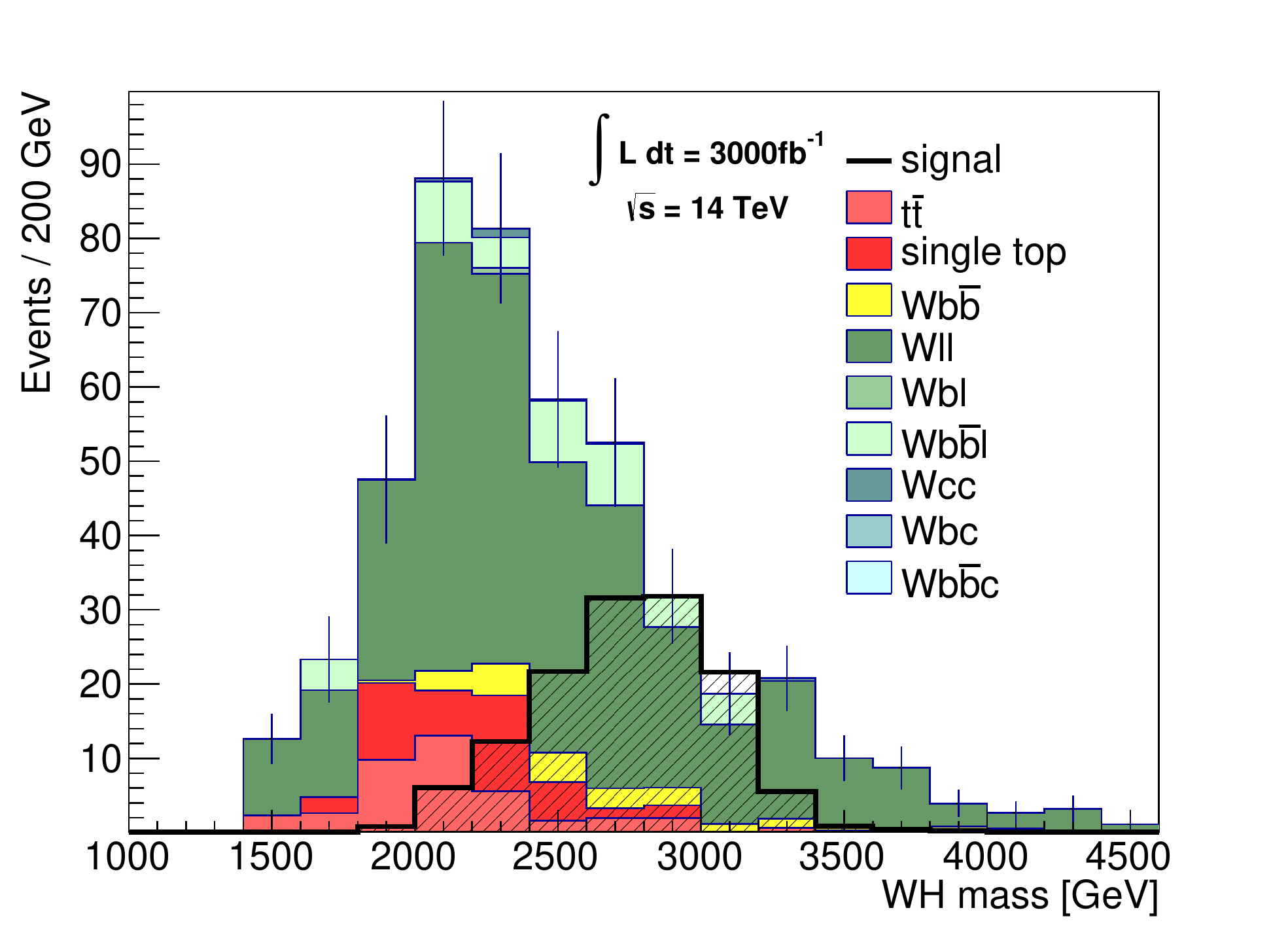}
  \includegraphics[width=0.49\textwidth]{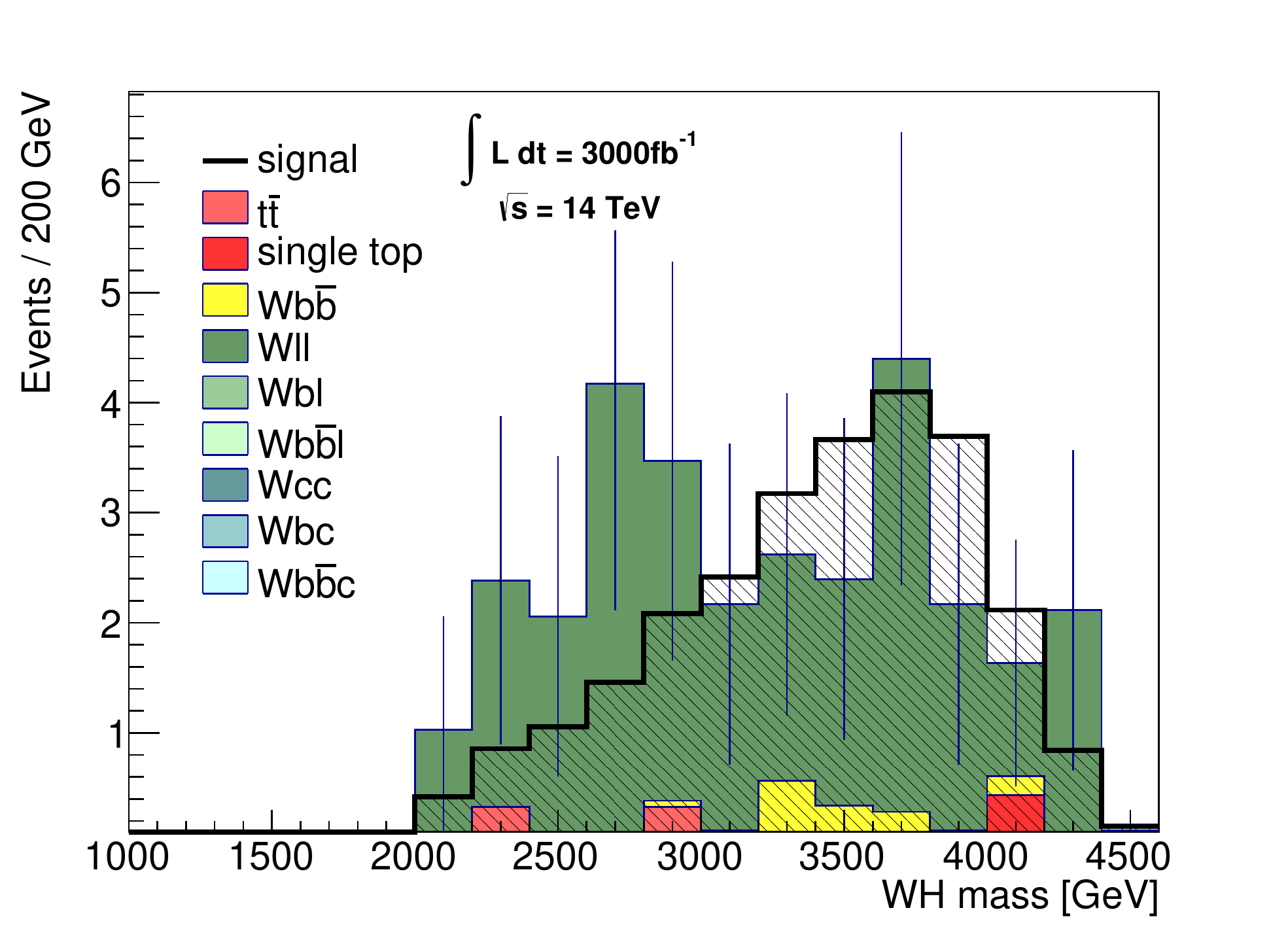}
  \caption{\label{fig:whMass}
The invariant mass distribution for the KK $W_{L}^{+}$ of mass 3TeV(left) and 4TeV(right) at $\sqrt{s}=14$ TeV
in the decay channel $pp\rightarrow W_{L}^+ \rightarrow  Wh \rightarrow \l^+\nu b\bar{b}$.
The neutrino momentum  is resolved using a $W$ mass constraint.
}
\end{figure}

Fig.\ref{fig:whMass} shows the invariant mass distribution of KK $W_{L}^+$ in the decay channel $pp\rightarrow \hbox{KK W} \rightarrow  Wh \rightarrow \l\nu b\bar{b}$ at the $\sqrt{s}=14$ TeV. The $W$ mass constraint is used to resolve the z-momentum component of the neutrino assuming it is the only source of the  missing energy. The higher $|p_{Z}^{\nu}|$ solution is selected from this quadratic equation although both solutions give similar $M_{l\nu b\bar{b}}$ distributions.

Table~\ref{tab:whres} presents the expected yields and observation significance at 3000~fb$^{-1}$ 14 TeV collisions.
The signal region is selected about one sigma of the mass resolution for each KK $W_L^+$ search hypothesis.
The mass resolution is driven 50\% from the natural width and another 50\% from the detector resolution.
The observation significance of the 3 TeV (4 TeV) KK $W_{L}^+$ is 5.4 (3.5) $\sigma$.

\begin{table}[h]
\centering
\begin{tabular}{|c||c|c|}
\hline
Process & $M_{\text{KK} W_L^+}=3$ TeV  & $M_{\text{KK} W_L^+}=4$ TeV \\
\hline 
w + light jets & 112.2 $\pm$ 10.7 & 13.4 $\pm$ 3.7 \\
w + mixed jets & 25.2 $\pm$ 10.1 & - \\
w + heavy jets & 10.2 $\pm$ 0.7 & 1.6 $\pm$ 0.3 \\
$t\bar{t}$ & 5.6 $\pm$ 1.4 & - \\ 
single top & 8.2 $\pm$ 3.4 & 0.4 $\pm$ 0.4 \\ 
\hline
total background & 161.4 $\pm$ 15.2 & 15.4 $\pm$ 3.8 \\ 
\hline
signal & 106.7 $\pm$ 3.0  & 19.2  $\pm$ 0.6  \\ 
\hline
significance & 5.4 & 3.5  \\ 
\hline
\end{tabular}
\caption{
 In the $pp \to \text{KK} W_L^{+} \to l \nu b\bar{b}$ process,
 the expected yields of signal and background process and the observation significance.
}
\label{tab:whres}
\end{table}

\subsection{Heavier KK fermions \protect\cite{Davoudiasl:2007wf}}

The KK fermions in the minimal model being $2$ TeV or heavier,
even single production of these particles can be very small (pair production is even smaller).

\subsection{Light KK fermions (a.k.a. ``Top partners") \protect\cite{Dennis:2007tv}}

As mentioned above, in non-minimal models, KK partners of top/bottom
can be light so that their
production (both pair and single, the latter 
perhaps in association with SM particles) can be significant. 
As these particles are ``top-like" with respect to their production at the LHC, the yields can be sizeable. For example, the pair production cross-section of a KK
bottom with mass of $500$ GeV is $\sim 1$ pb at $\sqrt{s} = 10$ TeV.
These particles decay into $t/b + W/Z/h$, where the  $W/Z$ can be 
somewhat
boosted at the LHC (even for fermionic KK
partners with masses as low as $\sim 500$ GeV). 
Some of these light KK fermions can have ``exotic" electric charges 
-- for example, $4/3$ and $5/3$. This makes them appealing with respect to a generic $b\prime/t\prime$ from, for example, a minimal extension of the number of
SM generations. 
For 
details, see references given above and Snowmass 2013 studies of ``Heavy fermions" under simplified models \cite{Snowmass}.

In addition, the other heavier (spin-1 or 2) KK modes can decay
into these light KK fermions, resulting in perhaps more distinctive
final states for the heavy KK's than the pairs of $W/Z$ or top quarks
that have been studied so far -- for 
such a study for KK gluon, see reference~\cite{Carena:2007tn}.

\section{Indirect KK effects}

In addition to signals from the {\em direct} production of the KK particles
at the LHC, there can also be effects from virtual exchange of
these KK particles on the properties of the SM
particles themselves. Again, details are given in the references.

\subsection{Top couplings to gauge bosons}

The shift in coupling of top quark 
to $Z$, {\em relative} to that in the SM, 
is given by (section 4.1 of \cite{Juste:2006sv})
\begin{eqnarray}
\delta \left( Z t \bar{t} \right) & \sim & \frac{ g_Z^2 k \pi r_c v^2 }{ M_{ \rm KK }^2 }, \;  \frac{ Y_5^2 v^2 }{ M_{ \rm KK }^2 }  \nonumber \\
& \stackrel{<}{\sim} & 10 \%, \; \hbox{for} \; M_{ \rm KK } \stackrel{>}{\sim} \; \hbox{a few TeV}  
\end{eqnarray}
Here, 
$Y_5$ denotes $5D$ Yukawa coupling in units of $k$.
Note that this holds for either 
RH or LH top quark, depending on case I or II, 
thereby allowing for experimental distinction between the two cases.

Similarly, we have 
\begin{eqnarray}
\delta \left( W t \bar{b} \right) & \sim & \frac{ g^2 k \pi r_c v^2 }{ M_{ \rm KK }^2 }, \;  \frac{ Y_5^2 v^2 }{ M_{ \rm KK }^2 }  \nonumber \\
& \stackrel{<}{\sim} & 10 \%, \; \hbox{for} \; M_{ \rm KK } \stackrel{>}{\sim} \; \hbox{a few TeV}
\end{eqnarray}
but only for case II (i.e., LH top/bottom being localized close to the TeV brane).

\subsection{Top-Higgs coupling}

The shift is given by \cite{Agashe:2009di}:
\begin{eqnarray}
\delta \left( h t \bar{t} \right)& \sim &  \frac{ Y_5^2 v^2 }{ M_{ \rm KK }^2 }
\end{eqnarray}
%

%\begin{eqnarray}
%\end{eqnarray}

\subsection{Higgs couplings to gauge bosons}

The shifts in Higgs couplings to 
$W$, $Z$ are as above for top quark (1st term there only). 
%
%\begin{eqnarray}
%\end{eqnarray}
%
For massless gauge bosons, we get
\cite{Azatov:2010pf}:
\begin{eqnarray}
\delta \left( h gg, \; h \gamma \gamma \right) & \sim &  \frac{ Y_5^2 v^2 }{ M_{ \rm KK }^2 }
\end{eqnarray}
It turns out that the above shifts (especially the latter ones) are [mostly $O(1)$]
different \cite{Falkowski:2007hz} for the case of ``gauge-Higgs unification" (where the Higgs boson is realized 
as $A_5$ \cite{Contino:2003ve}) as compared to the model where it is a mode of a $5D$ scalar field.

\subsection{Triple -gauge-couplings}

We estimate:
\begin{eqnarray}
\delta \left( WWZ \right) & \sim & \frac{g_Z^2 v^2 }{ M_{ \rm KK }^2 } \nonumber \\
& \stackrel{<}{\sim} & 0.1 \%, \; \hbox{for} \; M_{ \rm KK } \stackrel{>}{\sim} \; \hbox{a few TeV}  
\end{eqnarray}
the smallness of which is related (perhaps as expected in most theories) 
to that of precision EW observables (such as the $S$ parameter).

\section{Additional considerations}

Note that in general, 
brane-localized kinetic terms for $5D$ fields are allowed \cite{Carena:2002dz}, but we have assumed them
to be negligible here.
Similarly, the 
metric can deviate from pure AdS$_5$ near TeV brane \cite{Cabrer:2011qb}.
Both these cases result in O$(1)$ variations (relative to the above framework) of the phenomenology.
And, the UV-IR
hierarchy can be from flavor scale (only) of $\sim 1000$'s of TeV to EW scale \cite{Davoudiasl:2008hx}.
In this case, the phenomenology is qualitatively similar, but can be different by more than
$O(1)$.

The really/qualitatively different modifications are as follows.

\subsection{Flavor}

As mentioned above, we do need flavor symmetries to allow a few TeV KK scale
to be consistent with the relevant precision tests.
The 
signals outlined above are largely independent of the details of this implementation;
of course, any flavor-{\em violating} signals will depend on it.
In addition, flavor symmetries can allow the possibility of light fermions being localized 
near the TeV brane (while not explaining the flavor hierarchy as outlined in the above framework).
For these details, 
see companion note~\cite{flavor}.

\subsection{Dark Matter}

There is 
no candidate for dark matter (DM) in the minimal model presented above, for example, 
there is 
no KK--parity (unlike in UED).

However, in a well-motivated extension, it turns out that a DM candidate emerges naturally\footnote{The other possibility is to add a sector to the minimal model just for obtaining a DM candidate \cite{Ponton:2008zv}.}.
Namely, when we incorporate a GUT into the above framework, then suppressing proton decay 
gives a weakly-interacting massive particle (WIMP) DM candidate as a spin-off \cite{Agashe:2004ci}.
In this model, in addition to DM, there are (as is usual in WIMP models in general) 
{\em other} states charged under $Z_3$, but which are also charged SM: these have to 
decay into DM, plus SM: see section 15 of 2nd reference in \cite{Agashe:2004ci}.
In particular, the DM stabilization symmetry is
$Z_3$ (cf. $Z_2$ in UED or SUSY), which 
can have interesting consequences for DM signals: see references  \cite{Agashe:2010gt} for studies of this kind (in general).

\subsection{Gauge-Higgs unification}

Since this involves a further extension of the EW gauge symmetry,
one obtains
extra spin-1 particles, but these are typically heavier: see \cite{Agashe:2009bb} for a study\footnote{One also gets 
extra KK fermions which can result in quantitative changes to the phenomenology.}.

\section{Summary}
The framework of warped extra dimension is a well-motivated extension of the SM. In this whitepaper prepared for the Snowmass
2013 process, we have described searches at the LHC for the spin-1 and spin-2 KK particles in this framework. We find that the prospects
for discovery of these particles are quite promising, especially at the high-luminosity upgrade.
%
%In this Snowmass study, we have given an overview of LHC signals for a very well-motivated framework
%of SM fields propagating in a warped extra dimension. We have proposed two benchmarks with the right-handed (RH) or LH top quark, respectively,
%being localized very close to the TeV end of the extra dimension. We have presented some new results of spin-1 (gauge boson) and spin-2 (graviton) KK particles
%and their decays to top/bottom quarks (flavor-conserving) and W/Z and Higgs bosons, at the 14 TeV (with 300 fb$^{-1}$ and 3000 fb$^{-1}$) and 33 TeV LHC.
%In particular, we showed 3 and/or 5 sigma bounds on KK particles for high luminosity (14 TeV, 3000 fb$^{-1}$) compared to 300 fb$^{-1}$, as well as 33 TeV.
%It is also worth pointing out that some of these studies might be relevant for other types of models beyond the SM, which contain heavy particles decaying into top quarks, W, Z and Higgs.

\bigskip \noindent 
{\it Acknowledgments:} 
This work was supported in by the following grants (with respective authors): NSF Grant No.~PHY-0968854 (KA), Danish National Research Foundation under the contract number
DNRF90 (OA), Alexander von Humboldt Foundation under ``sgrants" (JE) ,the Department of Energy Office of Science and the Alfred P. Sloan Foundation and the Research Corporation for Science Advancement (TG), National Research Foundation of Korea(NRF) grant funded by the Korea government(MEST) N01120547 (SJL).

\end{document}